\documentclass[12pt]{article}
\usepackage{comment}
\usepackage{caption}
\usepackage{xspace}
\usepackage{graphicx}
\usepackage{amssymb}
 \usepackage{setspace}
\usepackage{xcolor}
\usepackage[margin=1.1in]{geometry}
\usepackage{booktabs}
\usepackage{amsmath}
\usepackage{multirow}
\usepackage[titletoc,title]{appendix}
\newcommand\reallywidehat[1]{%
\savestack{\tmpbox}{\stretchto{%
  \scaleto{%
    \scalerel*[\widthof{\ensuremath{#1}}]{\kern-.6pt\bigwedge\kern-.6pt}%
    {\rule[-\textheight/2]{1ex}{\textheight}}
  }{\textheight}%
}{0.5ex}}%
\stackon[1pt]{#1}{\tmpbox}%
}
\def\sym#1{\ifmmode^{#1}\else\(^{#1}\)\fi}

\usepackage{subcaption}

\usepackage[round]{natbib}


\usepackage{hyperref} 
\hypersetup{
colorlinks=true, breaklinks=true, bookmarksnumbered,
urlcolor=blue, linkcolor=blue, citecolor=blue, 
pdftitle={}, 
pdfauthor={\textcopyright}, 
pdfsubject={}, 
pdfkeywords={}, 
pdfcreator={pdfLaTeX}, 
pdfproducer={LaTeX with hyperref and ClassicThesis} 
}
\title{Religious Mayors, School Appointments, and Teenage Pregnancy}
\vspace{-.8cm}

\author{Marcela Mello\footnote{\scriptsize {Universidad de los Andes, Chile. Email: mmello@uandes.cl. }}  \and Jo\~ao Garcia\footnote{\scriptsize{Universidad de Santiago de Chile. Email: joao.garcia@usach.cl.  \newline
\textbf{Acknowledgement:} We are grateful to Dan Björkegren, Ernesto Dal Bó, Claudio Ferraz, Andrew Foster, Peter Hull, Brian Knight, Diana Moreira, and Bryce Steinberg for helpful comments and discussions. We are grateful to participants of RIDGE Political Economy Workshop, LACEA, GeFam, EESP-FGV seminar, PUC-Rio seminar, Insper seminar, PUC-Chile seminar, Universidad de Chile seminar, Universidad de Talca (Chile) seminar, Universidad de Santiago de Chile seminar, Brown bag seminar at Universidad de los Andes, and III Workshop on Economics of Education, Valle Nevado. 
} }}
\vspace{-.8cm}

\begin{document}

\maketitle
\vspace{-1.2cm}
\begin{abstract}
\thispagestyle{empty}
\footnotesize{
When religious movements win executive office, they can use bureaucratic levers to reshape public services along doctrinal lines. Using a regression discontinuity design on close mayoral elections in Brazil, we show that girls exposed to Pentecostal-party mayors during middle school experience birth rates 40\% higher, elevated STD rates, reduced HPV vaccination, and higher dropout rates; older cohorts already past school show no effects. The mechanism operates through personnel: these mayors replace school principals and reduce sexual education in municipal schools by 12.5 percentage points, with no changes in state schools outside their control. No effects emerge from other right-wing parties. }

 \noindent


\end{abstract}

\clearpage
\section{Introduction}

A growing body of evidence documents how religious institutions shape economic behavior through community norms, networks, and individual beliefs \citep{iannaccone1998introduction, guiso2003people, barro2003religion}. Less is known about what happens when religious movements gain political power and can use the tools of government to advance their agenda. A religious community that discourages premarital sex through social pressure operates through fundamentally different channels than a religious mayor who appoints school principals and sets the school curriculum. As Pentecostal and evangelical movements gain political ground across the developing world, understanding the governance consequences of religious political power becomes increasingly important.

As religious leaders take power, how do they enact policy? In decentralized systems, one answer is personnel discretion: local executives appoint school principals, clinic directors, and program coordinators who then alter the content of public services from the inside \citep{akhtari2022political, toral2024patronage, spenkuch2023ideology}. Yet existing work documents that political turnover of public servants affects service quality without identifying how specific ideological agendas are transmitted through personnel decisions. When a religious mayor replaces school principals who then remove sexual education from the curriculum, the appointment becomes a vehicle for translating religious doctrine into public policy, operating below the radar of legislative scrutiny or budgetary oversight. Causal evidence linking this full chain remains scarce, in part because separating the religious affiliation of political leaders from community characteristics is empirically challenging.

In this paper, we show that religious political power has significant consequences for adolescent health and human capital. Exploiting close elections in Brazil between 2008 and 2016, we find that electing a mayor from a Pentecostal-linked party causes a sharp increase in teenage pregnancy. For cohorts that spend their middle-school years under these administrations, birth rates rise by approximately 3 per 1,000—a 40 percent increase from a baseline of 7.5 per 1,000. These cohorts also exhibit elevated rates of syphilis infection, lower rates of HPV vaccination and higher school dropout rates. Crucially, we find no such effects for older girls who had already completed school when the administration began, suggesting the impact is driven by changes occurring within the school environment.

Brazil provides an unusually well-suited setting to address this challenge. Municipal mayors exercise substantial authority over local public schools, including appointing principals and determining the implementation of non-mandatory curriculum components such as sexual education. Two political parties, the \textit{Partido Social Crist\~ao} (PSC) and the \textit{Partido Republicano Brasileiro} (PRB), maintain explicit institutional ties to major Pentecostal denominations \citep{Lacerda2017, cammett2022religious} and are strongly associated with opposition to school-based sexual education \citep{Machado2017}, providing a transparent measure of religious political alignment. Brazil's large number of municipalities and competitive local elections produce many close races between Pentecostal and non-Pentecostal candidates, creating the variation needed for a credible regression discontinuity design.

We find evidence the mechanism operates through bureaucratic appointments. Pentecostal mayors replace school principals at higher rates, and municipal schools under their governance become 12.5 percentage points less likely to offer sexual education, while state schools in the same municipalities, where mayors lack administrative authority, show no change. Effects are concentrated in municipalities where schools were already offering sex education before the election, and do not vary with the local evangelical population share, pointing to institutional control of schools rather than diffusion of religious values in the community.

Other plausible channels find no support in the data. We find no effects on contraceptive availability at municipal health clinics, no reallocation of municipal budgets, no changes in primary care staffing, and no expansion of Pentecostal churches. The most visible change is in who runs the schools and what happens inside them, suggesting that policy change operates primarily through personnel discretion rather than through legislation, regulation, or spending.

These effects are specific to Pentecostal parties. Narrowly electing right-wing candidates from non-Pentecostal parties produces no effect on sexual education or teenage fertility, despite a sample four times larger. Brazil's party system is notoriously fragmented and ideologically fluid \citep{Mainwaring1999, Desposato2006}, but PRB and PSC represent a distinct institutional arrangement: they emerged directly from major Pentecostal denominations, recruit candidates through church networks, and function as political extensions of religious organizations rather than conventional parties \citep{oro2003church, Smith2019}. The combination of large effects on the outcomes these parties explicitly target, null effects for other conservative parties, and the institutional contrast between municipal and state schools supports the interpretation that Pentecostal party affiliation is a key element of the observed effects.

This paper contributes to the study of religion and governance. A growing quasi-experimental literature shows that the identity of political leaders affects policy choices and public goods provision \citep{chattopadhyay2004women, pande2003can, clots2012women, franck2012votes, brollo2016happens, iyer2016economics}. Within this strand, \citet{meyersson2014islamic} documents how Islamic mayors in Turkey shifted education infrastructure, and \citet{nellis2018secular} links secular party rule to reduced religious violence in Pakistan. We contribute a distinct mechanism: religious political power operating through \textit{bureaucratic appointments} that alter the content of public services, showing how religious ideology is transmitted through the administrative machinery of the state rather than through social influence alone. Our work complements \citet{costa2023stop, sola2023brother}, who document the electoral consequences of Pentecostal growth in Brazil, and \citet{mello2024religious}, who traces cultural channels linking Pentecostal media exposure to fertility.

A second contribution speaks to the political economy of decentralization. Classic arguments highlight both preference-matching benefits and capture risks when discretion is high \citep{oates1972fiscal, bardhan2000capture, bardhan2006decentralization}. Our findings reveal that personnel discretion can serve as an invisible channel of ideological influence, one that escapes the budgetary and legislative checks typically applied to local executives.

Finally, our results inform research on teenage fertility and sexual education. While experimental evidence from Kenya \citep{duflo2015education} and observational studies from the US \citep{carr2017teenage, stangerhall2011abstinence, fox2019funding} show that comprehensive sexual education reduces adolescent pregnancy and STIs, we provide quasi-experimental evidence on the reverse margin: politically induced removal of sexual education increases both, with cohort and school-sector comparisons that isolate the school-based channel.

\section{Background}

\subsection{Pentecostal Parties in Brazil}

Our identification strategy relies on classifying the \textit{Partido Republicano Brasileiro} (PRB, now Republicanos) and the \textit{Partido Social Cristão} (PSC) as Pentecostal parties. Among Brazilian political parties, these two stand out for the depth and institutional nature of their ties to Pentecostal churches, the disproportionate share of Pentecostal candidates and elected officials on their slates, and their role in advancing a distinctly Pentecostal political agenda. For comprehensive treatments of the relationship between Pentecostal churches and partisan politics in Brazil, see \citet{Lacerda2017}, \citet{Machado2012}, and \citet{cammett2022religious}.

The PRB is structurally linked to the \textit{Igreja Universal do Reino de Deus} (IURD, or Universal Church of the Kingdom of God), being widely described as the main partisan vehicle for IURD-backed candidacies. Although there are no formal legal links, the relationship works on a personalist basis. The party was founded in 2005 under direct leadership of IURD officials, and its organizational apparatus is tightly integrated with the church hierarchy \citep{Cerqueira2021}.

The PSC, in turn, has served as the primary electoral vehicle for the \textit{Assembleia de Deus} (Assembly of God), Brazil's largest Pentecostal denomination. While the Assembly of God's decentralized governance structure precludes the kind of top-down party control exercised by the IURD over the PRB, the connection operates through the systematic channeling of church-endorsed candidates through PSC slates, as well as through the party's explicit adoption of Pentecostal moral stances in its programmatic platform. The PSC's party statute enshrines positions closely aligned with Pentecostal doctrine, including the defense of life from conception \citep{PSCStatute2019}.

Both parties advance core elements of the Pentecostal political agenda in legislative arenas, particularly on issues related to sexual and reproductive morality, family structure, and religious education \citep{Machado2017, zilla2020evangelicals}. They also receive a comparatively high share of the Pentecostal vote, benefiting from church-based mobilization networks that provide electoral advantages unavailable to secular parties \citep{sola2023brother, cammett2022religious}.

To be clear, this classification does not imply that every PRB or PSC candidate is Pentecostal or fully aligned with Pentecostal ideology, nor that Pentecostal politicians are absent from other parties; indeed, individual Pentecostal legislators occupy prominent positions across the partisan spectrum \citep{Machado2017}. However, no other Brazilian party exhibits the same combination of institutional church linkages, systematic recruitment of Pentecostal candidates, programmatic alignment with Pentecostal doctrine, and privileged access to Pentecostal voter mobilization networks. The difference between PRB/PSC and other parties is both quantitative, in the concentration of Pentecostal members and candidates, and qualitative, in the organizational mechanisms that bind party and church. 

We provide direct evidence of this distinctiveness using TSE candidate registration data. Table \ref{tab:religion} reports the share of all municipal candidates (mayors, vice-mayors, and council members) who either list a religious ministry as their official occupation or use a religious title, such as \textit{Pastor}, \textit{Bispo}, or \textit{Reverendo}, in their campaign name. Approximately 2\% of PRB and PSC candidates display at least one of these religious signals, compared to 0.7\% among other right-wing parties and 0.4\% among PT and PMDB candidates, a three- to five-fold difference. Since many Pentecostal candidates do not adopt religious titles, these figures represent lower bounds on the true share of religiously motivated candidates, making the gap all the more striking. The pattern is even starker at the federal level: Table \ref{tab:congress} shows that approximately 70\% of PRB and PSC federal deputies elected between 2006 and 2018 are identified as evangelical by either \cite{lacerda2018assessing} or \citet{gomes2021deputados}, compared to about 10\% for other right-wing parties and 12.6 -- 13.3\% overall.

Both parties have repeatedly framed the regulation of sexual and moral education as a priority, often arguing that such topics should be governed primarily by families rather than schools. This framing is embedded not only in local discourse but also in party-linked legislative activity: at the federal level, PSC legislators held central agenda-setting roles in the main \textit{Escola sem Partido} proposal (PL 7180/2014), whose stated objective includes giving precedence to family values over school instruction in moral, sexual, and religious matters \citep{camara2017_audiencia_escolasempartido}. PRB legislators likewise actively promoted deliberation on the bill within the relevant special commission \citep{camaraBR_pl7180_ficha}. A representative municipal-level statement from PSC councilman Paulo Siufi illustrates the same logic: ``We want municipal schools to teach Portuguese, geography, mathematics, and science—and leave sexual education and religion to parents, and partisan politics to political actors. These topics don't need to be discussed in classrooms with children."\footnote{\textit{Esse projeto quer simplesmente manter a grade curricular, nós queremos que dentro das escolas municipais os professores ensinem português, geografia, matemática, ciências, e que eles deixem a educação sexual e a religião para os pais, e a política partidária para os agentes políticos. Esses temas não precisam ser discutidos em salas de aula com crianças.} \citep{camaraCG2016_siufi_escolasempartido}} Taken together, these party-linked statements and actions motivate clear predictions for local policy: when mayors affiliated with these parties take office, we should expect greater political pressure to restrict or displace school-based sexual education content toward parental control.

\subsection{School System and Mayoral Control}

Brazil's education system is highly decentralized and assigns distinct responsibilities to federal, state, and municipal governments. The federal government establishes broad guidelines through the \textit{Lei de Diretrizes e Bases da Educação} (LDB) and the \textit{Base Nacional Comum Curricular} (BNCC), but implementation varies substantially across levels of government. Municipal governments are responsible for early childhood, primary, and lower-secondary education (middle school, typically ages 10–15), while states are responsible for upper-secondary education (high school, typically ages 15–17).

The BNCC mandates core academic subjects but leaves significant discretion to local governments regarding ``transversal themes," which encompass health, citizenship, drug prevention, and sexuality. Sexual education falls within this non-mandatory category, meaning that its inclusion depends on municipal policy preferences and school-level decisions. As a result, municipalities differ markedly in whether and how they address sexual and reproductive health topics in the classroom, and these decisions can change with political turnover.

A central feature of the system and the key institutional lever for our analysis is the mayor's ability to appoint principals in municipal schools. Principal positions are classified as \textit{cargos de confiança} (positions of trust), granting mayors discretion to select individuals aligned with their political and ideological preferences. Approximately 40\% of principals in municipal schools reach their positions through political appointment rather than competitive selection or community election\footnote{Author's calculation using the \textit{Prova Brasil} survey.}. Since principals exert substantial influence over school programming, teacher assignments, and the interpretation of transversal themes, the appointment power provides a direct channel through which local political ideology can shape what is taught.

By contrast, mayors have no authority over state schools. Principal selection and curriculum decisions in state schools are handled exclusively by state-level authorities, insulating these schools from municipal political turnover. This institutional asymmetry is central to our identification strategy: any changes in the availability of sexual education programs following a Pentecostal party victory should be observed only in municipal schools, not in state schools located in the same municipality. We exploit this contrast directly in the analysis.

\section{Data}

\subsection{Electoral Data}

We use candidate-level electoral data from Brazil's \textit{Tribunal Superior Eleitoral} (TSE), covering all municipal elections from 2004 to 2016. For each municipality and election cycle, we observe every mayoral candidate's party affiliation, vote share, coalition composition, and personal characteristics including age, gender, education, and marital status. We identify Pentecostal candidates as those affiliated with the \textit{Partido Republicano Brasileiro} (PRB) or the \textit{Partido Social Cristão} (PSC), the two parties with institutional ties to major Pentecostal denominations.

Our sample is restricted to municipalities where exactly one Pentecostal candidate competed and elections were decided in a single round. Brazilian municipalities with more than 200,000 registered voters hold a runoff if no candidate exceeds 50 percent of votes, and the strategic dynamics of two-round elections differ fundamentally from single-round contests. We also exclude races where the elected candidate was subsequently removed, died, or was substituted, as well as ties between the second and third-place candidates (which would make the identity of the runner-up ambiguous). We further restrict the sample to the 2008, 2012, and 2016 election cycles, for which post-election outcome data are available. In races with more than two candidates, we retain the top two vote-getters and define the margin of victory between them; ties between the second and third-place candidates are dropped. We also exclude races where the non-Pentecostal candidate's coalition includes PRB or PSC, ensuring that the control side has no Pentecostal institutional involvement. The running variable is the Pentecostal candidate's margin of victory: positive when the Pentecostal candidate wins, negative when the Pentecostal candidate loses. 

Table \ref{tab:representativeness} in the Appendix compares the characteristics of municipalities in our RDD sample with the remaining Brazilian municipalities; the two groups are broadly similar in population, income, urbanization, literacy, sanitation, and evangelical population share. While some differences reach statistical significance owing to the large number of municipalities, the magnitudes are small, suggesting broad generalizability of our findings.

\subsection{Birth Outcomes}

Our primary outcome is the teenage birth rate, constructed from individual-level birth records maintained by Brazil's Ministry of Health (\textit{Sistema de Informações sobre Nascidos Vivos}, SINASC). These records cover the universe of registered births and include the mother's municipality of residence, age, and year of birth, allowing us to assign each birth to a specific maternal birth cohort and calendar year.

For each election cycle, we define the treatment cohort as girls born in the two years that make them aged 9–10 at the time of the election. We then count all births to women in this cohort occurring in years 2 through 5 after the election, a window that allows approximately one year for policy changes to take effect (accounting for principal replacement, curriculum adjustment, and the nine-month gestation period) and extends through the end of the mayoral term. We normalize birth counts by the estimated female population in the corresponding age group, obtained by interpolating between the 2000, 2010, and 2022 censuses, and express the outcome as births per 1,000 girls.

We construct the comparison cohort analogously, using girls aged 16–17 at the time of the election. Because these girls have already passed through middle school when the new administration begins, any effect on their birth outcomes, measured in the same post-election window, would indicate that Pentecostal mayors affect fertility through channels other than school curriculum.

For each cohort, we also compute a lagged dependent variable: the birth rate of the analogous cohort (same age relative to election) in the previous electoral cycle (four years earlier). This lagged outcome serves as a pre-treatment control and absorbs persistent municipality-level differences in cohort-specific fertility.

\subsection{School Curriculum and Principal Characteristics}

We measure the availability of sexual education in schools using the \textit{Prova Brasil} survey, a biennial national assessment administered by INEP (\textit{Instituto Nacional de Estudos e Pesquisas Educacionais}) to all public schools with sufficient enrollment. As part of the assessment, school principals complete a detailed questionnaire covering pedagogical practices, school governance, and curriculum content. The questionnaire includes items on whether the school addresses specific transversal themes, including sexual education and pregnancy prevention, drug use, racism, gender-based violence, and environmental education, in its programming.

We use data from the 2009, 2011, 2013, 2015, 2017, and 2019 waves, restricting the sample to schools offering the final years of \textit{ensino fundamental} (grades 6–9, equivalent to middle school). Each survey wave is matched to the concurrent mayoral term: 2009 and 2011 to the 2008 electoral cycle, 2013 and 2015 to the 2012 cycle, and 2017 and 2019 to the 2016 cycle. This matching captures the state of curriculum implementation during each administration.

Crucially, we observe curriculum choices separately for municipal and state schools within the same municipality. Since mayors control only municipal schools, any effect of Pentecostal mayors on sexual education should appear exclusively in municipal schools. The absence of effects in state schools located in the same municipality provides a powerful placebo test for the school-based mechanism.

The principal questionnaire also records how the principal was selected (political appointment, competitive examination, or community election) and the principal's tenure at the school. We use these variables to examine whether Pentecostal mayors exercise their appointment power more aggressively, a prerequisite for curriculum changes to operate through the personnel channel described in Section 2.

\subsection{Contraceptive Availability}

We measure the supply of contraceptives at municipal primary health clinics using the \textit{Program for the Improvement of Access and Quality of Primary Care} (PMAQ-AB), a federal evaluation program that surveyed health facilities in three waves: 2012, 2014, and 2018. In each wave, trained evaluators verified the physical availability of specific contraceptive methods at each clinic, including oral contraceptives, injectable contraceptives, male and female condoms, intrauterine devices (IUDs), and emergency contraception (morning-after pill). We match each wave to the concurrent mayoral term (2012 to the 2008 cycle, 2014 to the 2012 cycle, 2018 to the 2016 cycle) and collapse clinic-level indicators to municipality-cycle means representing the share of clinics stocking each method.

This variable tests the alternative hypothesis that Pentecostal mayors reduce fertility-related services at the health system level, through restricting contraceptive distribution, rather than through schools. If the school-based mechanism is primary, we should observe null effects on contraceptive availability.

\subsection{Health Outcomes: STDs, HPV Vaccination, Fetal Deaths, and Abortions}

We construct additional health outcomes to assess whether the consequences of reduced sexual education extend beyond fertility.

\textbf{Sexually transmitted diseases.} We obtain municipality-level case counts for syphilis and HIV from Brazil's national disease surveillance system (SINAN). For each election cycle, we construct cohort-specific incidence rates using the same age definitions as the birth outcome: treatment cohort cases among individuals aged 9–10 at the election, comparison cohort cases among individuals aged 16–17. Rates are expressed per 1,000 girls. Syphilis is approximately seven times more prevalent than HIV nationally, making it the more statistically powerful outcome.

\textbf{HPV vaccination.} Brazil's national HPV vaccination program, administered by the \textit{Programa Nacional de Imunizações} (PNI), began in 2014 targeting girls aged 9--14. The program is partly administered through school-based campaigns, making it a natural outcome to test whether Pentecostal mayors affect health service delivery through schools. We obtain municipality-level cumulative vaccination coverage by single year of age and sex from the PNI information system (SIPNI). Because coverage is cumulative (tracking all doses received by a birth cohort across their years of eligibility), we measure coverage at the last year of each mayoral term, when the cumulative measure best reflects the incumbent's influence. We focus on girls aged 9--11 at measurement, whose vaccination history falls mostly within the current administration, as coverage for older girls includes doses administered under the previous mayor.

\textbf{Fetal deaths.} We use records of fetal deaths from the Mortality Information System (SIM) to construct cohort-specific fetal death rates per 1,000 girls. Fetal deaths serve as a partial proxy for pregnancy complications and, in contexts where induced abortion is legally restricted, may capture some fraction of unsafe pregnancy terminations. However, the measure is noisy: fetal deaths also include spontaneous miscarriages and stillbirths unrelated to the policy channel of interest. Moreover, fetal deaths are only recorded after 20 weeks of gestation or if the fetus weighs above 500g.

\textbf{Hospital-recorded abortions.} We obtain data on hospital admissions coded as abortion-related from Brazil's hospital information system (SIH-SUS). Because induced abortion is illegal in Brazil except under narrow circumstances, hospital records predominantly capture post-abortion complications treated in emergency settings rather than the full incidence of induced abortions. This measurement limitation means we cannot distinguish spontaneous from induced cases. Null effects on this variable should therefore be interpreted cautiously: they may reflect either a true absence of changes in pregnancy termination or the inability of hospital records to capture clandestine abortions.

\subsection{Human Capital Accumulation}

To analyze effects on human capital accumulation, we use INEP data on dropout rates at the last two years of middle school (8th and 9th grades). These grades are the most relevant because fertility risk is highest at these ages, and our dynamic estimates show the largest birth rate effects emerging in the later years of the mayoral term. We match these data to the specific treatment cohort for each municipality/cycle. We are able to distinguish between dropout rates at state and municipal schools. We focus on dropout rates for girls.

More specifically, our measure of dropout includes in the numerator every student who did not complete the school year, regardless of whether they returned to school in subsequent years. In the denominator, we have every child who enrolled at the start of the year. Therefore, our measure would not pick up earlier dropouts who never enrolled in 8th or 9th grades. Similarly, there is some degree of noise related to students who repeat grades and may not be a part of the treated cohorts but still be included. Nevertheless, we expect this indicator to be a good measure of disruptions to schooling.

\subsection{Summary Statistics}

Table \ref{tab:descriptive} reports summary statistics for the full sample and for the subsample of municipalities within the optimal bandwidth. Panel A presents candidate characteristics: Pentecostal and non-Pentecostal candidates in close elections are comparable in age, gender composition, educational attainment, and marital status. Panel B describes municipality characteristics, including population, income per capita, urbanization, literacy, sanitation, the share of evangelical residents, and whether the incumbent mayor was from a Pentecostal party. Panel C reports the main outcome variables.

Within the bandwidth-restricted sample, the balance between Pentecostal-won and non-Pentecostal-won municipalities is strong: p-values from clustered t-tests fail to reject equality of means for all pre-treatment characteristics. The final analysis sample comprises 722 municipality-cycle observations across the three election cycles, of which 430 fall within the optimal bandwidth for the main outcome and specification (216 with Pentecostal winners and 214 with non-Pentecostal winners). This balance supports the identifying assumption that close Pentecostal victories approximate random assignment of political control near the threshold.

\section{Empirical Strategy}
Identifying the causal impact of religious political leadership on educational policy and health outcomes faces the fundamental challenge that politician ideology correlates with unobserved community characteristics that may independently influence these outcomes. To address this identification problem, we exploit close mayoral elections between Pentecostal and non-Pentecostal candidates using a sharp regression discontinuity design combined with a cohort-based approach that tracks specific groups of girls through their adolescent years. The RDD approach relies on the assumption that in sufficiently close elections, treatment assignment is as-good-as-random in a neighborhood around the electoral threshold. This assumption is plausible because politicians and voters cannot precisely control vote margins, making municipalities where Pentecostal candidates barely win comparable to those where they barely lose.

\subsection{Research Design: Cohort Construction and Timing}

Our identification strategy combines the regression discontinuity in close elections with a cohort-based design that leverages the timing of school exposure. The key idea is that different age cohorts experience different levels of exposure to a mayor's policies depending on when they pass through the school grades that mayors control.

\textbf{Treatment cohorts.} For each election cycle, we define the treatment cohort as girls aged 9–10 at the time of the election. These girls are about to enter middle school (grades 6–9, typically ages 10–14), and they will spend the entirety of the incoming mayor's four-year term within the middle school system. Early middle school is also where sexual and reproductive health education is most commonly offered, so this cohort receives maximal exposure to any policy changes. We measure their birth outcomes in the window starting two years after the election and extending to five years after, a window that allows roughly one year for policy changes to take effect (principal replacements, curriculum adjustments) and accounts for the nine-month gestation period.

\textbf{Comparison cohorts.} We define the comparison cohort as girls aged 16–17 at the time of the election. These girls are already in high school or have completed schooling by the time the new mayor takes office, placing them beyond the reach of municipal curriculum changes. When we observe their birth outcomes in the same post-election window, they are young adults (ages 18–21), bearing children at ages where school-based sexual education is no longer the relevant margin. If Pentecostal mayors affected fertility through channels other than schools, for instance, by reducing contraceptive access at municipal health clinics or shifting community norms, we would expect effects on this cohort as well. Null effects for the comparison cohort therefore support the school-based mechanism.

\begin{figure}[ht!]
\centering
\caption{Cohort Construction and Exposure Timing}
\label{fig:strategy_illustration}
\includegraphics[width=0.8\textwidth]{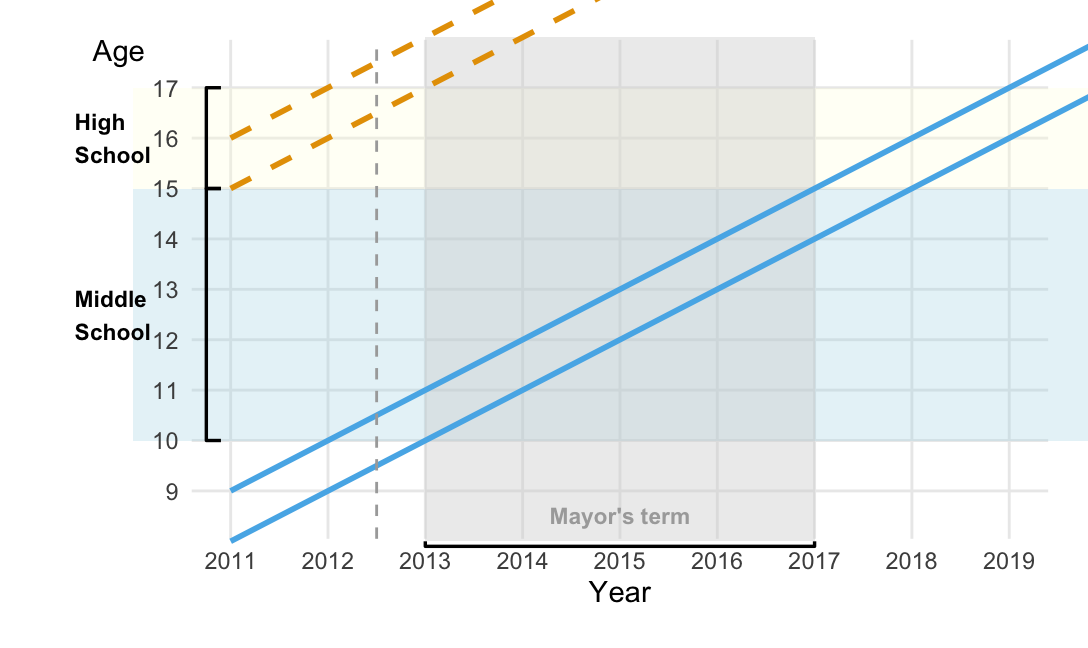}
\begin{minipage}{0.95\textwidth}
\vspace{0.2cm}
\small
\textbf{Notes:} This figure illustrates the cohort-based identification strategy using the 2012 election as an example. The blue lines track girls aged 9-10 at election (treatment cohort) as they age through middle school during the mayor's term (gray shaded area, 2013-2016). Orange dashed lines indicate comparison cohorts (16-17). The treatment cohort experiences maximum exposure to mayoral policies during their formative educational years, while the comparison cohort does not.
\end{minipage}
\end{figure}

Figure \ref{fig:strategy_illustration} illustrates this design for the 2012 election cycle. The solid blue lines trace the treatment cohort (girls aged 9–10 in 2012) as they age through middle school during the mayor's 2013–2016 term, the period of maximum exposure to any curriculum changes. The dashed orange lines trace the comparison cohort (girls aged 16–17 in 2012), who have already passed through middle school and are in high school or beyond when the mayor takes office. The shaded region marks the mayor's term. Because the treatment cohort spends all four years of the administration in middle school while the comparison cohort does not, any differential effect of Pentecostal mayors on birth outcomes between these cohorts can be attributed to school-based policy changes rather than to municipality-wide shifts in health services or cultural norms.

\subsection{Regression Discontinuity Specification}

We estimate the following equation for municipalities with close elections:
\begin{align*}
\centering
Y_{jmt} &= \alpha + \beta
\cdot \text{win}_{mt}  + \tau \cdot \text{X}_{mt}
+ \\
&\rho \cdot  \text{X}_{mt}  \times \text{win}_{mt}  + C_{jmt}  + \varepsilon_{jmt}
\end{align*}

\noindent where $Y_{jmt}$ represents the outcome of interest in unit $j$ in municipality $m$. The running variable $\text{X}_{mt}$ is the vote margin, calculated as the Pentecostal party's vote share minus the non-Pentecostal party's vote share. The treatment indicator $\text{win}_{mt}$ equals one when the Pentecostal candidate wins.

The control vector $C_{jmt}$ includes outcome lagged variables when available, election cycle fixed effects, and state fixed effects to account for time trends and state-level policies. When analyzing school-level outcomes, we include school characteristics. Standard errors are clustered at the municipality level to account for within-municipality correlation.

The parameter of interest, $\beta$, provides the estimated local average treatment effect of Pentecostal parties on the outcome of interest for cohorts exposed during their school-age years. This estimator assumes that municipalities are comparable close to the threshold and that agents (i.e., politicians and voters) are unable to precisely manipulate the running variable. We define ``close'' elections using the optimal bandwidth procedure of \citet{calonico2014robust}, which varies by outcome to optimize the bias-variance tradeoff. Section \ref{checks} presents validation tests supporting the identifying assumptions, including balance tests for predetermined covariates and density tests for electoral manipulation.

\section{Results}\label{results}
Using a regression discontinuity design that exploits close mayoral elections, we track cohorts of girls aged 9-10 at the time of each election through their adolescent years. We find that exposure to Pentecostal mayors during middle school causes teenage birth rates to increase by 3 per 1,000 girls, a roughly 40\% increase from baseline. These effects emerge gradually as cohorts age and are concentrated exclusively among school-age populations, with no effects for older cohorts who had already passed through school before policy implementation. Consistent with school-based mechanisms, municipal schools become 12.5 percentage points less likely to offer sexual education programs under Pentecostal governance, while state schools, where mayors lack administrative control, show no curriculum changes.

\subsection{Validity Checks}\label{checks}

Causal interpretation of the estimates requires that party identity assignment is effectively random in a neighborhood around the electoral threshold. This assumption is violated if politicians, municipalities, or voters can precisely manipulate election outcomes near the cutoff \citep{imbens2008regression}. Electoral manipulation would create systematic sorting around the threshold, undermining the validity of the regression discontinuity design.

\begin{figure}[ht!]
    \centering   
    \caption{Manipulation Test}
    \label{fig:manipulation}
    \includegraphics[width=0.9\textwidth]{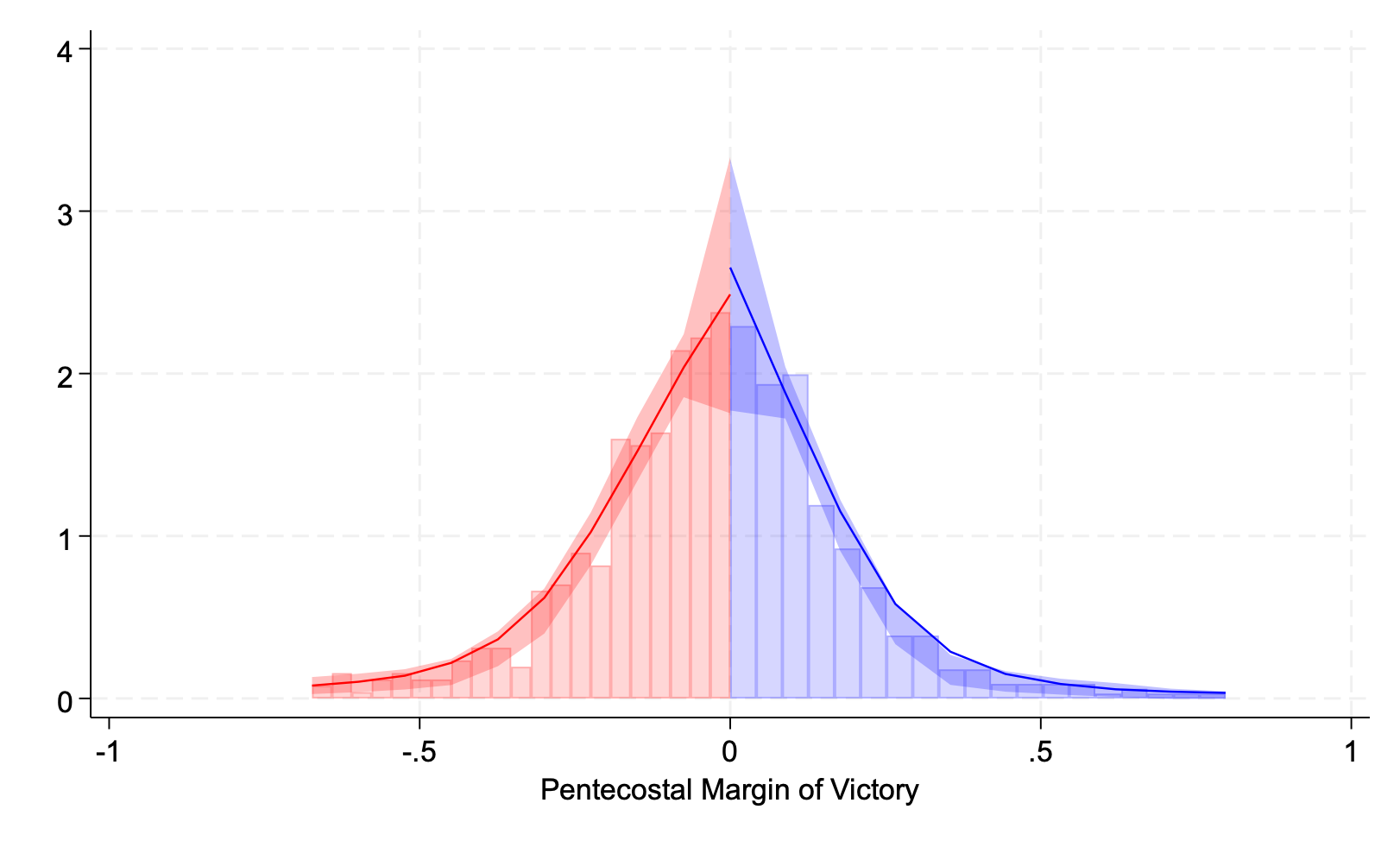}
       \begin{minipage}{\textwidth} 
{\footnotesize \textbf{Notes:} This figure shows the density plot for the McCrary test. We find no evidence of manipulation at the cutoff (T = 0.005, p = 0.996).}
\end{minipage}

\end{figure}

To test for such manipulation, we employ the density test proposed by \citet{mccrary2008manipulation}, which examines whether the distribution of the running variable exhibits a discontinuous jump at the threshold. Figure \ref{fig:manipulation} presents the results graphically, showing no evidence of systematic manipulation around the electoral threshold. The test fails to reject the null hypothesis of density continuity at the cutoff (p value: 0.996), providing strong support for the identifying assumption that treatment assignment is as-good-as-random near the threshold.

\begin{figure}[ht!]
    \centering
     \caption{Municipality Characteristics}
     \label{fig:municipality_balance}
     \begin{subfigure}[b]{0.48\textwidth}
     \caption{Log$_{10}$ Population}
    \includegraphics[width=0.98\textwidth]{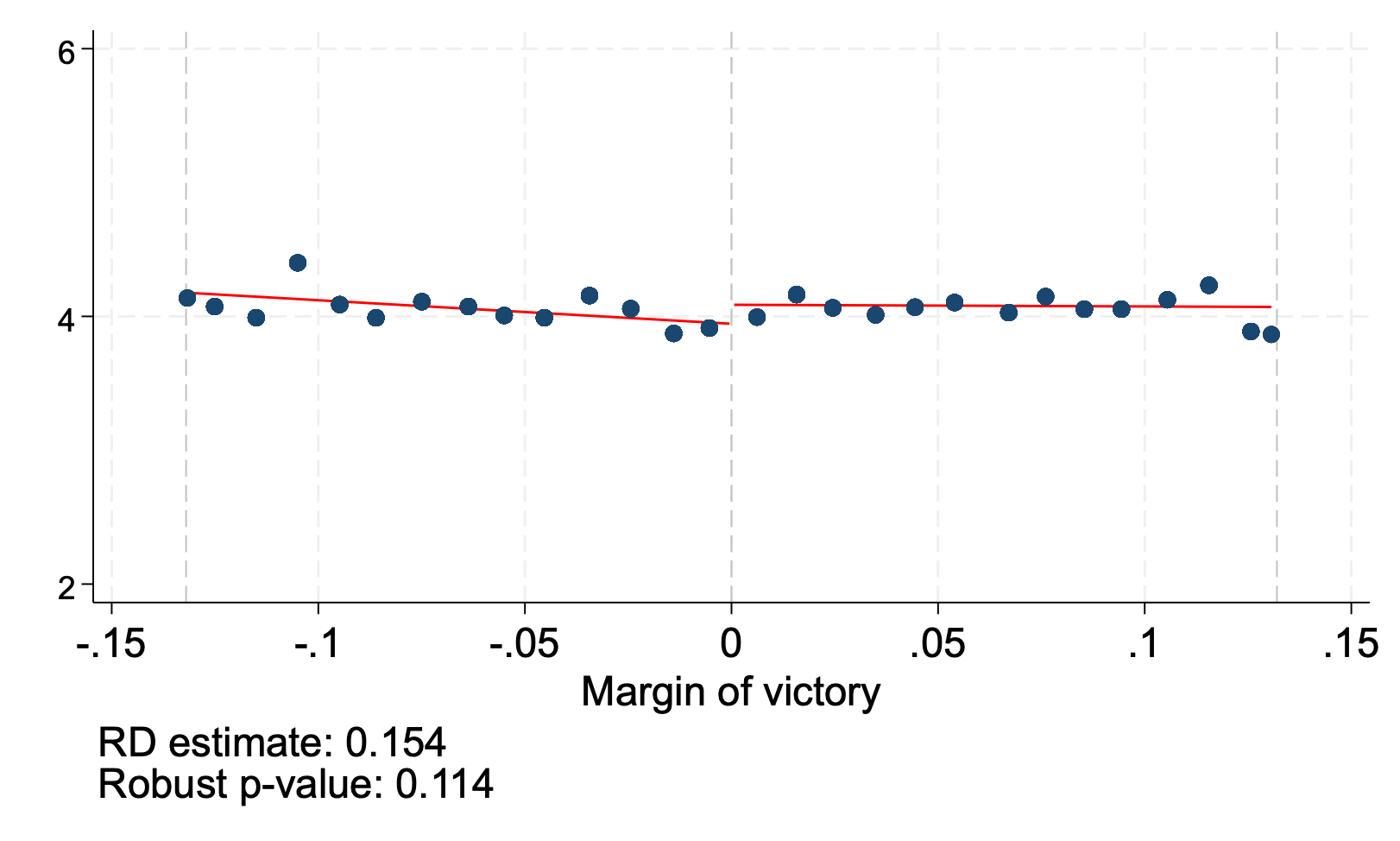} 
\end{subfigure}\begin{subfigure}[b]{0.48\textwidth}
     \caption{Urban Population}
    \includegraphics[width=0.98\textwidth]{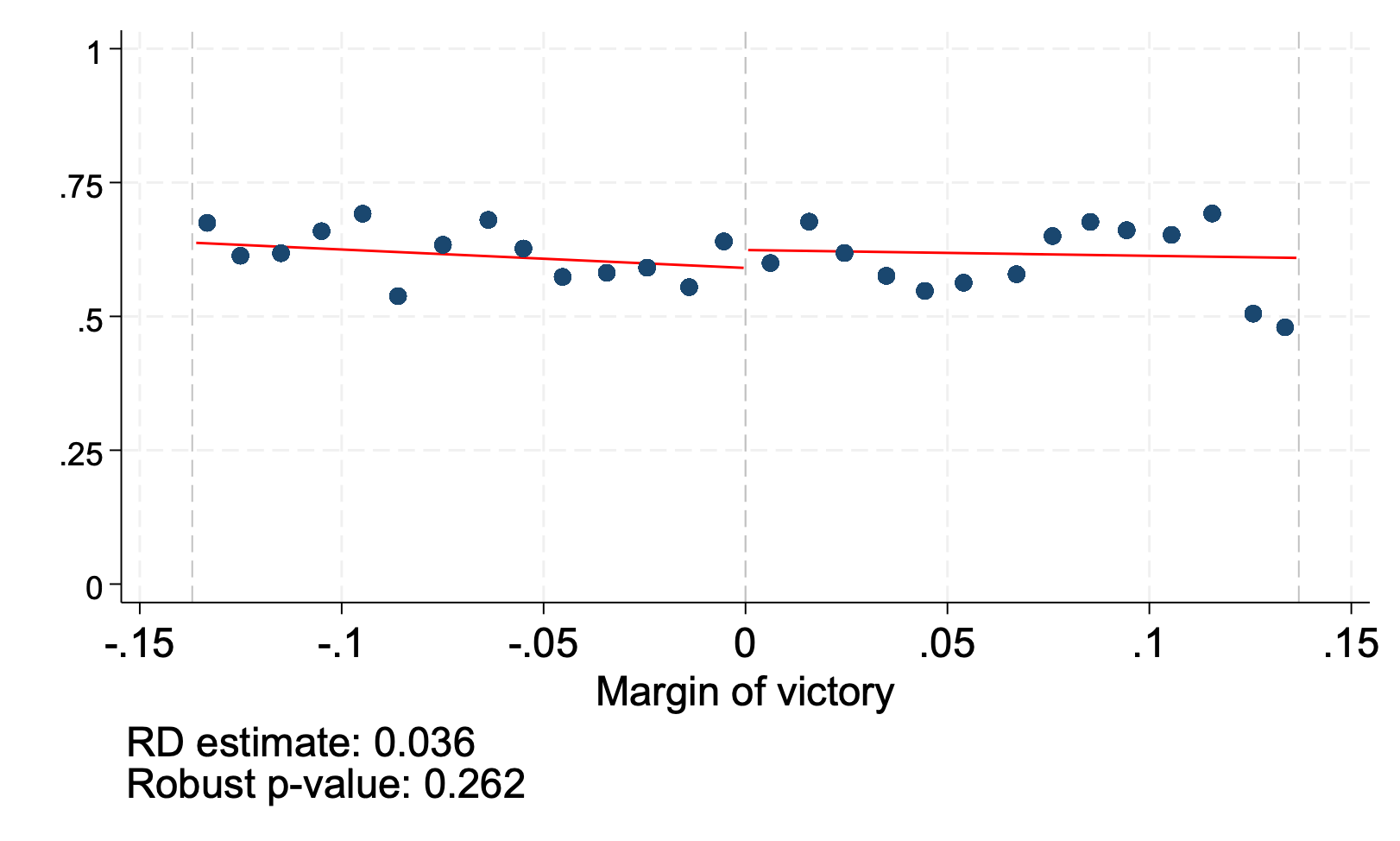}
\end{subfigure}
     \begin{subfigure}[b]{0.48\textwidth}
     \caption{Literacy}
    \includegraphics[width=0.98\textwidth]{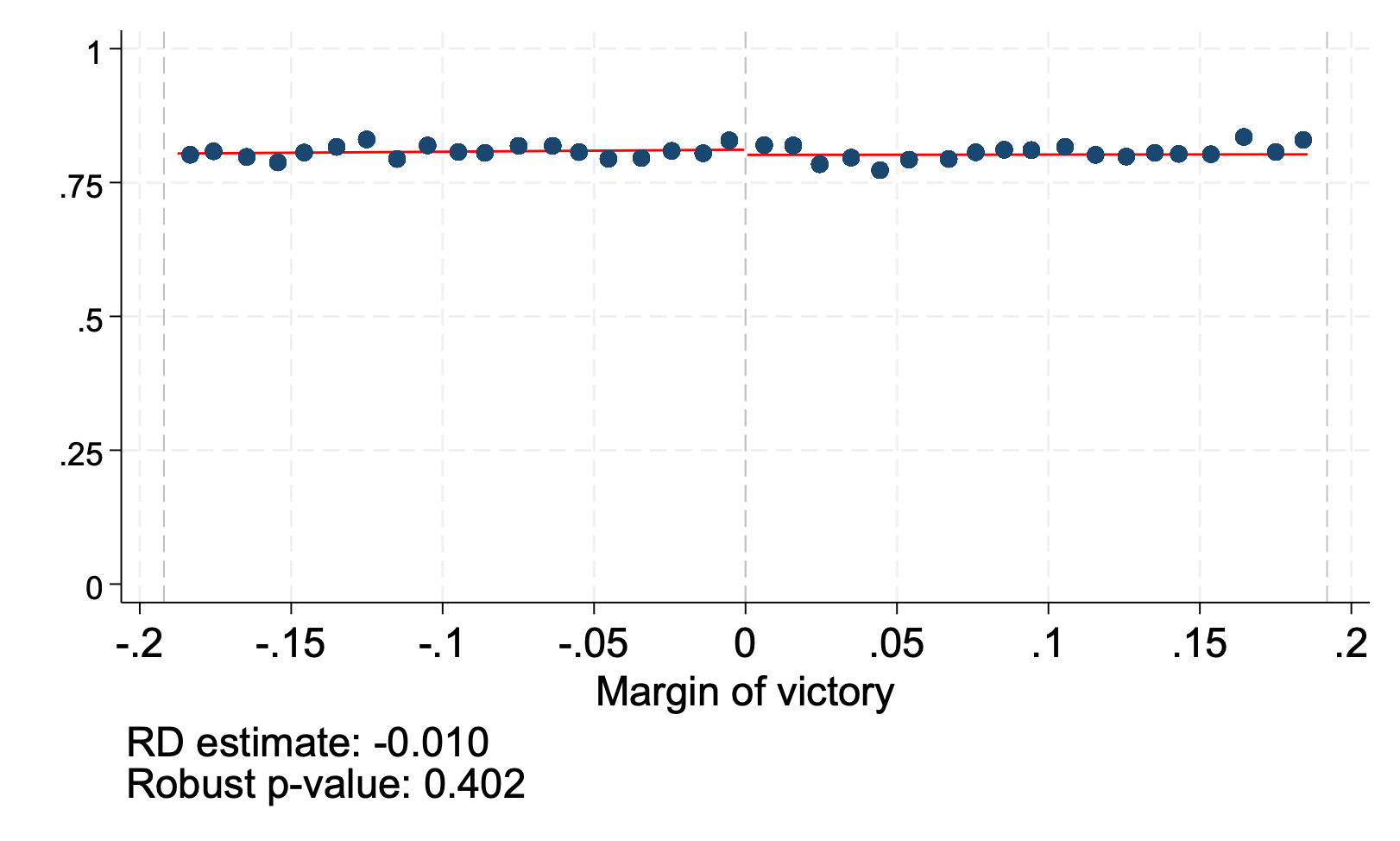} 
\end{subfigure}
\begin{subfigure}[b]{0.48\textwidth}
     \caption{Share of Pentecostals}
    \includegraphics[width=0.98\textwidth]{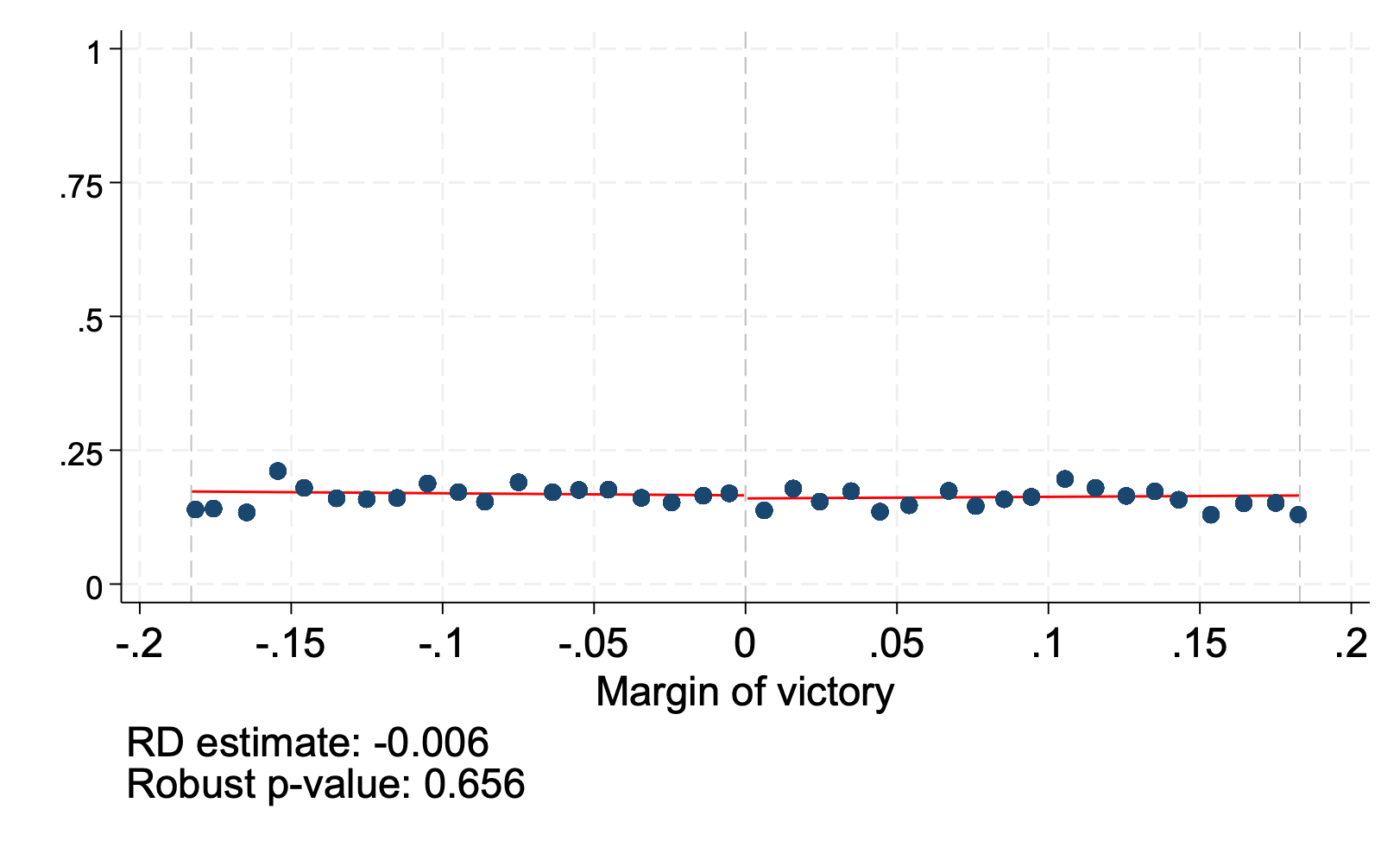}
\end{subfigure}
     \begin{subfigure}[b]{0.48\textwidth}
     \caption{Sanitation}
    \includegraphics[width=0.98\textwidth]{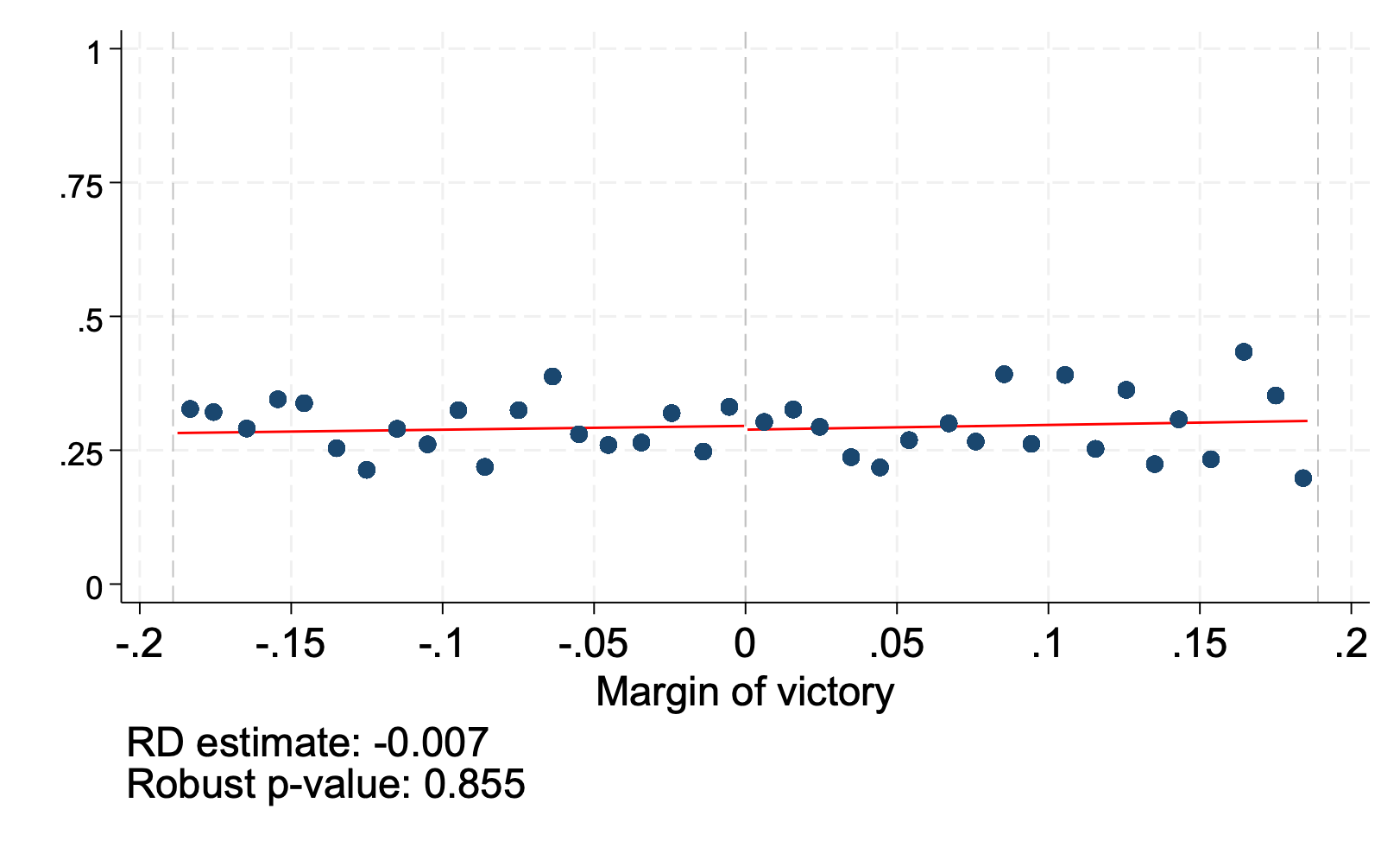} 
\end{subfigure}
\begin{subfigure}[b]{0.48\textwidth}
     \caption{Income per Capita}
    \includegraphics[width=0.98\textwidth]{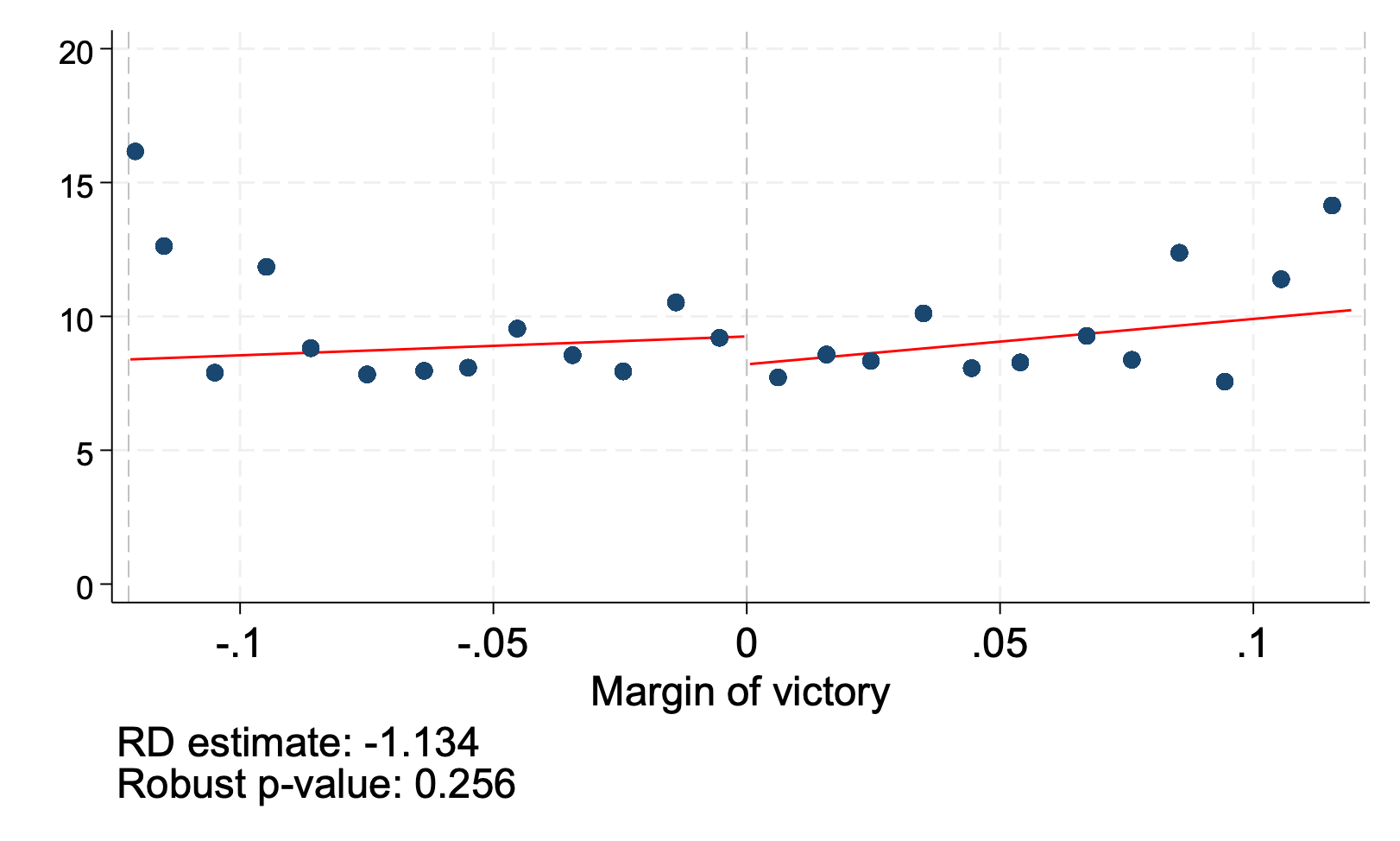}
\end{subfigure}
           \begin{minipage}{\textwidth} 
{\footnotesize \textbf{Notes:} This figure shows the RDD estimates for a) log$_{10}$ of population, b) share of population in urban areas, c) share of population that is literate, and d) share of Pentecostals, e) share of households have basic sanitation, f) income per capita in thousand BRL. The running variable is the margin of victory in 2008, 2012, or 2016 municipal elections, with positive values indicating Pentecostal victories. Adjusted for state fixed effects and electoral cycle fixed effects. Linear fit, triangular kernel, optimal bandwidth calculated following \citet{calonico2014robust}.
}
\end{minipage}
\end{figure}

\begin{figure}[ht!]
    \centering
     \caption{Candidate Characteristics}
     \label{fig:candidates_characteristics}
     \begin{subfigure}[b]{0.47\textwidth}
     \caption{Age}
    \includegraphics[width=0.98\textwidth]{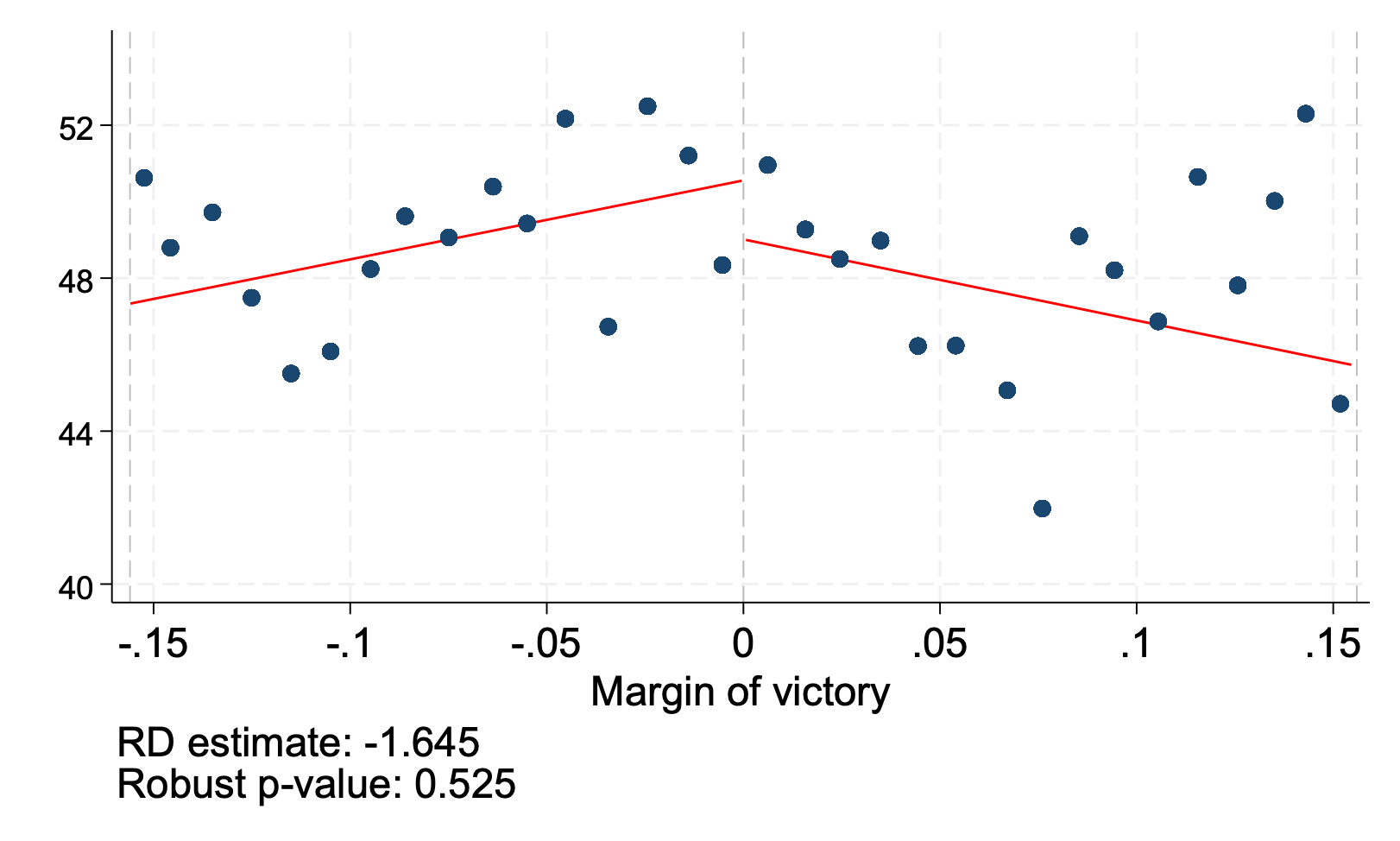} 
\end{subfigure}\begin{subfigure}[b]{0.47\textwidth}
     \caption{Female}
    \includegraphics[width=0.98\textwidth]{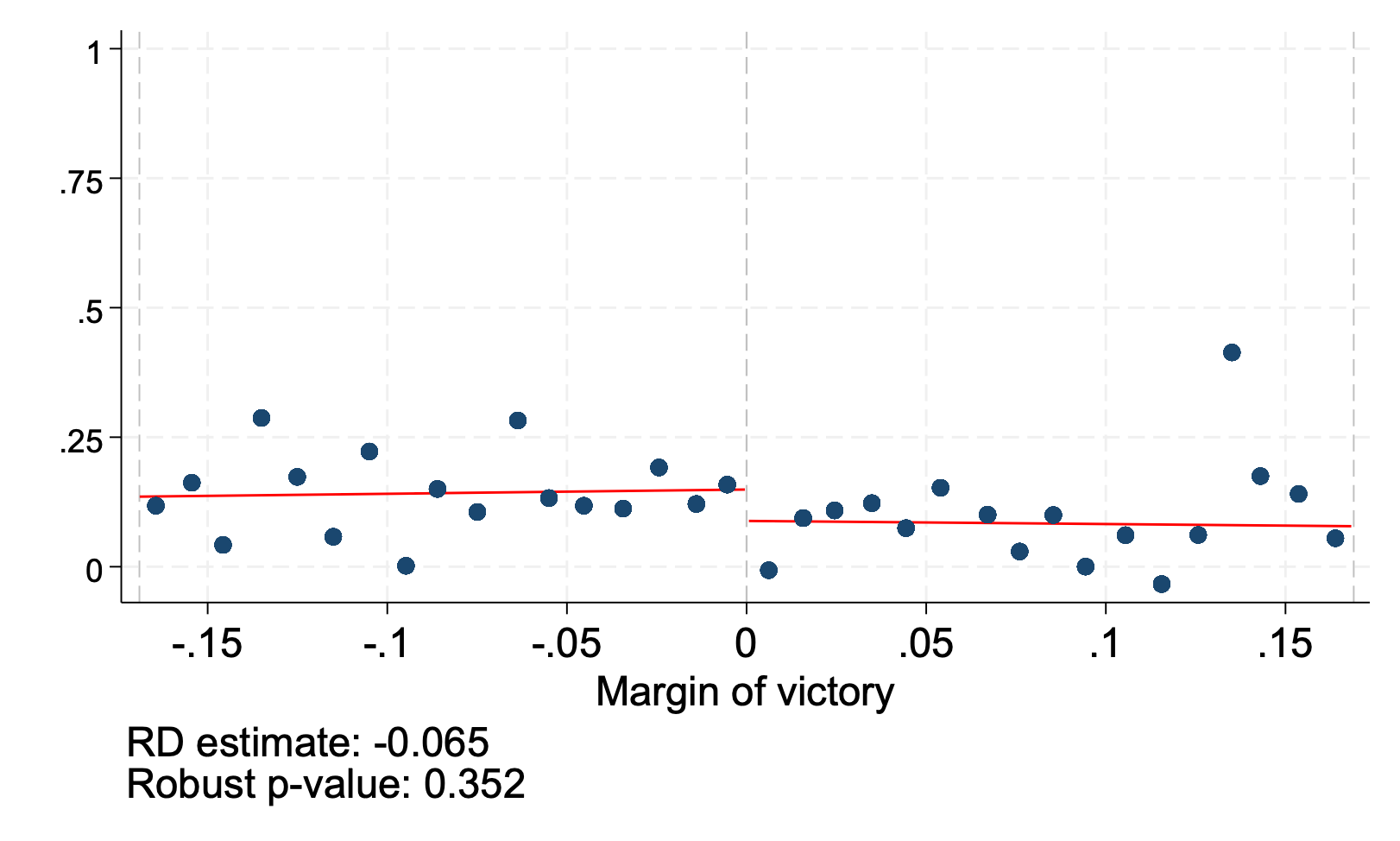}
\end{subfigure}
     \begin{subfigure}[b]{0.47\textwidth}
     \caption{College Education}
    \includegraphics[width=0.98\textwidth]{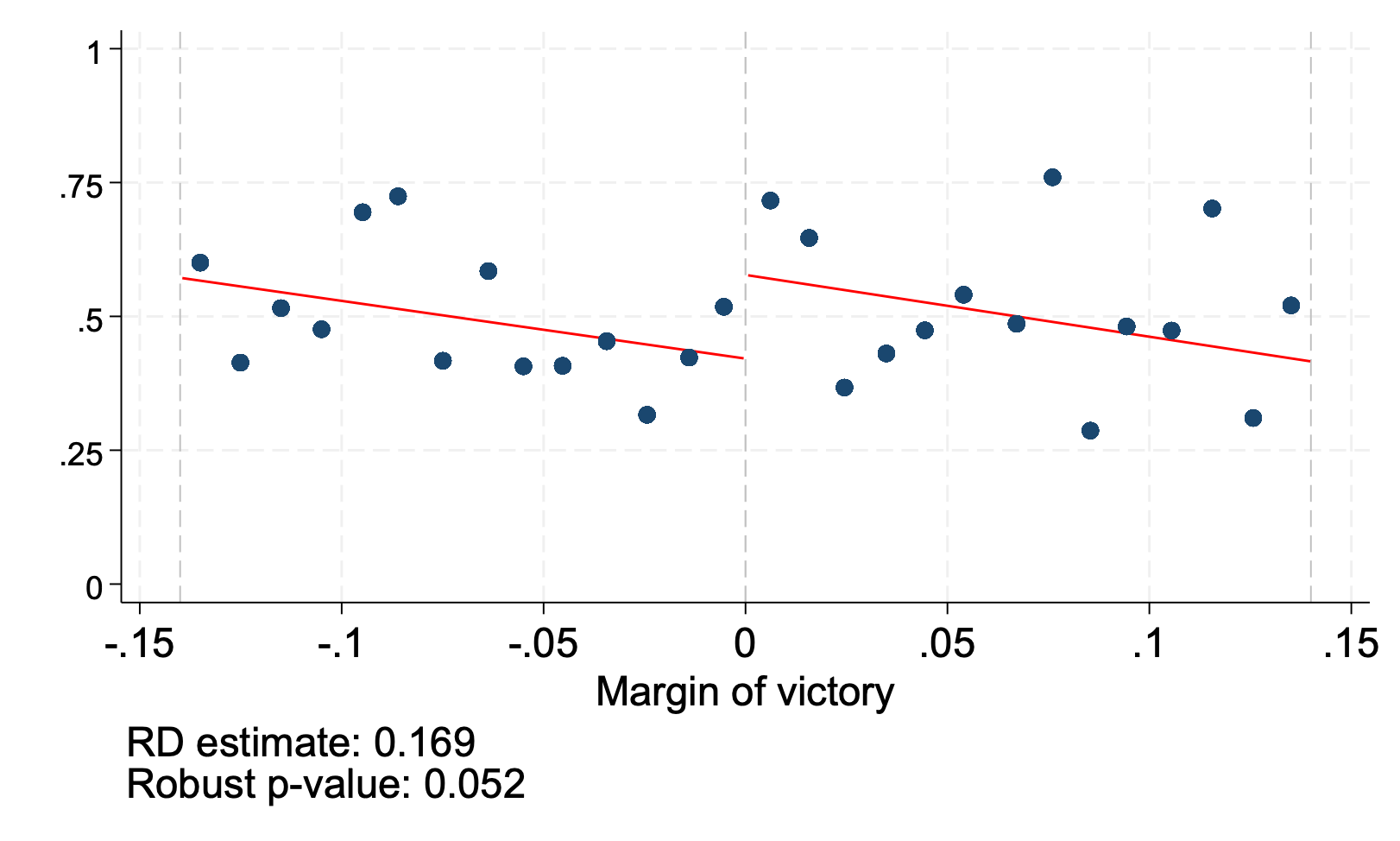} 
\end{subfigure}
\begin{subfigure}[b]{0.47\textwidth}
\caption{Married}
    \includegraphics[width=0.98\textwidth]{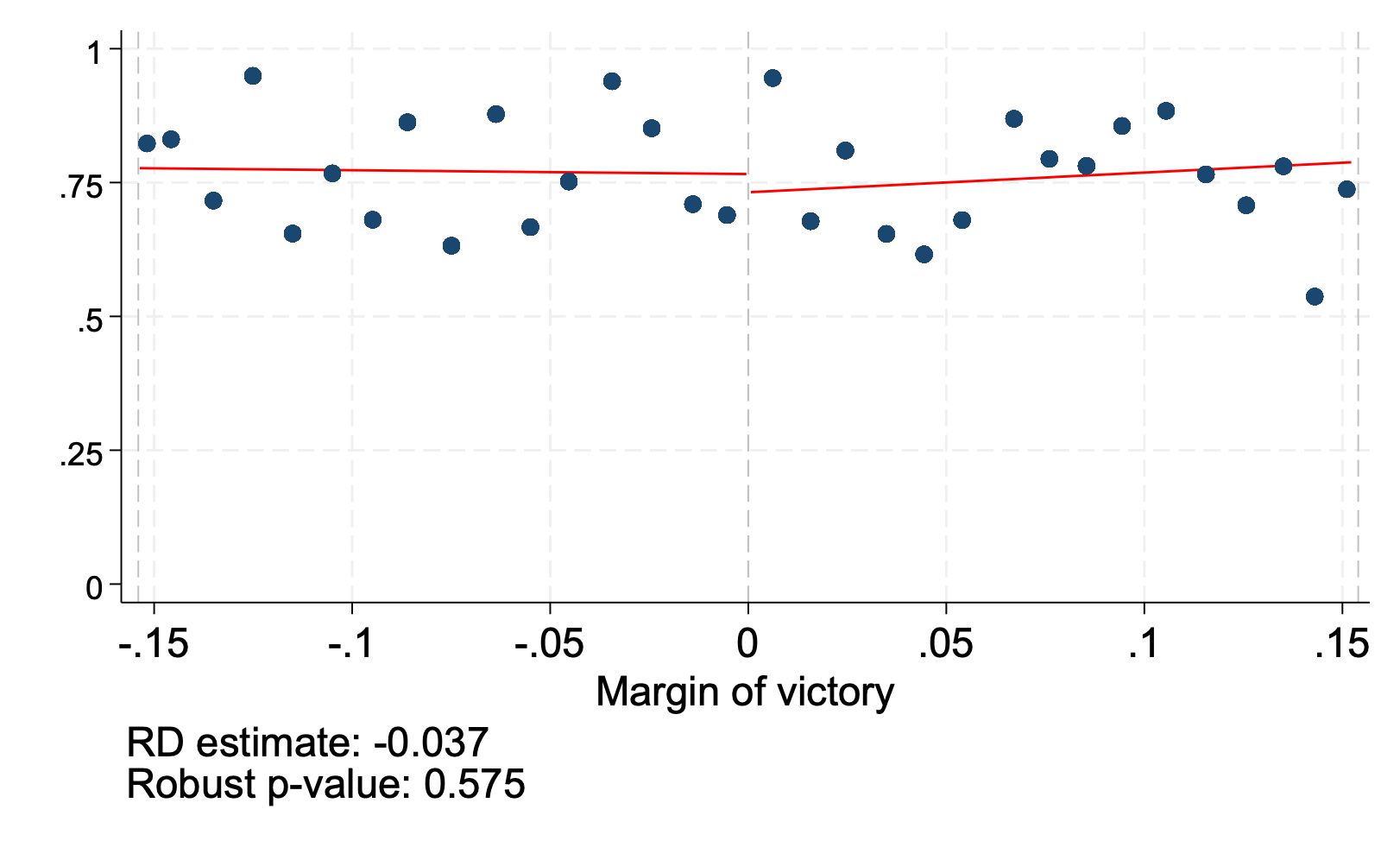}
\end{subfigure}
           \begin{minipage}{\textwidth} 
{\footnotesize \textbf{Notes:} This figure shows RD estimates of differences in characteristics of the winning candidate along the margin of victory. We show a) age, b) gender, c) some college education, d) marriage status. The running variable is the margin of victory in 2008, 2012, or 2016 municipal elections, with positive values indicating Pentecostal victories. Adjusted for state fixed effects and electoral cycle fixed effects. Linear fit, triangular kernel, optimal bandwidth calculated following \citet{calonico2014robust}.
}
\end{minipage}
\end{figure}

We also provide evidence that municipalities where Pentecostal parties narrowly won or lost the election are comparable in terms of baseline characteristics. Figure \ref{fig:municipality_balance} shows that population, urbanization, literacy rates, the share of Pentecostals, sanitation rates, and income per capita are similar between the two sets of municipalities. Similarly, Figure \ref{fig:candidates_characteristics} shows that Pentecostal candidates who narrowly win are similar to non-Pentecostal candidates who narrowly lose. There is a marginally significant difference in rates of college-level education (p value: 0.052), but no difference in age, gender or marriage rates. This balance in observable characteristics provides support for the regression discontinuity design's identifying assumption that treatment assignment is as-good-as-random in a neighborhood around the threshold.

\subsection{Teenage Pregnancy}\label{sec:fertility}

\subsubsection{Main Results}

\begin{figure}[ht!]
    \centering
        \caption{Effect of Pentecostal Mayors on Births from Middle School Aged Girls}
        \label{fig:birth_rate}
        
    \includegraphics[width=\textwidth]{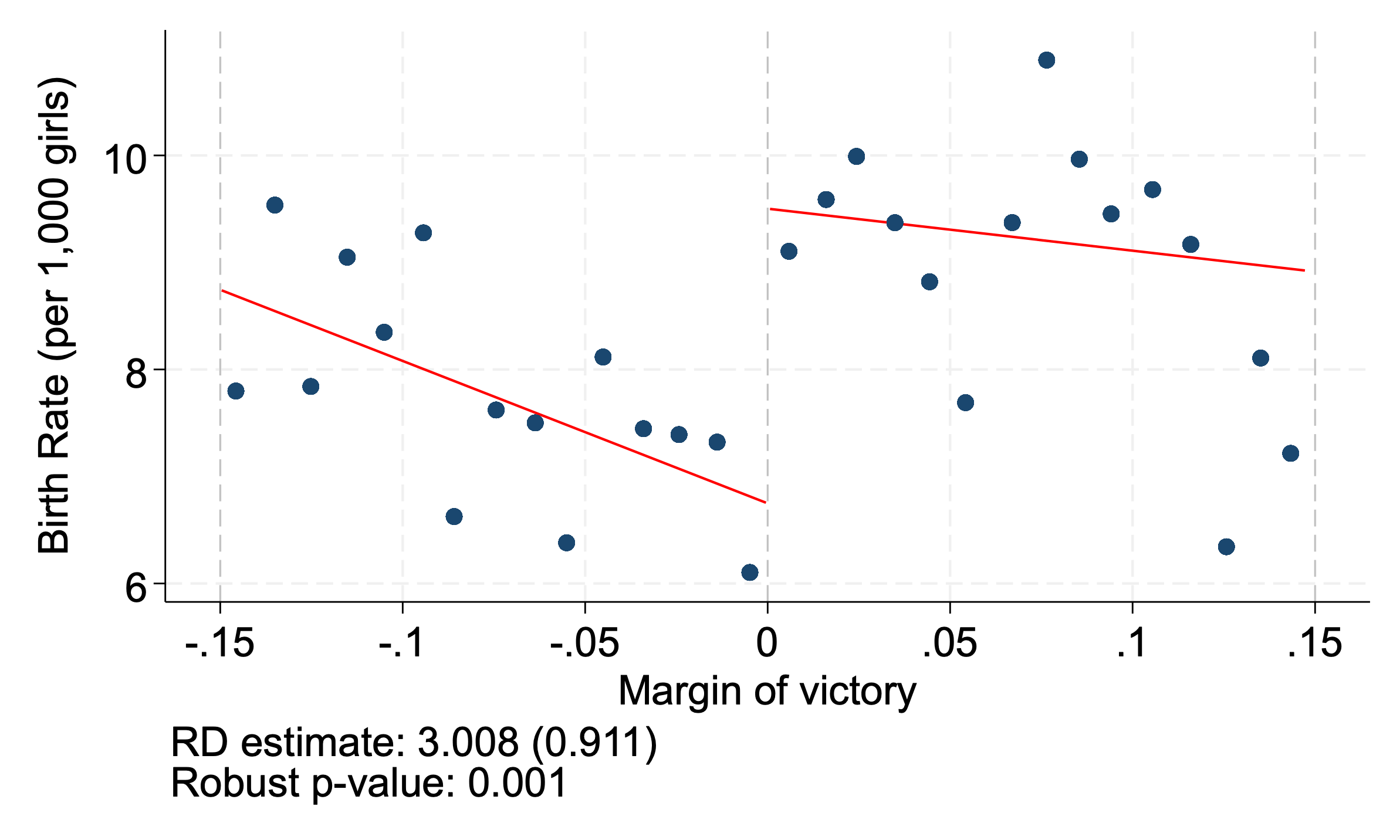}
         \begin{minipage}{1\textwidth} 
{\footnotesize \textbf{Notes:} This figure shows the birth rate for cohorts aged 9-10 at election against the margin of victory, averaged over years 2-5 post-election (when the cohort is aged 11-15). The running variable is the margin of victory in 2008, 2012, or 2016 municipal elections, with positive values indicating Pentecostal victories. Adjusted for state fixed effects, electoral cycle fixed effects and the lagged dependent variable. Linear fit, triangular kernel, optimal bandwidth calculated following \citet{calonico2014robust}.
}
\end{minipage}
\end{figure}

Figure \ref{fig:birth_rate} plots the covariate-adjusted teenage birth rate against the margin of victory, binned in one percentage point wide intervals. The outcome measures the average annual birth rate during years 2-5 post-election for girls who were aged 9-10 at the time of the election. The figure shows a clear discontinuity in the birth rate when the margin of victory becomes positive, indicating an increase of 3 births per 1,000 girls when a mayor affiliated with a Pentecostal party wins the election, corresponding to about a 40\% increase from the left-of-cutoff baseline ($\frac{3}{7.57}=0.3963$).

To contextualize the magnitude, consider that the average municipality within the bandwidth has 400 girls in the treated cohort, meaning the treatment leads to 1.2 extra births per year per municipality, on average (the median municipality has 200 girls, resulting in 0.6 births). These are large effects when considering the low baseline birth rate at these ages, but modest compared to overall fertility, or even teenage fertility at peak ages (over 16).

Table \ref{tab:births} panel A presents the regression discontinuity estimates across different specifications. Starting from no controls, we progressively add state fixed effects, electoral cycle fixed effects and control for the lagged dependent variable (i.e. the birth rate prevalent four years before, during the previous election). We then present estimates with a quadratic fit and Epanechnikov kernel (instead of triangular). The estimated coefficient remains relatively stable across all specifications, ranging from 3.52 to 2.86. All specifications are statistically significant at 5\%, and all but one at 1\%. Our preferred specification is column (4), corresponding to Figure \ref{fig:birth_rate}.

\subsubsection{Cohort Comparisons}

We present two comparisons to assess the validity of our design and the role of school-based exposure. First, we test for pre-existing differences: does our estimator pick up any differences in birth rates during the previous election cycle? Second, we compare effects across cohorts with different school exposure: do we find similar effects for young adults who were past school age when the new administration began?

\paragraph{Pre-Election Period for Treatment Cohort}

\begin{figure}[ht!]
    \centering
        \caption{Placebo: Effect of Pentecostal Mayors on Births}
        \label{fig:births_placebo}
    \begin{subfigure}[b]{0.98\textwidth}
     \caption{Middle School Aged Girls, Lagged}
    \includegraphics[width=0.9\textwidth]{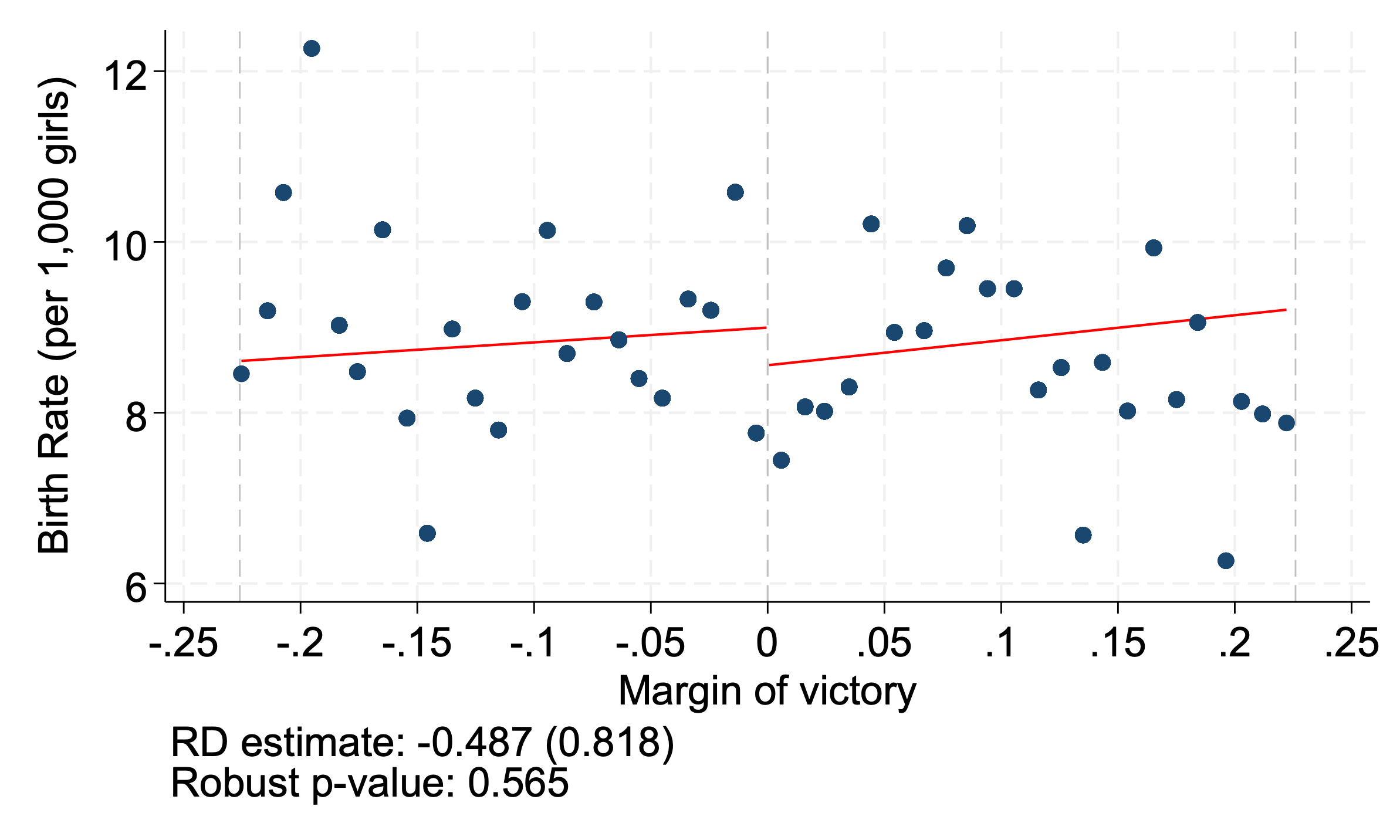} 
\end{subfigure}

\begin{subfigure}[b]{0.98\textwidth}
     \caption{Young Adult Women}
    \includegraphics[width=0.9\textwidth]{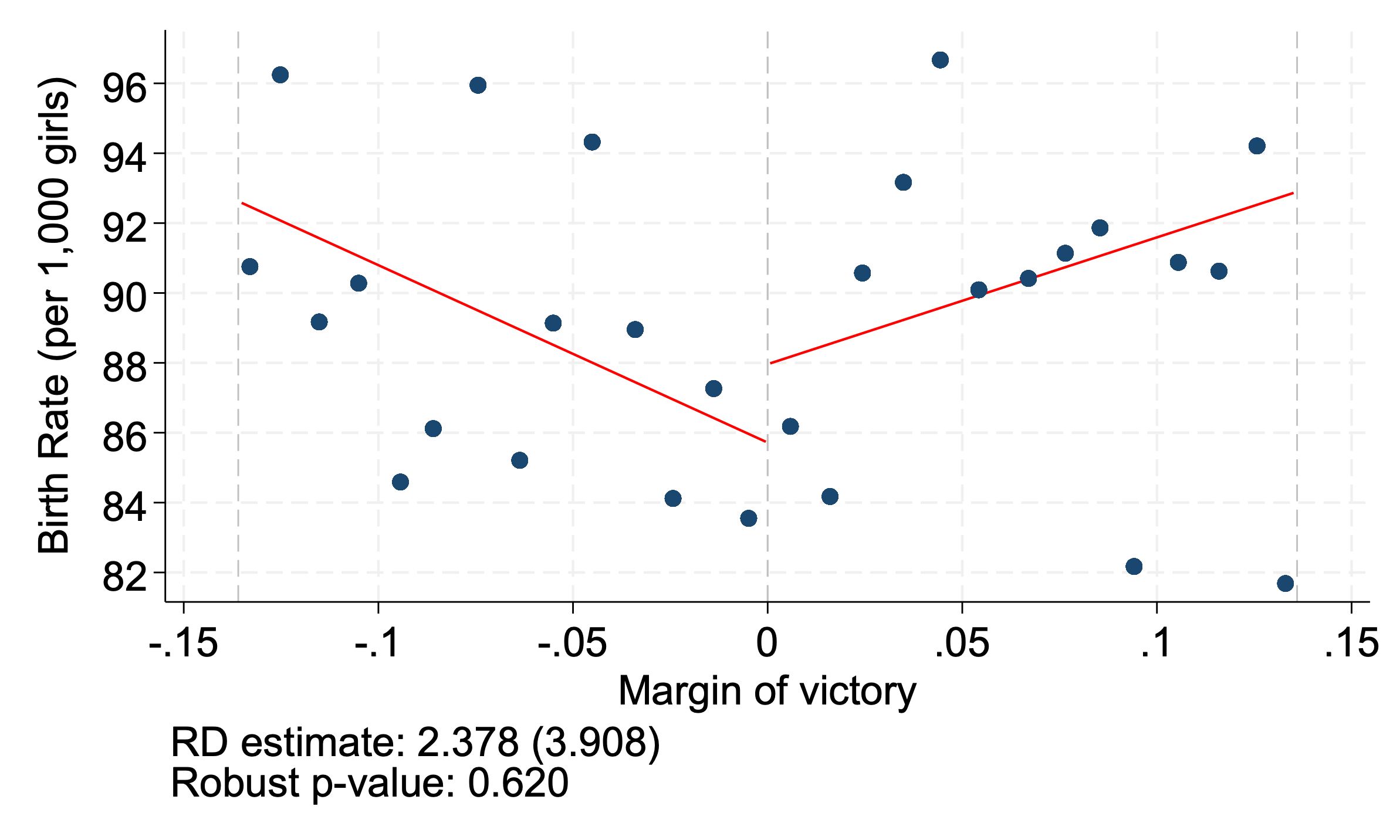}
\end{subfigure}
     \begin{minipage}{\textwidth} 
{\footnotesize \textbf{Notes:}  \textbf{(a)}: the  birth rate for cohorts aged 9-10 at the previous election, averaged over years 2-5 post-election (when cohort is 11-15);  \textbf{(b)}: the birth rate for cohorts aged 16-17 at election, averaged over years 2-5 post-election (when cohort is 18-22). The running variable is the margin of victory in 2008, 2012, or 2016 municipal elections, with positive values indicating Pentecostal victories. Adjusted for state fixed effects, electoral cycle fixed effects. Linear fit, triangular kernel, optimal bandwidth calculated following \citet{calonico2014robust}.
\par}\end{minipage}
\end{figure}

Figure \ref{fig:births_placebo}, top graph, examines the birth rates of the cohort four years younger than the treated cohort, at the same age. That is, the birth rate of girls who were 9-10 four years before the election (during the previous election cycle), from ages 11 to 15. At this earlier time point, the municipalities had different mayors, and the election we analyze had not yet occurred. Finding no discontinuity at the electoral threshold during this pre-election period confirms that municipalities where Pentecostal candidates would later narrowly win versus lose had similar baseline pregnancy patterns. We find no preexisting difference (p value (0.568)), and the point estimate is close to zero (-0.48).

\paragraph{Cohort Comparison: Young Adults}
If Pentecostal mayors affected fertility through channels beyond schools, such as restricting family planning clinic access, changing community norms, or implementing pro-natalist programs, we would expect effects across age groups, not only among school-enrolled cohorts. We therefore compare the treatment cohort results with those for young adults who were past school age when the new administration began.

Figure \ref{fig:births_placebo}, bottom graph, examines birth rates for women aged 16-17 at the time of election, a cohort that would be 18 to 22 during the target period, past school age. The figure shows no significant impact on birth rates for this older cohort (p value = 0.620). The point estimate is of a similar magnitude to that of the main outcome (2.38 versus 3.01), but the baseline birth rate is about 10 times higher. Therefore this effect represents only a 2.7\% increase. Furthermore, Table \ref{tab:births} panel B shows that the point estimate turns negative in some specifications, and does not reject zero effect in any specification. The 95\% confidence interval for the preferred specification is approximately $[-4.2, 8.9]$, meaning we can rule out effects larger than about 10\% of the baseline rate (89.7 per 1,000). The contrast between a 40\% proportional increase for school-age girls and the absence of detectable proportional effects for young adults suggests that the primary channel operates during adolescence, consistent with school-based exposure. However, the confidence intervals for the older cohort are wide enough that we cannot rule out modest effects through non-school channels.

\subsection{Robustness}

\begin{figure}
    \caption{Specification Plot}
    \centering
    \includegraphics[width=\linewidth]{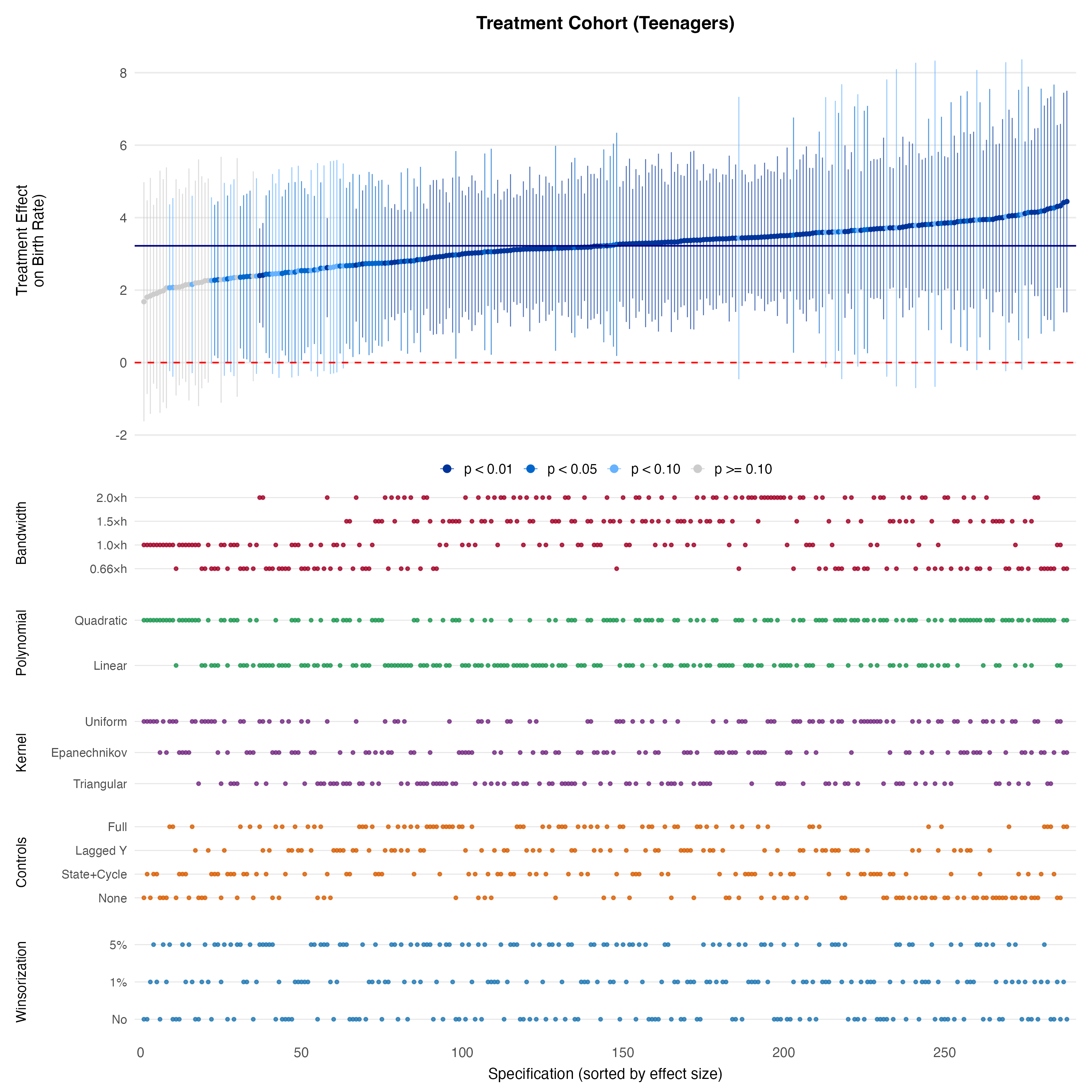}
    \label{fig:specs}
    \begin{minipage}{0.85\textwidth} 
    {\footnotesize \textbf{Notes:} This figure shows how the estimated coefficient changes over 288 RD specifications. The reference bandwidth is 0.15, with the bias-correction parameter rho fixed at 0.508. Dotted line indicates median estimate (3.2). Positive estimates: 100\%. Significant at 1\%: 60.8\%. Significant at 5\%: 81.6\%. Significant at 10\%: 92.7\%. }\end{minipage}
\end{figure}

As a further test of robustness, we present a specification plot in Figure \ref{fig:specs}. In it we plot the results from 288 different specifications, corresponding to 4 different bandwidths (0.66, 1, 1.5 and 2, expressed as multiples of the optimal bandwidth for the preferred specification, 0.15), 2 polynomial fits (linear, quadratic), 3 kernel choices (Uniform, Triangular and Epanechnikov), 4 sets of controls (None, State+Electoral Cycle Fixed Effects, Lagged Dependent Variable and Full), and 3 levels of winsorization of the outcome (none, 1\%, 5\%). The estimate is relatively stable, with no negative estimates and 60.8\% of results statistically significant at 1\%. Notably, our preferred specification (3.0) is very close to the sample median (3.2). We also note that the left tail, i.e. the smallest estimates, predominantly use quadratic fit with small bandwidth, a combination likely to result in noisy and unreliable estimates.

Table \ref{tab:sensitivity}, Panel A, further shows that the birth rate estimate is robust to trimming rather than winsorizing the outcome distribution, with relatively stable coefficients when dropping observations above the 99th or below the 1st percentile, or above the 95th or below the 5th percentile. We also conduct a placebo cutoff test, estimating the RD at artificial cutoffs ranging from $-0.10$ to $+0.10$ along the margin of victory. Figure \ref{fig:placebo_cutoff} shows that the estimated effect peaks sharply at the true cutoff and attenuates as the artificial cutoff moves away from zero, confirming that the discontinuity is localized at the Pentecostal victory threshold rather than reflecting a broader pattern in the data. Nearby placebo cutoffs show some significance because their estimation bandwidth overlaps with the true cutoff; the key pattern is the rapid decay of the effect with distance.

\subsubsection{Dynamic Effects: When Do Effects Emerge?}\label{sec:dynamic_effects}

To examine how effects unfold over time, we estimate the treatment effect separately for each year following the election. Instead of using separate RD estimators for each period, we stack them into a single regression. This procedure has the advantage of allowing clustering standard errors at the level of the municipality, thereby accounting for serial correlation in inference. On the other hand, we fix the bandwidth based on the optimal bandwidth in the baseline specification and we do not employ a bias-correction procedure. In practice, results are almost identical either way.

We can express the estimating equation as:
\begin{equation*}
Y_{mtk} = \sum_{k}\left[ \alpha_k + \beta_k \cdot \text{win}_{mt}  + \gamma_k \cdot  X_{mt} + \delta_k \cdot  X_{mt} \cdot \text{win}_{mt} + \tau_{tk} + \rho_{s(m)k}\right] + \varepsilon_{mtk}
\end{equation*}

\noindent where $m$ denotes the municipality, $t$ denotes the electoral cycle, $k$ is the year relative to election, $X_{mt}$ is the margin of victory, $\text{win}_{mt}$ indicates whether the Pentecostal candidate won. The levels (determined by $\alpha_k$ and $\beta_k$) and slopes ($\gamma_k$ and $\delta_k$) on either side of the cutoff are fully flexible with time from the election, as are election cycle fixed effects ($\tau_{tk}$) and state fixed effects ($\rho_{s(m)k}$). We use a triangular kernel and a bandwidth of 0.15.

\begin{figure}[ht!]
    \centering
        \caption{Dynamic Effect of Pentecostal Mayors on Births}
        \label{fig:event_study}
        \begin{subfigure}[b]{0.98\textwidth}
     \caption{Middle School Aged Girls}
    \includegraphics[width=0.9\textwidth]{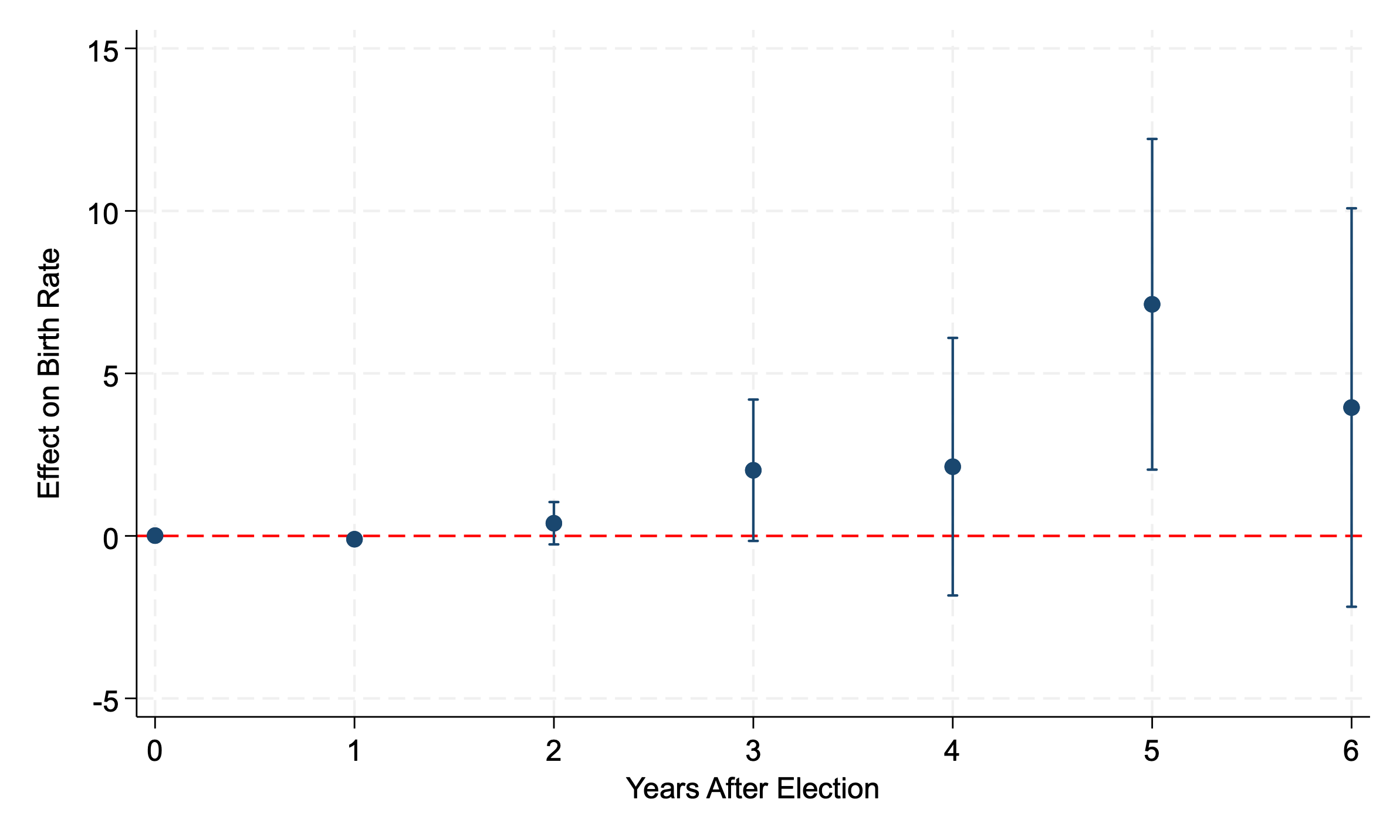} 
\end{subfigure}

\begin{subfigure}[b]{0.98\textwidth}
     \caption{Young Adults}
    \includegraphics[width=0.9\textwidth]{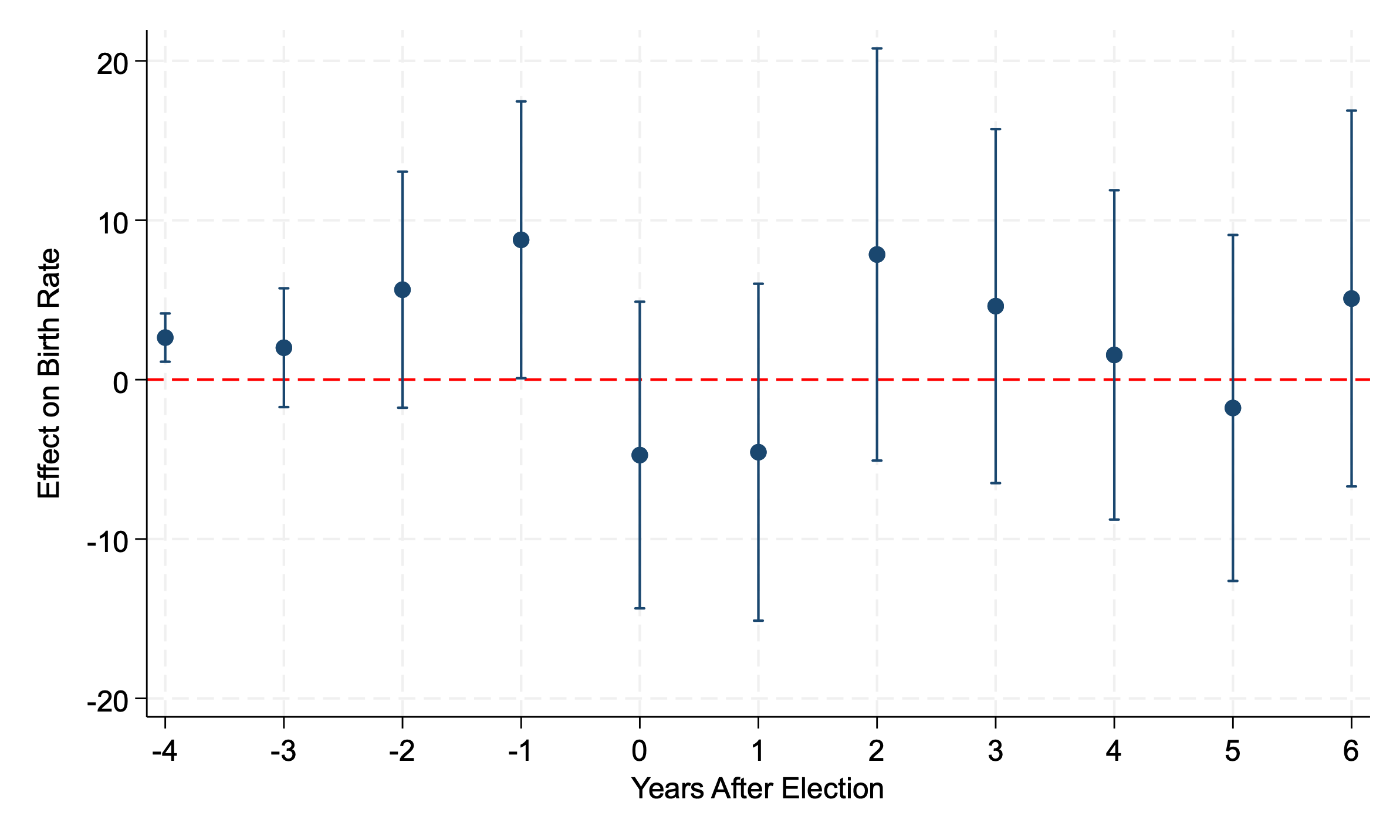}
\end{subfigure}
    \begin{minipage}{\textwidth} 
{\footnotesize \textbf{Notes:} This figure shows RD estimates of the effect on birth rate for cohorts aged (a) 9-10 at election, (b) 16-17 at election year, against the margin of victory, year by year from election year. The running variable is the margin of victory in 2008, 2012, or 2016 municipal elections, with positive values indicating Pentecostal victories. Adjusted for state fixed effects and electoral cycle fixed effects. See section \ref{sec:dynamic_effects} for exact specification.
}\end{minipage}
\end{figure}

Figure \ref{fig:event_study} panel (a) presents the dynamic evolution of treatment effects over time. The figure plots the estimates of the effect on birth rates for each year relative to the election, following girls who were aged 9-10 when the mayor took office. Unsurprisingly, the effects start at zero, since births at this age are rare. Effects remain small and statistically insignificant through years 1 and 2, when the cohort is aged 10-12 and just beginning middle school. Starting in year 3 (ages 12-13), point estimates begin to grow progressively larger and peak in years 5 and 6 (ages 14-16), when the cohort reaches peak adolescent fertility risk. By year 5, the estimated effect exceeds 6 additional births per 1,000 girls. The timing aligns precisely with when these cohorts would have maximum exposure to Pentecostal-appointed principals and reduced sexual education programming during their most formative educational years.

Figure \ref{fig:event_study} panel (b) shows the corresponding dynamic estimates for the older comparison cohort: girls aged 16-17 during the election year. The figure shows a strikingly different pattern. The estimates are negative during years 0 and 1 and positive elsewhere and follow no discernible pattern.

\subsection{Sexual Education and School-Based Mechanisms}\label{sex_educ}

The concentration of effects among school-age cohorts, combined with null effects for older cohorts and gradual emergence over time, provides compelling evidence that Pentecostal mayors influence outcomes through school-based interventions. To investigate which specific policies change, we examine sexual education provision in schools, exploiting institutional variation in mayoral authority: mayors control municipal schools through principal appointments but have no authority over state schools.

\subsubsection{Sexual Education Provision: Municipal vs. State Schools}

\begin{figure}[ht!]
    \centering
        \caption{Effect of Pentecostal Mayors on Sexual Education}
        \label{fig:sex_grav_mun}
    \begin{subfigure}[b]{0.95\textwidth}
     \caption{Municipal Schools}
    \includegraphics[width=0.98\textwidth]{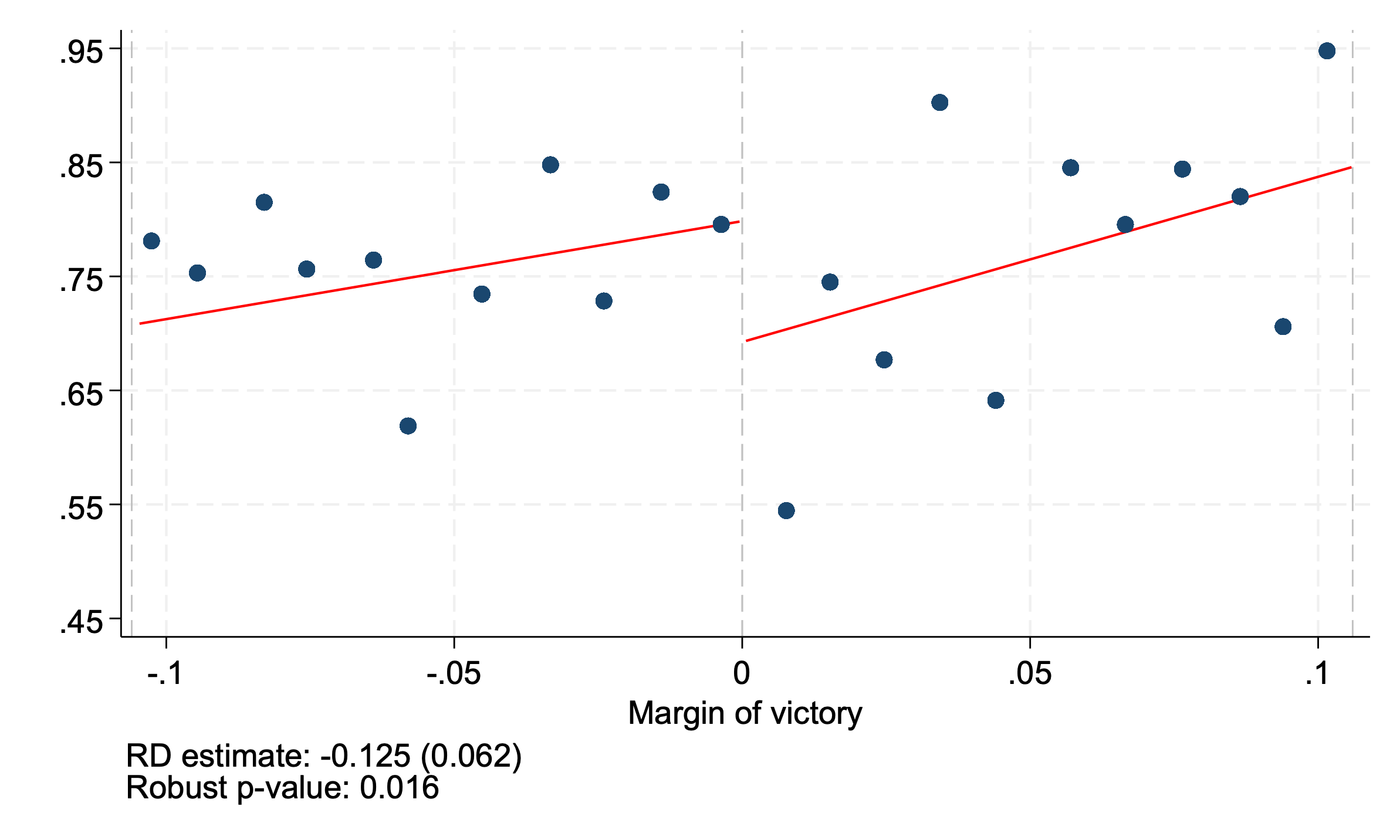} 
\end{subfigure}

\begin{subfigure}[b]{0.95\textwidth}
     \caption{State Schools}
    \includegraphics[width=0.98\textwidth]{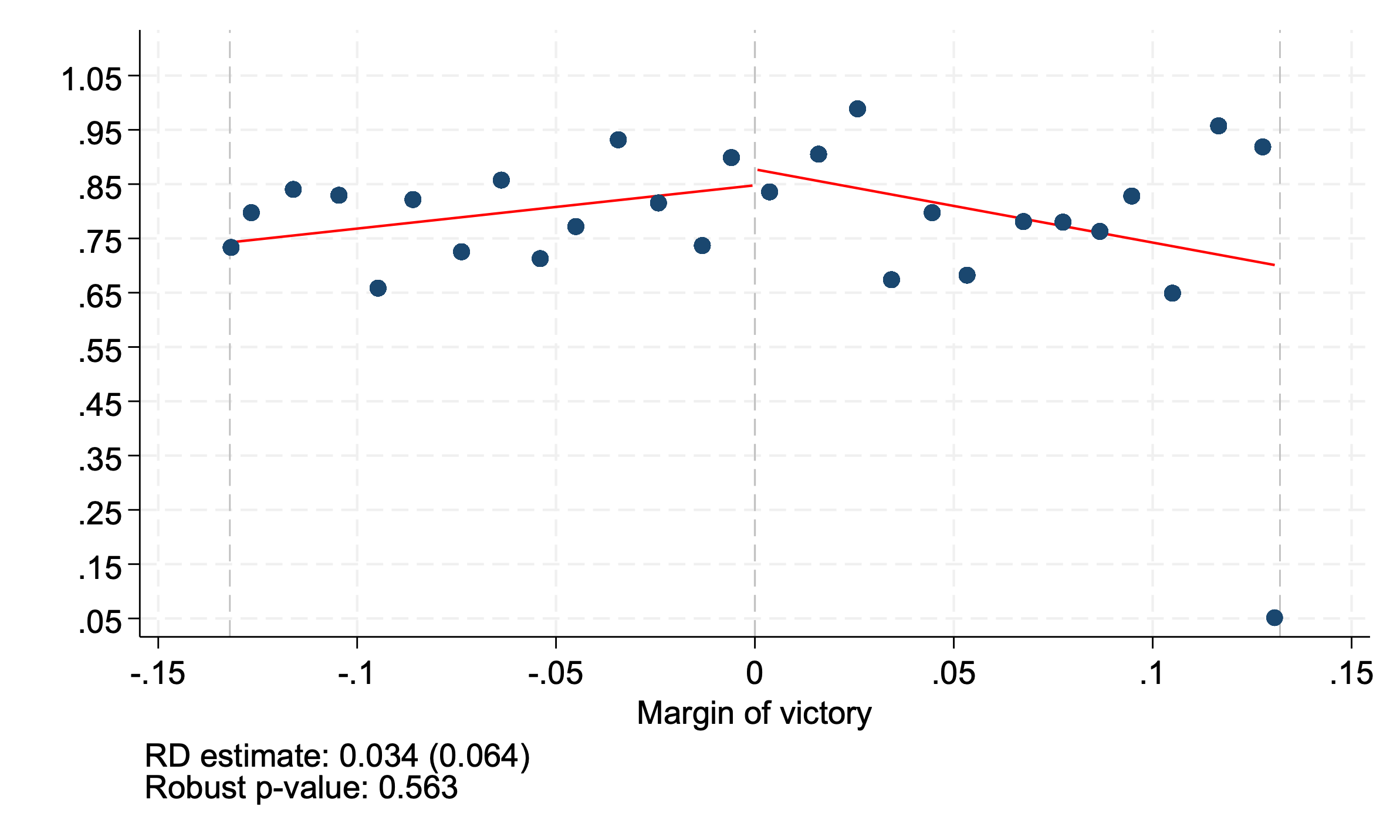}
\end{subfigure}
\begin{minipage}{\textwidth}{\footnotesize \textbf{Notes:} 
This figure shows the presence of sexual education courses in municipal (a) and state (b) schools. The running variable is the margin of victory in 2008, 2012, or 2016 municipal elections, with positive values indicating Pentecostal victories. Adjusted for state fixed effects, electoral cycle fixed effects. Linear fit, triangular kernel, optimal bandwidth calculated following \citet{calonico2014robust}.}\end{minipage}
\end{figure}

Figure \ref{fig:sex_grav_mun}, top panel, shows the covariate-adjusted share of municipal schools offering sexual education activities against the margin of victory. Results indicate that municipal schools are 12.5 percentage points less likely to include sexual education in the curriculum when Pentecostal parties win elections, representing an 18\% reduction (p = 0.016).

Critically, Figure \ref{fig:sex_grav_mun}, bottom panel, shows no similar changes in state schools, where mayors lack authority over principal selection and curriculum decisions. The absence of effects in state schools, which serve similar populations within the same municipalities, demonstrates that curriculum changes result specifically from mayoral control over municipal school administration rather than from broader political or social trends.

Table \ref{tab:sex_educ_muni} confirms the robustness of these findings across specifications. For municipal schools, the estimated reduction in sexual education provision is stable at 12-15 percentage points across all but one specification (Table \ref{tab:sex_educ_muni}, Columns 2-4), with statistical significance at conventional levels. For state schools, the estimate is near zero and statistically insignificant (column 5), with a 95\% confidence interval of $[-0.09, 0.16]$, allowing us to rule out reductions larger than 9.1 percentage points, well below the 12.5 pp reduction observed in municipal schools. This confirms that mayors do not influence schools beyond their administrative control.

To interpret these magnitudes, note that roughly 52\% of students in our sample are enrolled in state schools and 44\% are in municipal schools.\footnote{The remainder are in a combination of federal and private schools.} Under the extreme assumption that all effects operate exclusively through municipal schools, our main estimate of 3 births per 1,000 girls (cumulative over years 2--5 post-election) rescales to $\frac{3}{0.44}=6.8$ births per 1{,}000 \textit{municipal-school} girls over the same four-year window, or 1.7 births per 1{,}000 municipal-school girls per year. Combined with the 12.5 percentage point reduction in sexual education provision in municipal schools, this implies an increase of roughly $6.8/12.5=0.55$ births per 1{,}000 girls over four years (0.055 percentage points) for each 1 percentage point reduction in sexual education provision, under the strong assumption that sexual education is the sole operative channel.

\subsubsection{Other Curricular Changes}

One concern is that Pentecostal mayors might have broader effects on school content, which may influence fertility patterns in other ways. To address this, we examine several other optional courses: Environment, Inequality, Racism, Sexism, Violence, and Drug Prevention. Table \ref{tab:other_courses} presents the findings. We find no evidence of changes in most courses, with the notable exception of a reduction in drug prevention (p $<$ 0.01) and violence prevention (p $<$ 0.05) activities in municipal schools. This result suggests Pentecostal mayors selectively intervene in curriculum areas related to moral behavior, like sexuality, violence, and substance abuse, where their religious doctrine provides clear guidance, rather than broadly reducing optional programming.

Consistent with this interpretation, Table \ref{tab:other_courses} shows no changes in courses in state schools, except for a marginal increase in violence prevention course (p $<$ 0.10), confirming that effects operate specifically through mayoral control of municipal schools rather than through broader trends affecting all educational institutions.

\subsubsection{Principal Appointments}

To understand how mayors influence curriculum, we examine principal selection processes. Mayors have authority to appoint principals in municipal schools, and principals wield substantial control over which activities and programs schools offer.

\begin{figure}[ht!]
    \centering
        \caption{Effect of Pentecostal Mayors on School Principals}
        \label{fig:principals}
     \begin{subfigure}[b]{0.88\textwidth}
     \caption{Less than 2 years of experience \\ Municipal Schools}
    \includegraphics[width=0.88\textwidth]{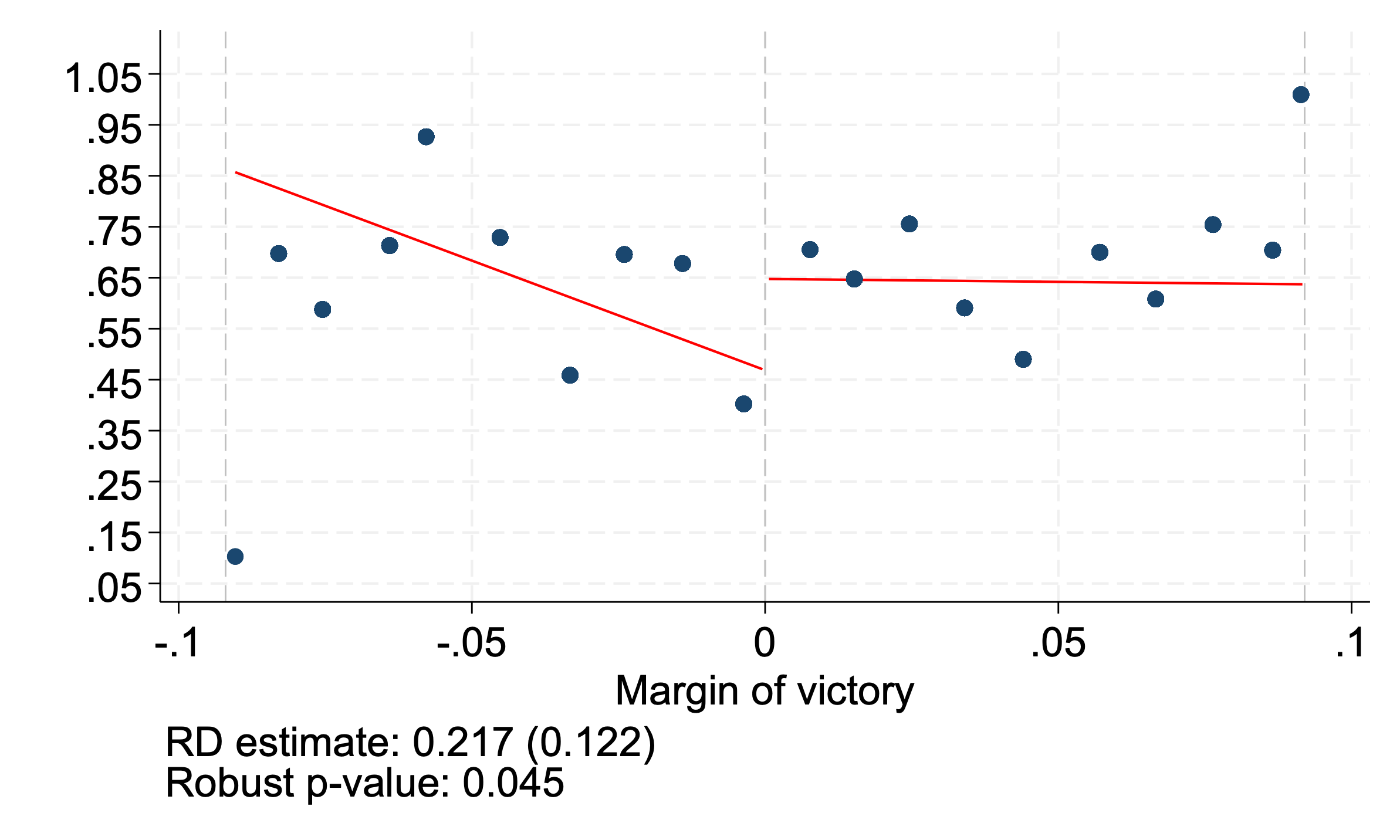} 
\end{subfigure}

\begin{subfigure}[b]{0.88\textwidth}
     \caption{Less than 2 years of experience \\ State Schools}
    \includegraphics[width=0.88\textwidth]{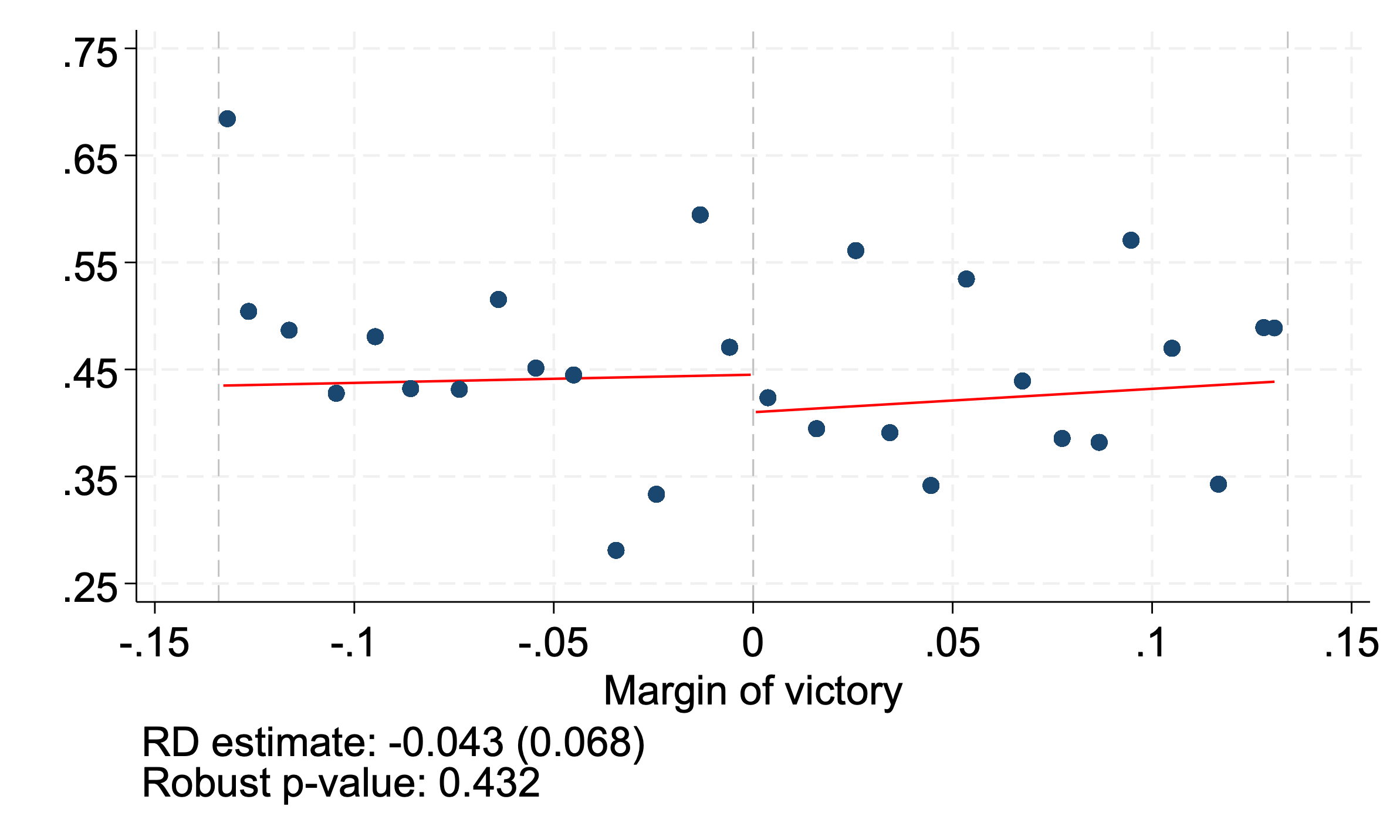} 
\end{subfigure}

     \begin{minipage}{0.85\textwidth} 
    {\footnotesize \textbf{Notes:} This figure shows rates of turnover of school principals for municipal and state schools. The running variable is the margin of victory in 2008, 2012, or 2016 municipal elections, with positive values indicating Pentecostal victories. Adjusted for state fixed effects, electoral cycle fixed effects. Linear fit, triangular kernel, optimal bandwidth calculated following \citet{calonico2014robust}.}
    \end{minipage}
\end{figure}

Figure \ref{fig:principals} shows that, under Pentecostal mayors, principals in municipal schools are more likely to be replaced and report external interference in school matters. Panel (a) shows a 21.7 percentage point increase in the probability that principals have less than 2 years of experience (p value = 0.045).\footnote{The data were collected in 2009, 2013, and 2019, in surveys as part of \textit{Prova Brasil}. Since this is the year after the elections, less than 2 years of experience means the principal was replaced in the first year of the new administration.} Panel (b) shows no increase in principal turnover in state schools. This evidence demonstrates that political appointment of principals represents one mechanism through which mayors can shape school policies to align with their ideological preferences.

\subsection{Sexually Transmitted Diseases}

To assess whether effects extend beyond pregnancy to other dimensions of sexual health, we analyze sexually transmitted disease diagnoses. Using the same cohort-based approach, we track syphilis diagnoses for girls aged 9–10 at the time of election during years 2–5 post-election (when they are aged 11–15), applying the same observation window as for birth outcomes. We focus on syphilis because reporting is mandatory. 

\begin{figure}[ht!]
    \centering
        \caption{Effect of Pentecostal Mayors on Sexually Transmitted Diseases}
        \label{fig:stds}
     \begin{subfigure}[b]{0.88\textwidth}
    \centering
     \caption{Syphilis cases among teenagers}
    \includegraphics[width=0.88\textwidth]{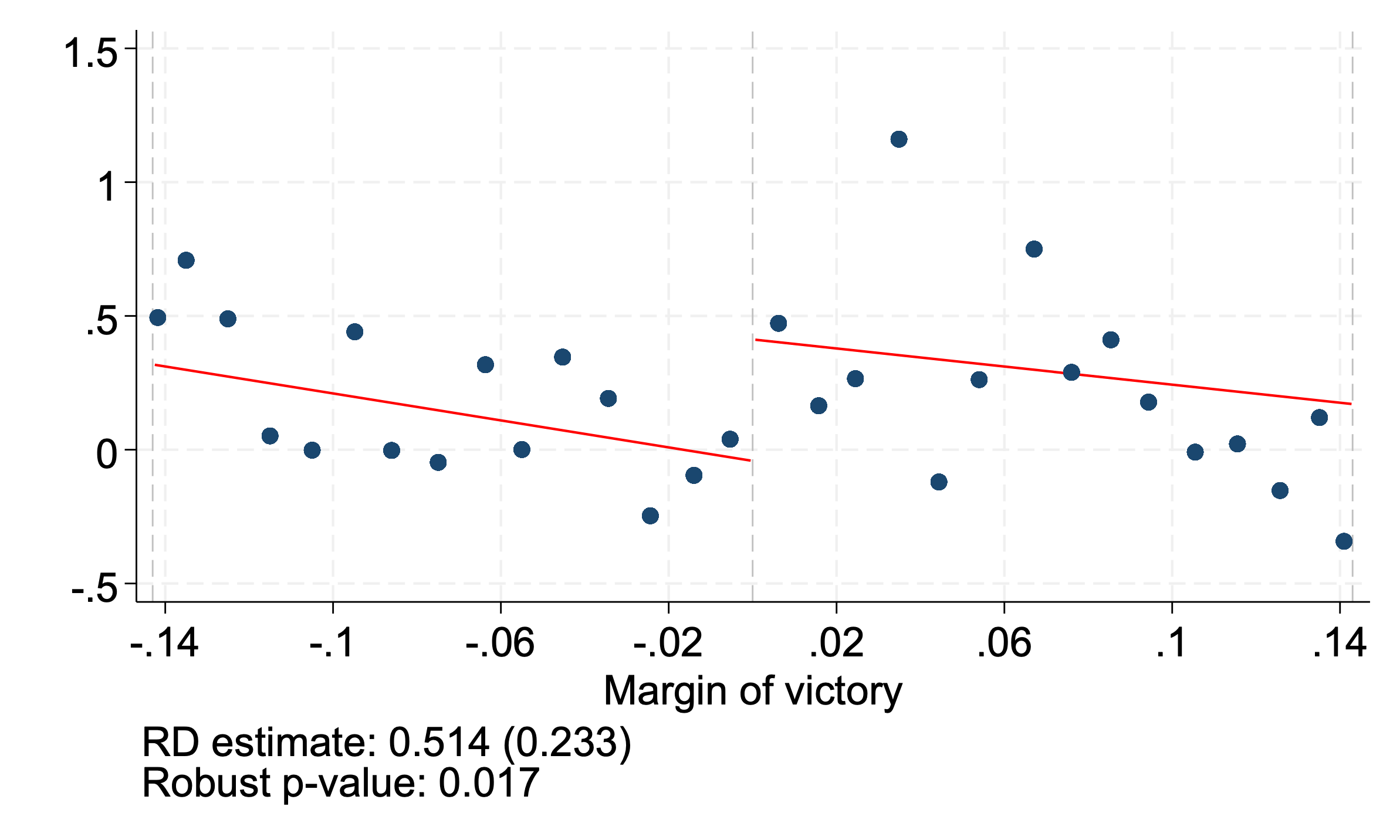} 
\end{subfigure}

\begin{subfigure}[b]{0.88\textwidth}
    \centering
     \caption{Syphilis cases among young adults}
    \includegraphics[width=0.88\textwidth]{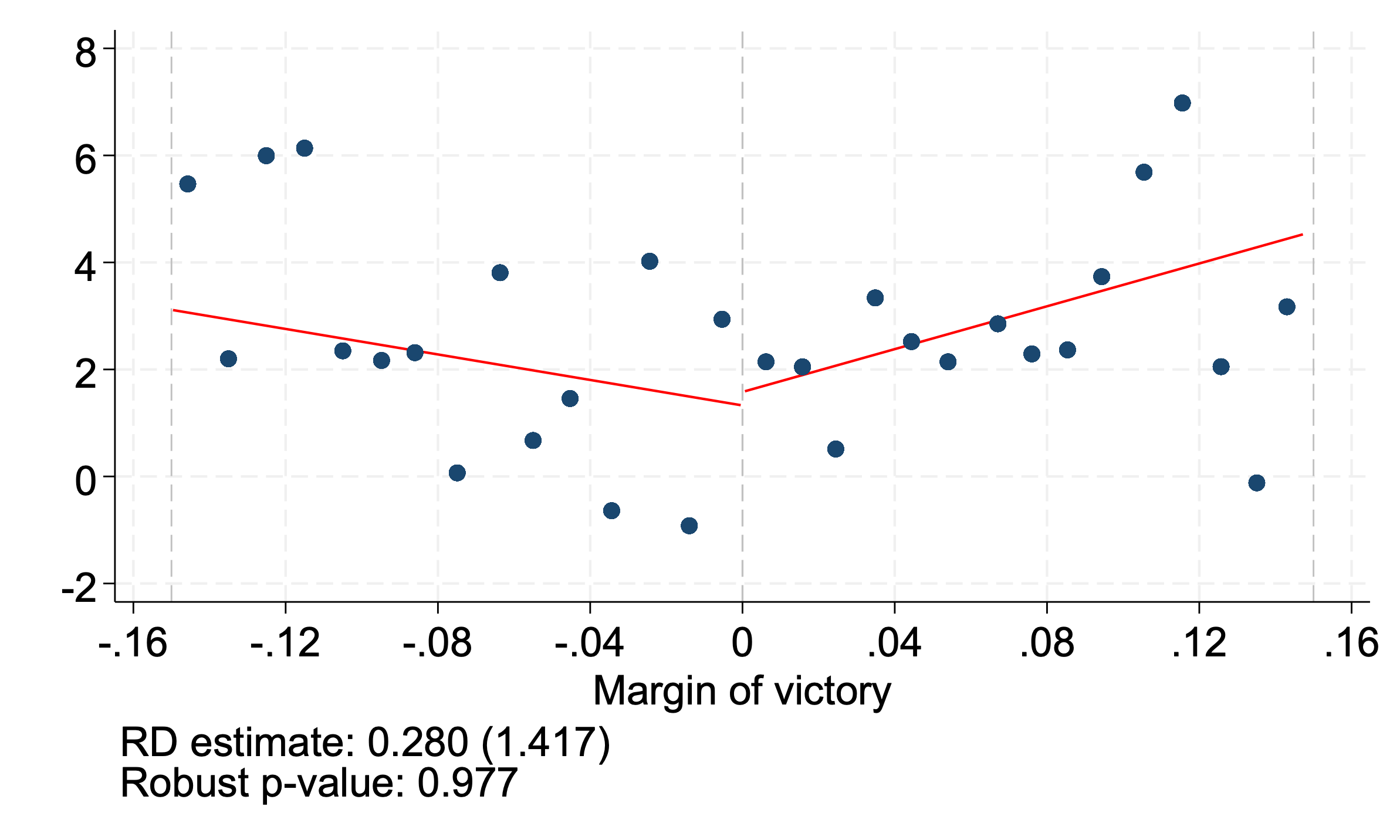}
\end{subfigure}
    
     \begin{minipage}{0.85\textwidth} 
    {\footnotesize \textbf{Notes:} This figure shows rates of syphilis among populations: (a) teenagers (9-10 years-old during election year), and (b) young adults (16-17 during election year). The running variable is the margin of victory in 2008, 2012, or 2016 municipal elections, with positive values indicating Pentecostal victories. Adjusted for state fixed effects, electoral cycle fixed effects. Linear fit, triangular kernel, optimal bandwidth calculated following \citet{calonico2014robust}.}\end{minipage}

\end{figure}

Panel (a) of Figure \ref{fig:stds} reveals a statistically significant increase of 0.514 syphilis cases per 1,000 cohort members (p = 0.017). Given the low baseline STD risk for this population (0.17 per 1,000), this increase represents a roughly 300\% increase. Panel (b) presents results for young adults (16-17 during election year). Syphilis rates in this population did not change following the election of Pentecostal mayors (RD estimate = 0.28, p = 0.977), suggesting that whatever policies Pentecostal mayors altered, they affected only the school-age population. We also analyze HIV cases, but we find no significant differences (teenagers RD estimate 0.01, p value 0.46; young adults RD estimate: 0.53, p-value 0.17), possibly because baseline rates are roughly one order of magnitude lower than those of syphilis in both populations.

To contextualize the findings, it should be noted that syphilis in teenagers is rare and the distribution is heavily skewed, with the large majority of municipality-cycles reporting zero cases among the treated cohort. This skewness is a predicted feature of the epidemiology rather than a statistical concern: STD transmission follows nonlinear dynamics in which the basic reproduction number depends on local prevalence and contact patterns. A treatment that increases unprotected sexual activity should therefore produce a distribution with many zeros and occasional local outbreaks, municipalities where the pathogen is already circulating see amplified transmission, while municipalities without local prevalence see no change, rather than uniform small increases everywhere.

Consistent with this interpretation, the extensive margin (whether a municipality records \textit{any} syphilis cases) is not statistically significant, and neither is the syphilis rate after trimming the top and bottom 1\% of the distribution (Table \ref{tab:sensitivity}, Panel B). These results confirm that the syphilis effect is driven by the intensive margin, larger outbreaks in the subset of municipalities where cases occur, rather than by a broad shift affecting all municipalities. This is the expected pattern for an infectious disease: the policy effect does not create outbreaks from nothing, but amplifies transmission where conditions for spread are already present. A leave-one-out exercise (Figure \ref{fig:leave_one_out}) shows that no single municipality drives the result: dropping the most influential treated municipality reduces the estimate from 0.51 to 0.37, while the remaining leave-one-out coefficients range from 0.41 to 0.57, with 90\% falling between 0.50 and 0.53.

\subsection{HPV Vaccination}

Brazil introduced HPV vaccination in 2014 for girls aged 9--14, with school-based campaigns serving as a primary delivery channel. Because Pentecostal mayors control the school system through which these campaigns are partly administered, vaccination coverage provides a direct test of whether these mayors affect health service delivery through schools.

\begin{figure}[ht!]
    \centering
    \caption{Effect of Pentecostal Mayors on HPV Vaccination Coverage (\%)}
    \label{fig:hpv}
    \includegraphics[width=0.88\textwidth]{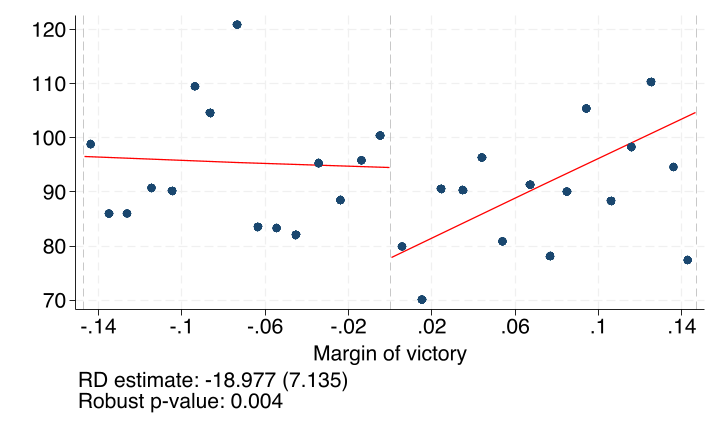}

    \begin{minipage}{0.85\textwidth}
    {\footnotesize \textbf{Notes:} This figure shows cumulative HPV vaccination coverage for girls aged 9--11, measured at the last year of each mayoral term (2016 and 2020). The running variable is the margin of victory in 2012 or 2016 municipal elections, with positive values indicating Pentecostal victories. Adjusted for state fixed effects, electoral cycle fixed effects. Linear fit, triangular kernel, optimal bandwidth calculated following \citet{calonico2014robust}.}
    \end{minipage}
\end{figure}

Figure \ref{fig:hpv} presents RD estimates for cumulative HPV vaccination coverage measured at the end of each mayoral term. We focus on girls aged 9--11, whose vaccination history falls predominantly within the current administration (coverage for older girls includes doses administered under the previous mayor). Pentecostal mayors reduce average HPV coverage for this group by 19.0 percentage points (p = 0.004). Examining each age separately reveals a clear exposure gradient: ages 9, 10, and 11 show large negative effects, while ages 12--14, whose cumulative coverage reflects mostly the previous administration, show no effect (Table \ref{tab:hpv}). The result is robust to alternative bandwidths, polynomial orders, and kernel choices (Table \ref{tab:hpv}).

\subsection{Educational Attainment}

\begin{figure}
    \centering
    \caption{Effect of Pentecostal Mayors on Girls' Dropout Rates (\%)}
    \label{fig:dropout}
     \begin{subfigure}[b]{0.88\textwidth}
     \caption{Municipal Schools}
    \includegraphics[width=0.98\textwidth]{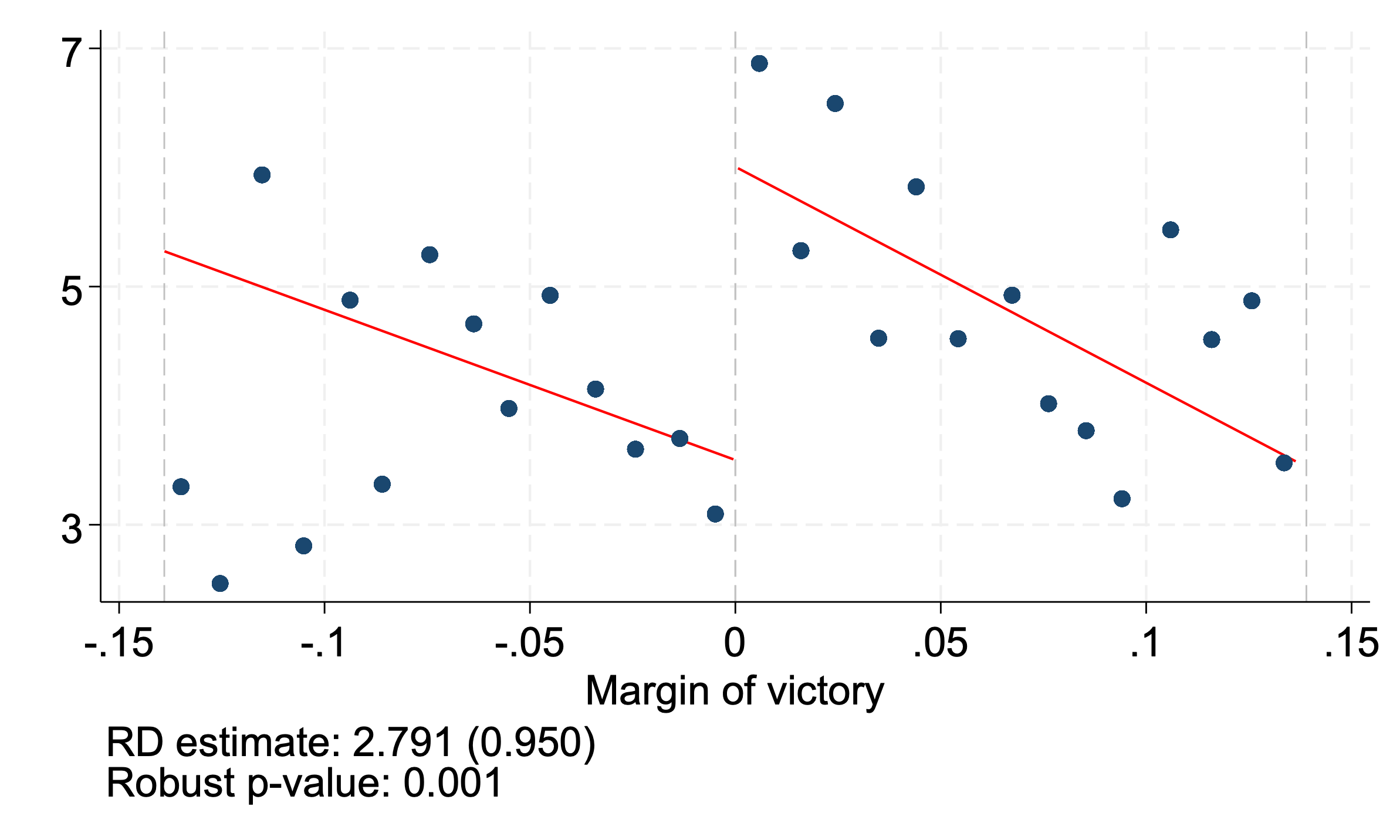} 
\end{subfigure}

\begin{subfigure}[b]{0.88\textwidth}
     \caption{State Schools}
    \includegraphics[width=0.98\textwidth]{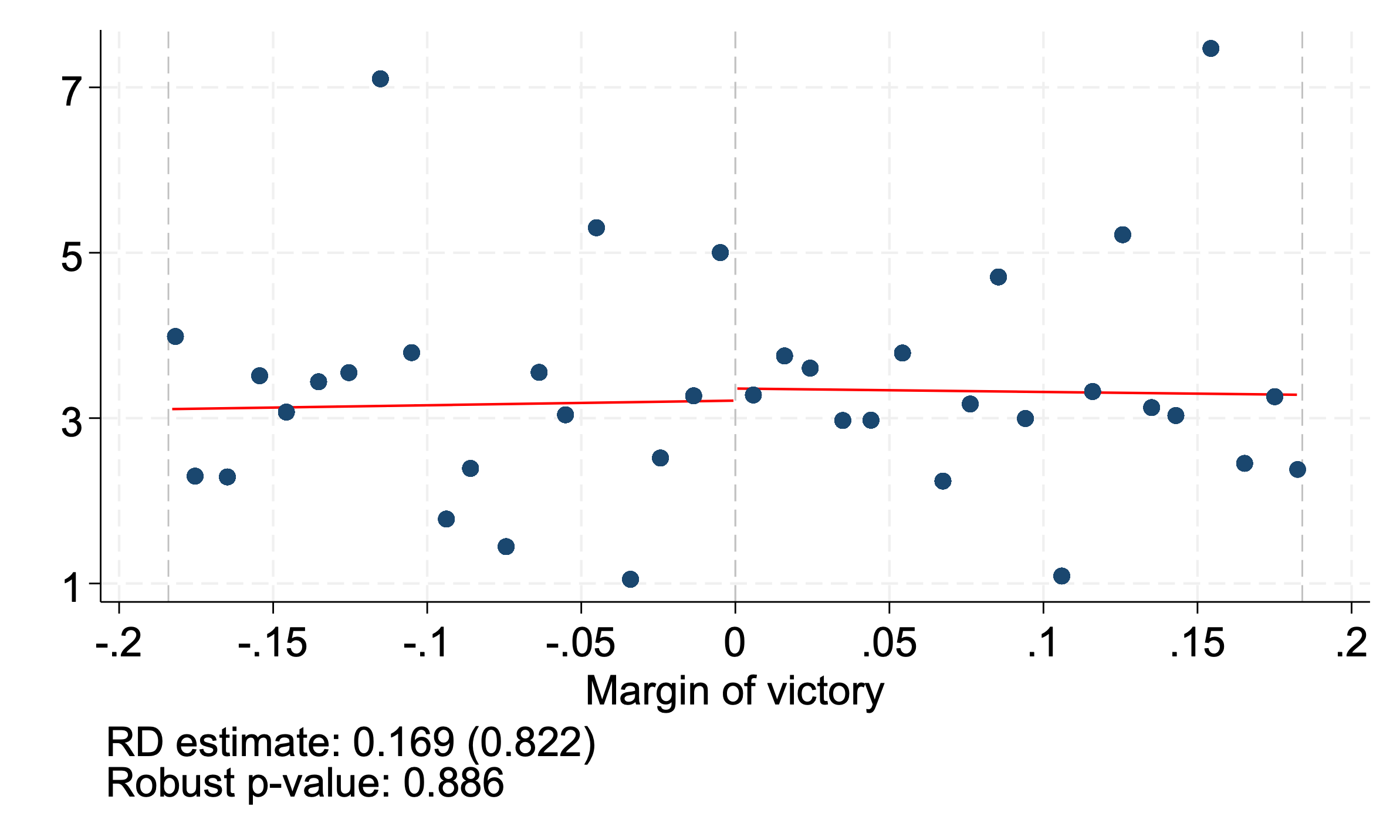}
\end{subfigure}

     \begin{minipage}{0.85\textwidth} 
        {\footnotesize \textbf{Notes:} This figure shows the dropout rate for girls at the end of middle school for cohorts 9-10 during election year, in percentage points. The running variable is the margin of victory in 2008, 2012, or 2016 municipal elections, with positive values indicating Pentecostal victories. Adjusted for state fixed effects, electoral cycle fixed effects. Linear fit, triangular kernel, optimal bandwidth calculated following \citet{calonico2014robust}.  \par}
        \end{minipage}

\end{figure}

Beyond curriculum changes, Pentecostal mayors may affect educational attainment if teenage pregnancy leads girls to leave school. Figure \ref{fig:dropout} examines dropout rates at the end of middle school (8th-9th grade) for girls in the treatment cohort. Panel (a) shows that municipal schools experience a 2.8 percentage point increase in female dropout rates under Pentecostal mayors (p = 0.001), representing a substantial increase from a baseline of approximately 4 percent. Panel (b) confirms that state schools, outside mayoral control, show no corresponding change (estimate: 0.17 pp, p = 0.886). This pattern mirrors the sexual education findings: effects emerge only in schools where mayors exercise administrative authority.

The dropout effect is consistent with teenage pregnancy forcing girls out of school, though the causal direction could also run partly in reverse if reduced educational engagement increases pregnancy risk. Regardless of the precise pathway, these results suggest that Pentecostal governance generates lasting human capital consequences beyond the immediate fertility effects.

\subsection{Heterogeneity}

If Pentecostal mayors affect fertility primarily through the removal of school-based sexual education, the effect should be larger in municipalities where more schools offered such programming before the election. In contrast, if the mechanism operates through broader cultural or normative shifts associated with Pentecostal political power, the effect should vary with the local evangelical population share.

We split the sample at the median of the municipality-level share of municipal schools that reported offering sexual education in the previous electoral cycle, a measure of ex-ante exposure to the content that Pentecostal mayors seek to suppress. The two subsamples are strikingly different: in the high-provision group, 98.1\% of municipal schools offered sex education at baseline, compared to 49.0\% in the low-provision group. The point estimate is substantially larger in municipalities with high baseline sex education provision (3.2 per 1,000) than in municipalities with low baseline provision ($-$0.4 per 1,000), though neither subsample estimate is individually significant due to reduced power (Figure \ref{fig:het_sexed}). This pattern is consistent with a mechanism that operates through the removal of existing school-based programming rather than through broader cultural or normative shifts.

In contrast, splitting the sample by local evangelical population share yields nearly identical effects in both subgroups (Figure \ref{fig:het_evangelical}), suggesting that the channel runs through institutional control of schools rather than through the diffusion of Pentecostal values in the community.

\subsection{Alternative Mechanisms}

\textbf{Access to contraceptives} The birth rate effects could arise through channels other than sexual education if Pentecostal mayors restricted access to contraceptives or abortion services. Brazilian mayors control municipal health centers (\textit{Unidades Básicas de Saúde}), which serve as the primary source of free contraceptives for low-income populations. If Pentecostal mayors reduced contraceptive availability, whether through policy directives, funding decisions, or staffing choices, this could independently increase birth rates regardless of school curriculum changes.

Table \ref{tab:contraceptives} examines the availability of six categories of contraceptives in municipal health centers: oral contraceptives, injectable contraceptives, male condoms, female condoms, IUDs, and emergency contraception (morning-after pill). We find no evidence that Pentecostal mayors affect the availability of any contraceptive method. Point estimates are uniformly small and statistically insignificant, with p-values ranging from 0.139 to 0.927. For male condoms, the most widely available method (baseline: 98\%), the 95\% confidence interval is $[-0.02, 0.04]$, ruling out reductions larger than 2 percentage points. For oral contraceptives (baseline: 80\%), we can rule out reductions larger than 15 percentage points. Even emergency contraception, which might be a particular target of religious opposition, shows no reduction (estimate: 0.01, p = 0.927). The high baseline availability rates (70--98 percent for most methods) suggest that contraceptive access is nearly universal in these health centers and remains so regardless of the mayor's religious affiliation.

\begin{figure}
    \centering
    \caption{Abortion-related hospitalization and fetal death rates}
    
\begin{subfigure}[b]{0.8\textwidth}
     \caption{Abortion}
    \includegraphics[width=\textwidth]{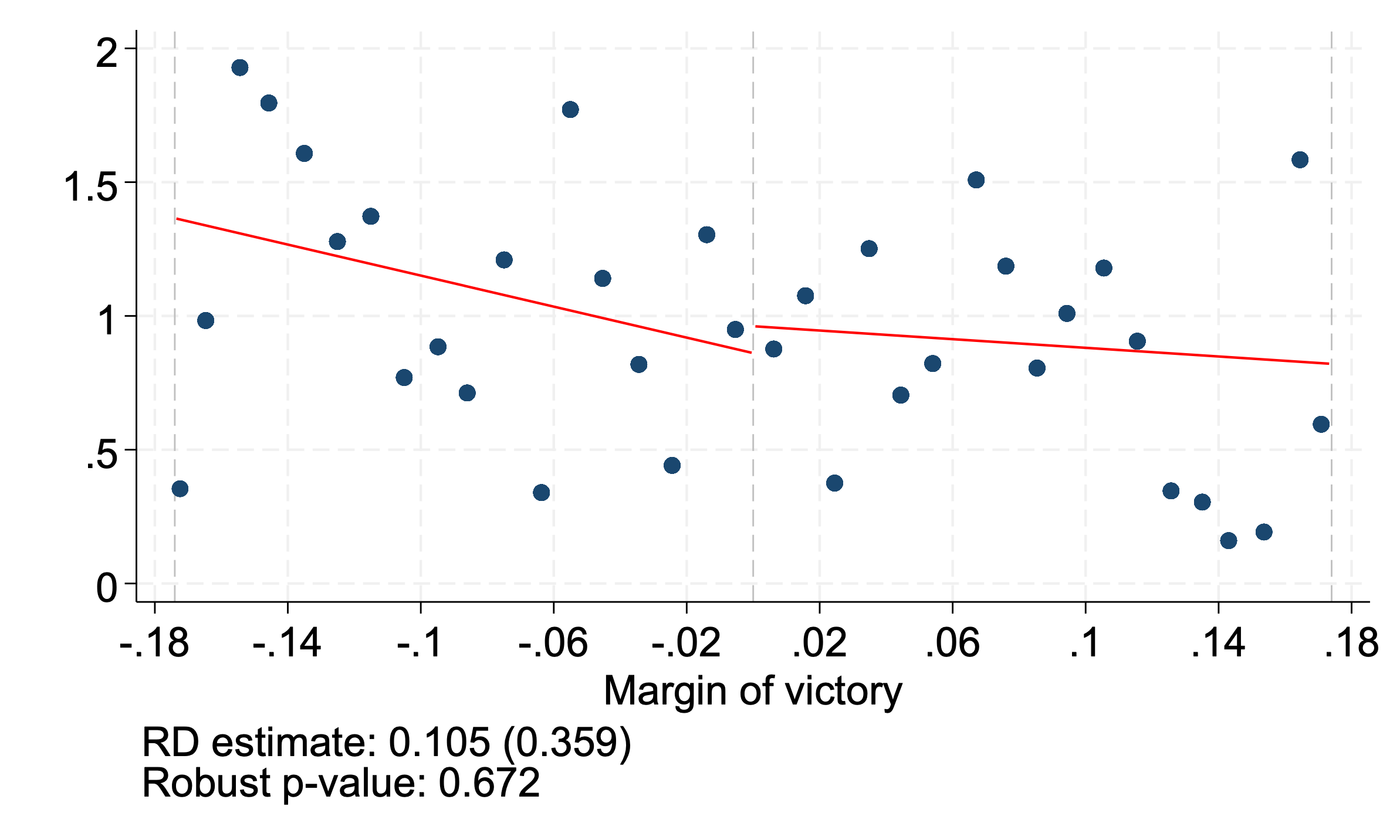} 
\end{subfigure}

\begin{subfigure}[b]{0.8\textwidth}
     \caption{Fetal Deaths}
    \includegraphics[width=\textwidth]{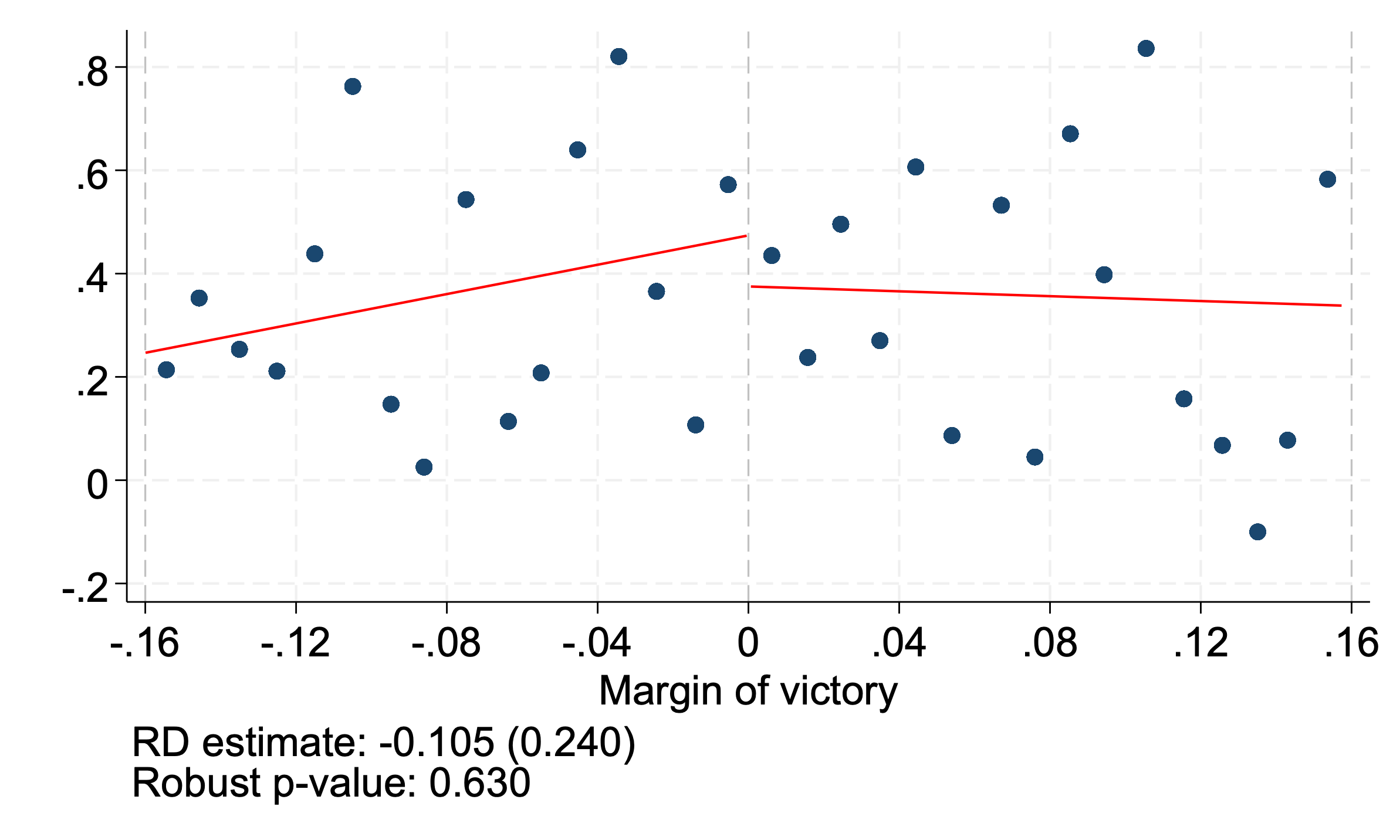}
\end{subfigure}

    \label{fig:abortion}
         \begin{minipage}{0.85\textwidth} 
        {\footnotesize \textbf{Notes:} This figure shows rates of abortion-related hospitalizations of teenagers and fetal deaths from teenage mothers, both by 1,000. The running variable is the margin of victory in 2008, 2012, or 2016 municipal elections, with positive values indicating Pentecostal victories. Adjusted for state fixed effects, electoral cycle fixed effects. Linear fit, triangular kernel, optimal bandwidth calculated following \citet{calonico2014robust}.}
        \end{minipage}
\end{figure}

\textbf{Abortion} We also examine whether Pentecostal governance affects abortion access or pregnancy outcomes conditional on conception. Although abortion is illegal in Brazil except in narrow circumstances, abortion-related hospitalizations provide a proxy for the incidence of pregnancy termination, as complications from clandestine procedures frequently result in medical care. Figure \ref{fig:abortion} shows no discontinuity in either abortion-related hospitalizations (estimate: 0.105 per 1,000, p = 0.672) or fetal deaths (estimate: -0.105 per 1,000, p = 0.630) among teenage mothers. The absence of effects on these margins reinforces that Pentecostal mayors do not appear to operate through health system restrictions. Combined with the contraceptive findings, these results support the interpretation that school-based sexual education, rather than reproductive health access, is the primary mechanism through which Pentecostal governance affects teenage fertility.

\textbf{General Conservativeness} A potential concern is that results reflect general conservative ideology rather than specific religious orientation. Pentecostal parties' policy positions overlap substantially with non-religious conservative parties, making it difficult to distinguish the effects of Pentecostal religiosity per se from conservativeness. To address this, we examine close races between non-Pentecostal right-wing parties and other parties. 

We use the party classifications based on an average of expert surveys from \citet{tarouco2015partidos}. We classify as ``right-wing parties" all parties with a mean score of 5 or more, in a measure where 1 is the most left-wing and 7 the most right-wing.\footnote{The resulting right-wing parties are PSDC, PR, PRP, PP, and DEM.} For comparison, PRB and PSC, the parties we consider linked to Pentecostalism, score 5.1 and 5.2, respectively. We exclude any elections where both candidates were from either a right-wing party or a Pentecostal party. The resulting dataset contains 2{,}872 elections. 

\begin{figure}[ht!]
     \centering
         \caption{Effect of Right-Wing Parties on Teenage Pregnancy and Sexual Education in Schools}
         \begin{subfigure}[b]{0.95\textwidth}\centering
     \caption{Teenage Birth Rate}
    \includegraphics[width=0.8\textwidth]{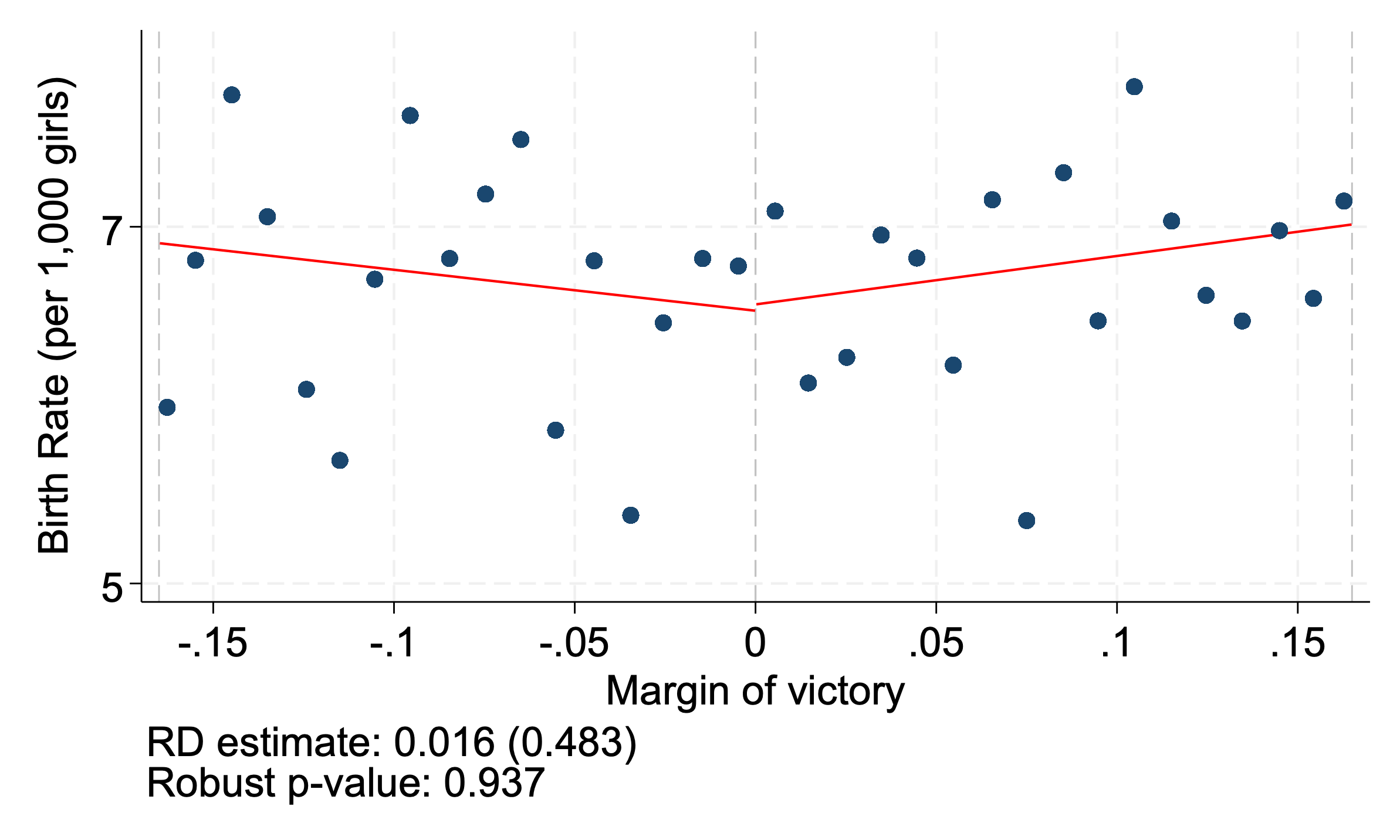} 
\end{subfigure}

\begin{subfigure}[b]{0.95\textwidth}\centering
     \caption{Sexual Education in Municipal Schools}
    \includegraphics[width=0.8\textwidth]{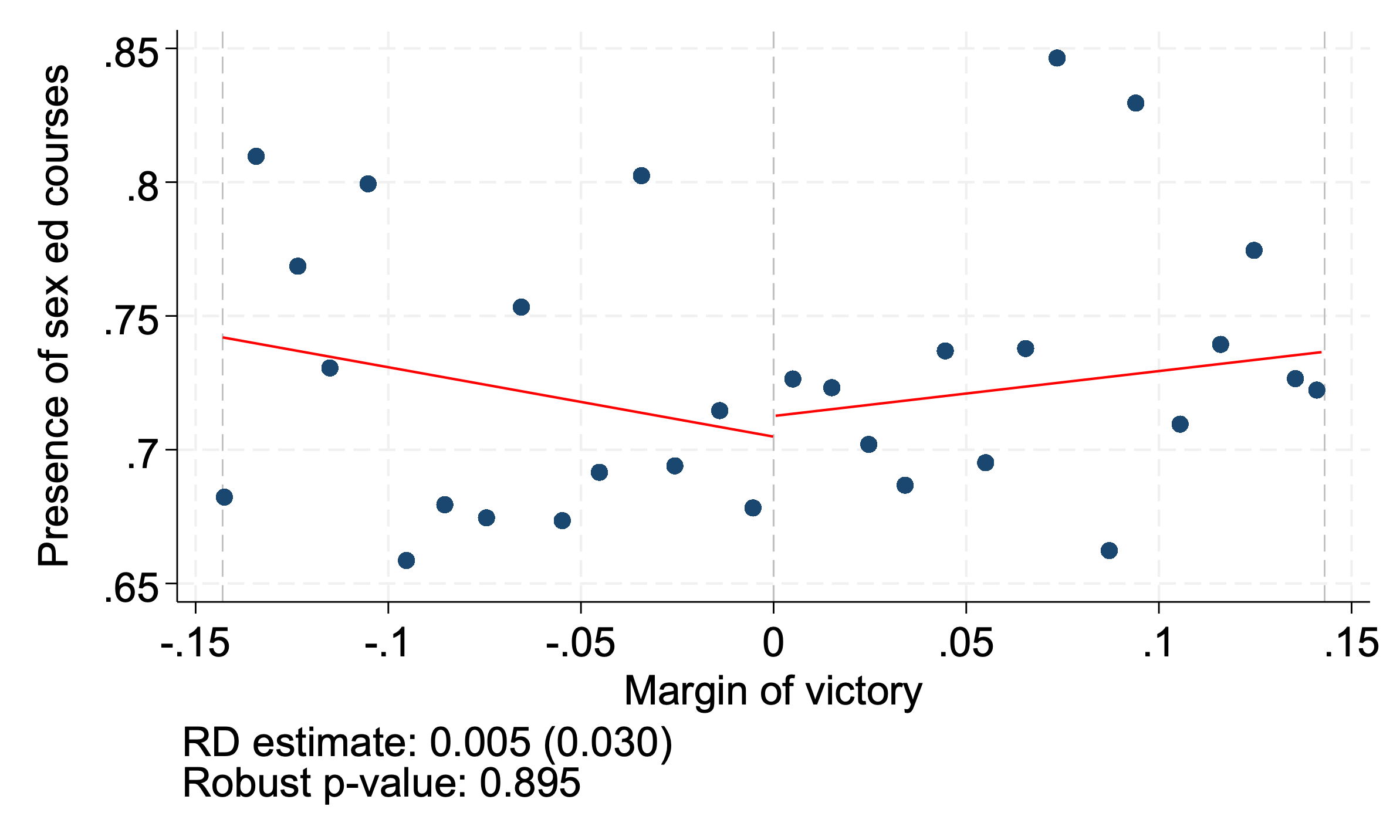}
\end{subfigure}
     \label{fig:right_wing}
      \begin{minipage}{0.85\textwidth} 
     {\footnotesize \textbf{Notes:} This figure shows (a) the birth rate for cohorts aged 9-10 at election against the margin of victory, cumulative over years 2-5 post-election (when the cohort is aged 11-15); (b) the presence of sexual education courses in municipal schools. The running variable is the margin of victory in 2008, 2012, or 2016 municipal elections, with positive values indicating right-wing victories. All elections where either of the top two candidates are from Pentecostal parties are excluded from the sample. Adjusted for state fixed effects, electoral cycle fixed effects and the lagged dependent variable. Linear fit, triangular kernel, optimal bandwidth calculated following \citet{calonico2014robust}}\end{minipage}

 \end{figure}

Figure \ref{fig:right_wing} shows no statistically significant effects on teenage pregnancy or sexual education provision when non-Pentecostal conservative parties win. The point estimate on teenage births is 0.016 (p = 0.937), and on sexual education in municipal schools is 0.003 (p = 0.930), both essentially zero and orders of magnitude smaller than for Pentecostal parties. Notably, the right-wing sample is four times larger (2{,}872 elections), providing substantially more statistical power than the Pentecostal analysis. This pattern suggests that effects are specifically tied to religious orientation rather than general conservative ideology, though we cannot completely rule out that Pentecostal parties differ from other conservatives in dimensions beyond religion.

\textbf{Other mechanisms} We conduct several additional tests to rule out alternative channels, with full results reported in the Appendix. Pentecostal mayors do not affect staffing of nurses, doctors or community health workers in the \textit{Estrat\'{e}gia Sa\'{u}de da Fam\'{i}lia} primary care program (Table \ref{tab:esf_staffing}). We find no net entry of Pentecostal or Catholic churches (Table \ref{tab:churches}). We do not find differences in rates of civil marriage among youth or average age at marriage (Table \ref{tab:marriages}). We see no differences in the allocation of municipal budget shares across health, education, social assistance, and culture (Table \ref{tab:expenditures}). We also find no discontinuity in population growth rates using actual census counts from 2000, 2010, and 2022, providing evidence against selective migration affecting the municipal population. The null effect on marriages is particularly informative for welfare interpretation: the additional births are not accompanied by transitions into formal partnerships, making it harder to frame the fertility increase as a revealed preference for earlier family formation, though we cannot observe informal cohabitation (\textit{uni\~{a}o est\'{a}vel}).

Examining the effect on community social assistance centers (\textit{Centros de Referência de Assistência Social}, CRAS), we find null effects on the number of community activities targeted towards the youth, the number of families receiving special social services attention (\textit{Programa de Atenção Integral à Família}, PAIF). The one notable finding is that CRAS social assistance centers report increased articulation with education services under Pentecostal mayors, with no change in health-sector articulation, consistent with targeted coordination through the school system rather than a general shift in municipal governance (Table \ref{tab:cras}).

\section{Discussion}

An important limitation of our analysis is that we cannot directly link individual girls to the schools they attended. Birth records identify mothers' municipality of residence but not their educational history. However, the combination of changes confined to institutions under mayoral control, the concentration of fertility effects among school-age cohorts, and the gradual emergence consistent with cumulative exposure, provides compelling support for a school-based pathway.

Moreover, what we measure captures only the extensive margin of sexual education provision: whether schools report offering any programming at all. We do not observe curriculum content, instructional quality, or frequency. The measured reduction in formal sexual education is best understood as one visible marker of a broader reorientation of the school environment under Pentecostal governance, not as a sufficient statistic for the full mechanism. That other curriculum topics, environment, social inequality, racism, sexism and homophobia, violence, and drugs, are unaffected suggests this reorientation is targeted specifically at sexuality-related content, rather than reflecting general school quality deterioration.

Our evidence identifies the school system as an important mechanism, but it does not rule out complementary channels operating outside schools. Pentecostal mayors may also shift community norms, facilitate church-school connections, or alter the broader informational environment in ways that affect adolescent behavior. Such channels are harder to isolate empirically: to the extent they affect all age groups, they would be absorbed by our cohort comparison, which shows null effects. The large and cohort-specific effects we document are therefore most parsimoniously explained by school-based changes, though we interpret our estimates as capturing the total effect of Pentecostal governance, which may include diffuse cultural shifts that happen to be concentrated among school-age populations exposed to institutional changes.

Our results are identified from close elections in municipalities where Pentecostal-party candidates were competitive. The average municipality in our sample has approximately 20,000 inhabitants, which is close to the median Brazilian municipality. However, because Brazil's population distribution is highly concentrated, roughly 60 percent of the national population lives in the larger municipalities excluded by our sample restrictions. The effects could differ in these larger urban centers, where institutional constraints on mayoral discretion are stronger, a greater share of students attend state or private schools, and alternative sources of information about sexual health are more available. Our findings speak most directly to smaller municipalities where mayors exercise substantial control over educational institutions and where Pentecostal churches are most embedded in local governance, arguably the setting where this form of religious political influence is most consequential.

\section{Conclusion}

This paper provides some of the first quasi-experimental evidence that the ascent of religious movements to political executive office has immediate and significant consequences for public service delivery and demographic outcomes. By exploiting close mayoral elections in Brazil, we demonstrate that Pentecostal mayors use their administrative discretion, specifically the power to appoint school principals, to de-emphasize sexual education in municipal schools. This ideological shift in the classroom is not merely symbolic; it results in a 40 percent increase in teenage birth rates among exposed cohorts and a sharp rise in sexually transmitted infections.

To illustrate the practical magnitude, the average municipality in our sample has approximately 400 girls in the treated cohort. Each Pentecostal mayoral victory therefore generates roughly 4.8 additional teen births over the four-year term, approximately 5 additional female middle school dropouts from municipal schools, and about 0.8 additional adolescent syphilis cases. Across the 342 Pentecostal-won municipality-cycles in our sample, these per-municipality effects imply on the order of 1,600 additional teen births and 1,700 additional female dropouts over the 2008--2016 period, though extrapolation from close-election estimates to all Pentecostal victories requires assuming similar effects away from the threshold.

Our findings contribute to the political economy of decentralization by illustrating that local discretion can be a double-edged sword. While decentralization allows policies to reflect local values, it also enables organized ideological groups to reshape essential services in ways that may negatively impact human capital and public health. That these changes occur without any observable shift in budgets, legislation, or health-system access underscores how personnel discretion can serve as an invisible channel of ideological influence, one that escapes the formal checks typically applied to local executives. Crucially, the absence of similar effects following the election of non-Pentecostal conservative mayors suggests that these outcomes are driven by specific religious institutional commitments rather than general right-wing ideology.

The significant increase in school dropouts among affected girls suggests that the long-term economic ``tail" of religious political power may be substantial. The reduction in HPV vaccination adds a further dimension: because HPV vaccination protects against cervical and other cancers, lower coverage among exposed cohorts may translate into elevated cancer risk decades later. Future research should track these cohorts into adulthood to measure the impact on labor force participation, lifetime earnings, and the intergenerational transmission of religiosity and socioeconomic status. Additionally, more work is needed to understand the ``bureaucratic friction" within schools, specifically, how career teachers respond to politically appointed principals who seek to restrict curriculum. Finally, as Pentecostalism continues to grow globally, examining whether these patterns hold in other decentralized democracies in Sub-Saharan Africa and Latin America remains an urgent task for scholars of development and political economy.

\clearpage
\bibliography{ref}
\clearpage

\section*{Figures}

 \clearpage

\clearpage
\section*{Tables}

\begin{table}[h]
    \centering\footnotesize
    \caption{Summary}
    \label{tab:descriptive}
\begin{tabular}{l cc c cc c}
\hline\hline
 & \multicolumn{3}{c}{Full Sample} & \multicolumn{3}{c}{Within Bandwidth} \\
\cmidrule(lr){2-4} \cmidrule(lr){5-7}
 & Pentec. & Non-Pent. & p-val. & Pentec. & Non-Pent. & p-val. \\
 & (1) & (2) & (3) & (4) & (5) & (6) \\
\hline
\multicolumn{7}{l}{\textit{Candidate Characteristics}} \\
Age & 47.82 & 48.66 & 0.293 & 48.12 & 49.33 & 0.248 \\
Female (\%) &  8.5 & 13.4 & 0.043 &  9.7 & 13.6 & 0.230 \\
College degree (\%) & 46.2 & 48.2 & 0.615 & 44.9 & 45.3 & 0.933 \\
Married (\%) & 72.5 & 77.9 & 0.106 & 74.1 & 77.6 & 0.398 \\
\hline
\multicolumn{7}{l}{\textit{Municipality Characteristics}} \\
Population (log10) &  4.12 &  4.10 & 0.528 &  4.10 &  4.10 & 0.983 \\
Income per capita & 12.01 & 10.21 & 0.084 & 10.66 &  9.52 & 0.396 \\
Urban (\%) & 62.0 & 61.9 & 0.981 & 61.5 & 60.3 & 0.552 \\
Literacy (\%) & 80.7 & 81.4 & 0.335 & 80.1 & 81.0 & 0.279 \\
Sanitation (\%) & 31.3 & 32.3 & 0.647 & 28.1 & 28.2 & 0.967 \\
Evangelicals (\%) & 16.1 & 16.4 & 0.604 & 16.4 & 16.4 & 0.988 \\
Pentec. incumbent (\%) & 13.7 &  8.7 & 0.031 & 12.5 & 12.6 & 0.971 \\
Treated cohort size & 401.9 & 395.1 & 0.896 & 381.9 & 397.9 & 0.779 \\
Placebo cohort size & 413.7 & 407.4 & 0.907 & 392.9 & 411.4 & 0.750 \\
\hline
\multicolumn{7}{l}{\textit{Outcomes}} \\
Birth rate (teen cohort) &  8.57 &  7.70 & 0.029 &  9.09 &  7.57 & 0.005 \\
Birth rate (adult cohort) & 88.78 & 85.95 & 0.128 & 92.15 & 87.26 & 0.053 \\
Sex education taught (\%) & 73.1 & 71.4 & 0.545 & 74.6 & 69.9 & 0.162 \\
Principal $<$2 years (\%) & 51.2 & 47.0 & 0.114 & 51.7 & 50.8 & 0.788 \\
Abortions (per 1,000) &  1.02 &  1.23 & 0.250 &  0.99 &  1.24 & 0.275 \\
Fetal deaths (per 1,000) &  0.49 &  0.32 & 0.127 &  0.51 &  0.40 & 0.474 \\
Syphilis (per 1,000) &  0.42 &  0.15 & 0.077 &  0.30 &  0.17 & 0.211 \\
HIV (per 1,000) &  0.03 &  0.03 & 0.937 &  0.01 &  0.02 & 0.167 \\
\hline
N (muni-cycles) & 342 & 380 & & 216 & 214 & \\
N (municipalities) & 306 & 365 & & 206 & 208 & \\
\hline\hline
\end{tabular}

  \begin{minipage}{\textwidth} 
{\footnotesize \textbf{Notes:} This table presents summary statistics for the full sample (left panel) and the sample within the optimal bandwidth of 0.15 for the main specification (right panel). Columns (1) and (4) show means for municipality-cycles where Pentecostal candidates won; columns (2) and (5) show mean for cases when they lost; columns (3) and (6) present p values for the test of no difference. Birth rates are measured during years 2-5 after the election and expressed per 1,000 girls.}
\end{minipage}
\end{table}

\newpage

\begin{table}[ht!]
    \centering
    \caption{Effect of Pentecostal Mayors on Birth Rate}
\begin{tabular}{lcccccc}
\hline\hline
 & (1) & (2) & (3) & (4) & (5) & (6) \\
\hline
\multicolumn{7}{l}{\textit{Panel A: Treatment Cohort (Ages 9--10 at Election)}} \\
[0.5em]
Birth Rate &  3.52*** &  3.12*** &  3.07*** &  3.01*** &  2.86** &  2.97*** \\
  & (1.153) & (0.964) & (0.977) & (0.911) & (1.106) & (0.909) \\
Mean dep. var. &  8.26 &  8.32 &  8.35 &  8.34 &  8.21 &  8.39 \\
Effective N &   481 &   453 &   443 &   430 &   526 &   419 \\
Robust p-value & 0.001 & 0.001 & 0.001 & 0.001 & 0.013 & 0.001 \\
Bandwidth & 0.175 & 0.164 & 0.159 & 0.150 & 0.210 & 0.145 \\
[0.75em]
\multicolumn{7}{l}{\textit{Panel B: Placebo Cohort (Ages 16--17 at Election)}} \\
[0.5em]
Birth Rate &  5.74 & -1.45 & -2.36 &  2.38 &  1.32 &  2.75 \\
  & (5.822) & (4.612) & (4.610) & (3.335) & (4.502) & (3.311) \\
Mean dep. var. & 89.93 & 89.04 & 88.73 & 89.69 & 89.30 & 89.99 \\
Effective N &   410 &   326 &   318 &   399 &   482 &   402 \\
Robust p-value & 0.555 & 0.487 & 0.365 & 0.620 & 0.879 & 0.515 \\
Bandwidth & 0.140 & 0.107 & 0.102 & 0.136 & 0.175 & 0.138 \\
[0.5em]
\hline
\multicolumn{7}{l}{\textit{Specification}} \\
[0.25em]
State FE & N & Y & Y & Y & Y & Y \\
Cycle FE & N & N & Y & Y & Y & Y \\
Lagged Dep. Var. & N & N & N & Y & Y & Y \\
Polynomial Order & 1 & 1 & 1 & 1 & 2 & 1 \\
Kernel & Tri. & Tri. & Tri. & Tri. & Tri. & Epan. \\
\hline\hline
\end{tabular}

    \label{tab:births}
  \begin{minipage}{\textwidth} 
{\footnotesize \textbf{Notes:}  RDD estimates of Pentecostal electoral victory on birth rates per 1,000 girls in specific cohorts. The outcome is measured as births per 1,000 girls aged 9 or 10 during election year, averaged over years 2-5 post-election. The running variable is the margin of victory in 2008, 2012, or 2016 municipal elections, with positive values indicating Pentecostal victories. Standard errors in parentheses, clustered at the municipality level. Specifications vary by kernel type (Triangular, Epanechnikov), polynomial order (1st/2nd), and included controls. Column (1) uses no controls, column (2) includes state fixed effects; column (3) adds electoral cycle fixed effects; column (4) adds lagged birth rates; column (5) uses a quadratic polynomial; column (6) uses a Epanechnikov kernel. Optimal bandwidth selected following \citet{calonico2014robust}. Robust p-values use bias-corrected inference. Effective N is the sample size within the optimal bandwidth on both sides of the threshold.}
\end{minipage}
\end{table}

\newpage

\begin{table}[h]
    \centering
        \caption{Effect of Pentecostal Mayor on Sexual Education in Schools} 
\begin{tabular}{lccccc}
\hline\hline
 & (1) & (2) & (3) & (4) & (5) \\
\hline
Schools: &  \multicolumn{4}{c}{Municipal} & State \\
\cmidrule(r){2-5} \cmidrule(r){6-6}
Sex Ed & -0.063 & -0.120** & -0.125** & -0.150** & 0.034 \\
  & (0.072) & (0.063) & (0.062) & (0.078) & (0.064) \\
\hline
State FE & N & Y & Y & Y & Y \\
Cycle FE & N & N & Y & Y & Y \\
Order polyn. & 1 & 1 & 1 & 2 & 1 \\
Mean dep. var. &  0.77 &  0.76 &  0.77 &  0.77 &  0.81 \\
Effective N. Obs. &  1131 &   896 &   924 &  1216 &   789 \\
Robust p-value &   0.247 &   0.023 &   0.016 &   0.039 &   0.563 \\
Optimal Bandwidth & 0.131 & 0.096 & 0.106 & 0.158 & 0.132 \\
\hline\hline
\end{tabular}
 
   \label{tab:sex_educ_muni}
                     \begin{minipage}{\textwidth} 
{\footnotesize \textbf{Notes:} RDD estimates of Pentecostal electoral victory on the presence of sexual education courses in schools. The running variable is the margin of victory in 2008, 2012, or 2016 municipal elections, with positive values indicating Pentecostal victories. Standard errors in parentheses, clustered at the municipality level. Specifications vary by polynomial order (1st/2nd), and included controls, and whether schools are controlled by the municipality or the state. Column (1) has no state fixed effects; column (2) adds state fixed effects; column (3) adds electoral cycle; column (4) uses a quadratic polynomial; column (5) uses state schools. Optimal bandwidth selected following \citet{calonico2014robust}. Robust p-values use bias-corrected inference. Effective N is the sample size within the optimal bandwidth on both sides of the threshold.}
\end{minipage}
\end{table}

\newpage

\begin{table}[h]
    \centering
        \caption{Effect of Pentecostal Mayor on Other Courses in Schools} 
\begin{tabular}{lcccccc}
\hline\hline
 & (1) & (2) & (3) & (4) & (5) & (6) \\
\hline
Subjects & Environment & Inequality & Racism & Sexism & Violence & Drugs \\
\hline
Municipal Schools &  0.02 & -0.06 &  0.03 & -0.02 & -0.11** & -0.20*** \\
  & (0.042) & (0.074) & (0.055) & (0.078) & (0.059) & (0.060) \\
\hline
Mean dep. var. &  0.92 &  0.52 &  0.65 &  0.32 &  0.71 &  0.78 \\
Effective N. Obs. &   801 &  1069 &  1207 &  1093 &   950 &   855 \\
Robust p-value &   0.573 &   0.250 &   0.598 &   0.737 &   0.034 &   0.000 \\
Optimal Bandwidth & 0.085 & 0.122 & 0.160 & 0.128 & 0.109 & 0.093 \\
\hline\hline
State Schools &  0.03 &  0.01 &  0.02 &  0.03 & -0.11* &  0.01 \\
  & (0.044) & (0.063) & (0.057) & (0.066) & (0.060) & (0.050) \\
\hline
Mean dep. var. &  0.93 &  0.59 &  0.78 &  0.49 &  0.73 &  0.84 \\
Effective N. Obs. &   651 &   759 &   772 &   707 &   875 &   866 \\
Robust p-value &   0.321 &   0.869 &   0.601 &   0.454 &   0.088 &   0.922 \\
Optimal Bandwidth & 0.104 & 0.127 & 0.129 & 0.111 & 0.154 & 0.149 \\
\hline\hline
\end{tabular}
 
   \label{tab:other_courses}
                 \begin{minipage}{\textwidth} 
{\footnotesize \textbf{Notes:} RDD estimates of Pentecostal electoral victory on the presence of various subjects in municipal and state schools. The running variable is the margin of victory in 2008, 2012, or 2016 municipal elections, with positive values indicating Pentecostal victories. Standard errors in parentheses, clustered at the municipality level. All estimates use degree 1 polynomial fit, triangular kernel and include state and electoral cycle fixed effects. Optimal bandwidth selected following \citet{calonico2014robust}. Robust p-values use bias-corrected inference. Effective N is the sample size within the optimal bandwidth on both sides of the threshold.}
\end{minipage}
\end{table}

\newpage

\begin{table}[ht!]
    \centering
    \small
        \caption{Effect of Pentecostal Mayors on Contraceptive Availability in Municipal Health Centers}
\begin{tabular}{lcccccc}
\hline\hline
 & (1) & (2) & (3) & (4) & (5) & (6)  \\
\hline
\multirow{2}{*}{Type}  & \multirow{2}{*}{Oral} & \multirow{2}{*}{Injectable} & Male  & Female & \multirow{2}{*}{IUD} & Morning \\
  &  &  &  condom & condom &  & After \\
\hline
\hline
 & -0.04 & -0.03 &  0.01 & -0.11 &  0.02 &  0.01 \\
  & (0.057) & (0.072) & (0.016) & (0.096) & (0.081) & (0.087) \\
\hline
Mean dep. var. &  0.80 &  0.70 &  0.98 &  0.74 &  0.30 &  0.39 \\
Effective N. Obs. &   353 &   344 &   331 &   247 &   351 &   406 \\
Robust p-value &   0.599 &   0.672 &   0.585 &   0.139 &   0.769 &   0.927 \\
Optimal Bandwidth & 0.161 & 0.152 & 0.176 & 0.117 & 0.196 & 0.188 \\
\hline\hline
\end{tabular}

    \label{tab:contraceptives}
                \begin{minipage}{1\textwidth} 
{\footnotesize \textbf{Notes:} RDD estimates of Pentecostal electoral victory on availability of different types of contraceptive in municipal health centers. The running variable is the margin of victory in 2008, 2012, or 2016 municipal elections, with positive values indicating Pentecostal victories. Standard errors in parentheses, clustered at the municipality level. Oral includes either oral Noretisterone or Ethinylestradiol plus Levonorgestrel; Injectable includes injectable Norethisterone Enanthate plus Estradiol or Medroxyprogesterone acetate. All estimates use degree 1 polynomial fit, triangular kernel and include state and electoral cycle fixed effects. Optimal bandwidth selected following \citet{calonico2014robust}. Robust p-values use bias-corrected inference. Effective N is the sample size within the optimal bandwidth on both sides of the threshold.}
\end{minipage}
\end{table}

\clearpage
\appendix
\setcounter{table}{0}
\renewcommand{\thetable}{A\arabic{table}}
\setcounter{figure}{0}
\renewcommand{\thefigure}{A\arabic{figure}}

\section{Additional Data Sources}\label{app:data}

\textbf{Primary care staffing.} We use administrative records from the National Registry of Health Establishments (CNES) to measure staffing levels in the \textit{Estrat\'{e}gia Sa\'{u}de da Fam\'{i}lia} (ESF), Brazil's primary care program staffed by municipal employees. We estimate RDD effects on three categories of ESF professionals per 1,000 population: community health workers, doctors, and nurses.

\textbf{Social assistance.} We use administrative data from the annual \textit{Censo SUAS} (2012--2017) to examine social assistance services delivered through municipal \textit{Centros de Refer\^{e}ncia de Assist\^{e}ncia Social} (CRAS). Outcomes include youth programming availability, number of families receiving ongoing attention (\textit{Prote\c{c}\~{a}o e Atendimento Integral \`{a} Fam\'{i}lia}) and articulation with health or education services.

\textbf{Religious infrastructure.} We construct a municipality-year panel of church openings and closings using Brazil's national business registry (CNPJ), classifying entities as Pentecostal or Catholic based on their legal name. We aggregate net entry (openings minus closings) per 10,000 population across years 1--4 of each mayoral term.

\textbf{Civil marriages.} We use data from IBGE's \textit{Estat\'{i}sticas do Registro Civil}, which records the universe of civil marriages by municipality, year, and age group of the bride. We construct marriage rates per 1,000 girls for three age brackets (under 15, 15--19, and 20--24) and the average age at marriage, aggregated across the mayoral term.

\textbf{Municipal expenditures.} We use SICONFI fiscal data to compute the share of total municipal expenditure allocated to health, education, social assistance, and culture, averaged across years 1--4 of each mayoral term.

\begin{table}[ht!]
    \centering
    \small
        \caption{Religious Signals among Municipal Candidates by Party}
\begin{tabular}{lcccc}
\hline\hline
 & Religious & Religious & Either & N \\
 & Occupation & Alias & Signal & Candidates \\
\hline
PRB & 0.575 & 1.846 & 2.074 & 24,156 \\
PSC & 0.465 & 1.765 & 1.897 & 29,470 \\
Other Right-Wing & 0.138 & 0.608 & 0.655 & 185,042 \\
PT & 0.081 & 0.343 & 0.377 & 80,395 \\
PMDB & 0.090 & 0.387 & 0.410 & 91,919 \\
PSDB & 0.135 & 0.502 & 0.545 & 71,945 \\
Other Parties & 0.147 & 0.776 & 0.834 & 883,542 \\
\hline\hline
\end{tabular}

    \label{tab:religion}
                \begin{minipage}{\textwidth}
{\footnotesize \textbf{Notes:} Share (\%) of municipal candidates displaying religious signals, by party family. Sample includes all candidates for mayor, vice-mayor, and city council in the 2008, 2012, and 2016 elections. ``Religious Occupation'' indicates the candidate listed a religious ministry as their official profession, excluding candidates whose campaign name includes ``Padre'' (Catholic priest). ``Religious Alias'' indicates the candidate's campaign name includes a Pentecostal or evangelical title such as \textit{Pastor}, \textit{Bispo}, \textit{Mission\'{a}rio}, \textit{Ap\'{o}stolo}, \textit{Irm\~{a}o} or \textit{Reverendo}. ``Either Signal'' is the union of both indicators. ``Other Right-Wing'' includes PSDC, PR, PRP, PP, and DEM, classified as right-wing (mean score $\geq$ 5) based on the expert survey in \citet{tarouco2015partidos}.}
\end{minipage}
\end{table}

\begin{table}[ht!]
    \centering
    \small
        \caption{Evangelical Federal Deputies by Party Family}
\begin{tabular}{lcccccc}
\hline\hline
 & \multicolumn{3}{c}{Gomes (2021)} & \multicolumn{3}{c}{Lacerda (2018)} \\
 & \multicolumn{3}{c}{Legislatures 2007--2019} & \multicolumn{3}{c}{Legislature 2015--2019} \\
\cmidrule(lr){2-4} \cmidrule(lr){5-7}
 & Evang. & Total & Share & Evang. & Total & Share \\
 & Dep. & Dep. & (\%) & Dep. & Dep. & (\%) \\
\hline
PRB & 25 & 37 & 67.6 & 15 & 21 & 71.4 \\
PSC & 29 & 40 & 72.5 & 9 & 13 & 69.2 \\
Other Right-Wing & 47 & 454 & 10.4 & 10 & 99 & 10.1 \\
PT & 12 & 193 & 6.2 & 1 & 68 & 1.5 \\
PMDB & 25 & 207 & 12.1 & 4 & 65 & 6.2 \\
PSDB & 15 & 142 & 10.6 & 5 & 54 & 9.3 \\
Other Parties & 70 & 698 & 10.0 & 24 & 193 & 12.4 \\
\hline
All Parties & 223 & 1771 & 12.6 & 68 & 513 & 13.3 \\
\hline\hline
\end{tabular}

    \label{tab:congress}
                \begin{minipage}{\textwidth}
{\footnotesize \textbf{Notes:} Number and share of evangelical federal deputies (C\^{a}mara dos Deputados) by party family, from two independent sources. Left panel: \citet{gomes2021deputados} identifies evangelical deputies across the 53rd--55th legislatures (elected 2006, 2010, 2014); total deputy counts are from the C\^{a}mara dos Deputados open data API and include all who served (including \textit{suplentes}). Right panel: \citet{lacerda2018assessing} identifies evangelical candidates among elected federal deputies in 2014. ``Other Right-Wing'' includes PP, DEM, PR, PSDC, and PRP, classified as right-wing (mean score $\geq$ 5) based on the expert survey in \citet{tarouco2015partidos}. Party names are harmonized to account for mergers (PFL$\to$DEM, PL$\to$PR).}
\end{minipage}
\end{table}

\begin{table}[ht!]
    \centering
    \small
        \caption{Sample Representativeness}
\begin{tabular}{lccccc}
\hline\hline
 & \multicolumn{2}{c}{Other Municipalities} & \multicolumn{2}{c}{RDD Sample} & \\
 & Mean & SD & Mean & SD & p-value \\
\hline
Log$_{10}$ Population & 4.086 & 0.507 & 4.111 & 0.438 & 0.206 \\
Income per Capita (1,000 BRL) & 12.971 & 14.052 & 11.292 & 14.784 & 0.008 \\
Urban Share & 0.640 & 0.221 & 0.625 & 0.214 & 0.102 \\
Literacy Rate & 0.841 & 0.098 & 0.814 & 0.096 & 0.000 \\
Sanitation & 0.365 & 0.301 & 0.325 & 0.289 & 0.001 \\
Share Evangelical & 0.172 & 0.096 & 0.165 & 0.087 & 0.078 \\
\hline
N Municipalities & \multicolumn{2}{c}{4952} & \multicolumn{2}{c}{612} & \\
\hline\hline
\end{tabular}

    \label{tab:representativeness}
                \begin{minipage}{\textwidth}
{\footnotesize \textbf{Notes:} Comparison of municipality characteristics between non-sample Brazilian municipalities and the RDD analysis sample. ``Other Municipalities'' includes all municipalities with data in the 2010 IBGE Census that are not part of the RDD sample. ``RDD Sample'' includes municipalities where exactly one PRB or PSC candidate competed in the top two positions in a single-round mayoral election (2008, 2012, or 2016), excluding races where the opponent's coalition included PRB or PSC. The $p$-value column reports the significance level from a regression of each characteristic on a sample membership indicator with heteroskedasticity-robust standard errors. Population is in log$_{10}$ scale. Income per capita is in thousands of 2010 BRL. Urban share, literacy, sanitation, and share evangelical are expressed as proportions (0--1).}
\end{minipage}
\end{table}

\begin{table}[ht!]
    \centering
    \small
        \caption{Sensitivity Analysis: Trimming and Extensive Margin}
\begin{tabular}{lccc}
\hline\hline
 & (1) & (2) & (3) \\
\hline
\multicolumn{4}{l}{\textit{Panel A: Teen Birth Rate (per 1,000 girls)}} \\
 & Baseline & Trimmed 1\% & Trimmed 5\% \\
[0.5em]
RD Estimate & 3.005*** & 2.100*** & 1.991*** \\
  & (0.904) & (0.844) & (0.795) \\
\hline
Mean dep. var. & 8.343 & 8.863 & 7.547 \\
Effective N &   434 &   363 &   384 \\
Robust p-value & 0.000 & 0.007 & 0.007 \\
Bandwidth & 0.151 & 0.139 & 0.140 \\
[0.75em]
\multicolumn{4}{l}{\textit{Panel B: Syphilis}} \\
 & Baseline & Extensive & Trimmed 1\% \\
 & (rate per 1,000) & (any cases) & (rate per 1,000) \\
[0.5em]
RD Estimate & 0.514** & 0.070 & 0.169 \\
  & (0.233) & (0.065) & (0.327) \\
\hline
Mean dep. var. & 0.197 & 0.075 & 2.111 \\
Effective N &   312 &   441 &    23 \\
Robust p-value & 0.017 & 0.182 & 0.756 \\
Bandwidth & 0.143 & 0.134 & 0.101 \\
Share zero (treated) & 0.917 & & \\
Share zero (control) & 0.932 & & \\
\hline\hline
\end{tabular}

    \label{tab:sensitivity}
                \begin{minipage}{\textwidth}
{\footnotesize \textbf{Notes:} Panel A reports RDD estimates of the effect of Pentecostal mayors on teenage birth rates (per 1,000 girls aged 9--10 at election, years 2--5 post-election) under different sample restrictions. Column (1): baseline specification with winsorization at the 99th percentile. Column (2): observations above the 99th or below the 1st percentile are dropped. Column (3): observations above the 95th or below the 5th percentile are dropped. Panel B reports sensitivity analyses for syphilis. Column (1): baseline syphilis rate per 1,000 (winsorized at 99th percentile). Column (2): extensive margin---a binary indicator for whether the municipality-cycle recorded any syphilis cases in the treated cohort. Column (3): syphilis rate trimmed at the 1st and 99th percentiles. ``Share zero'' reports the fraction of municipality-cycles with zero syphilis cases among treated and control municipalities within the baseline bandwidth. All specifications use degree 1 polynomial fit, triangular kernel, state and electoral cycle fixed effects, lagged dependent variable control, and optimal bandwidth following \citet{calonico2014robust}. Standard errors in parentheses, clustered at the municipality level. Robust p-values use bias-corrected inference.  $^{*}$ $p<0.10$, $^{**}$ $p<0.05$, $^{***}$ $p<0.01$.}
\end{minipage}
\end{table}

\begin{table}[ht!]
    \centering
    \small
        \caption{Effect of Pentecostal Mayors on ESF Health Worker Staffing}
\begin{tabular}{lccc}
\hline\hline
 & (1) & (2) & (3) \\
\hline
 & Community & Doctors & Nurses \\
 & Health Workers &  &  \\
\hline
  &  0.12 &  0.00 &  0.18* \\
  & (0.111) & (0.042) & (0.104) \\
\hline
Mean dep. var. &  2.87 &  0.56 &  1.00 \\
Effective N. Obs. &   494 &   534 &   359 \\
Robust p-value &   0.320 &   0.941 &   0.051 \\
Optimal Bandwidth & 0.159 & 0.175 & 0.105 \\
\hline\hline
\end{tabular}

    \label{tab:esf_staffing}
                \begin{minipage}{\textwidth}
{\footnotesize \textbf{Notes:} RDD estimates of Pentecostal electoral victory on the number of \textit{Estrat\'{e}gia Sa\'{u}de da Fam\'{i}lia} (ESF) professionals per 1,000 population. The running variable is the margin of victory in municipal elections, with positive values indicating Pentecostal victories. Standard errors in parentheses, clustered at the municipality level. All estimates use degree 1 polynomial fit, triangular kernel and include state and electoral cycle fixed effects, lagged dependent variable, and log population. Optimal bandwidth selected following \citet{calonico2014robust}. Robust p-values use bias-corrected inference.  $^{*}$ $p<0.10$, $^{**}$ $p<0.05$, $^{***}$ $p<0.01$.}
\end{minipage}
\end{table}

\begin{table}[ht!]
    \centering
    \small
        \caption{Effect of Pentecostal Mayors on Church Entry}
\begin{tabular}{lcc}
\hline\hline
 & (1) & (2) \\
\hline
 & Pentecostal & Catholic \\
\hline
  & -0.036 & -0.032 \\
  & (0.067) & (0.234) \\
\hline
Mean dep. var. & 0.034 & 0.279 \\
Effective N. Obs. &   405 &   451 \\
Robust p-value &   0.502 &   0.780 \\
Optimal Bandwidth & 0.166 & 0.185 \\
\hline\hline
\end{tabular}

    \label{tab:churches}
                \begin{minipage}{\textwidth}
{\footnotesize \textbf{Notes:} RDD estimates of Pentecostal electoral victory on net church entry (openings minus closings) per 10,000 population during the mayoral term (years 1--4 after election). Column (1): Pentecostal churches. Column (2): Catholic churches. Church openings and closings are identified from Brazil's national business registry (CNPJ) using economic activity codes (CNAE) that distinguish religious organizations by denomination. The running variable is the margin of victory in municipal elections, with positive values indicating Pentecostal victories. Standard errors in parentheses, clustered at the municipality level. All estimates use degree 1 polynomial fit, triangular kernel and include state and electoral cycle fixed effects and lagged dependent variable. Optimal bandwidth selected following \citet{calonico2014robust}. Robust p-values use bias-corrected inference.  $^{*}$ $p<0.10$, $^{**}$ $p<0.05$, $^{***}$ $p<0.01$.}
\end{minipage}
\end{table}

\begin{table}[ht!]
    \centering
    \small
        \caption{Effect of Pentecostal Mayors on Early Marriage}
\begin{tabular}{lcccc}
\hline\hline
 & (1) & (2) & (3) & (4) \\
\hline
 & Avg. Age & Under 15 & 15--19 & 20--24 \\
 & at Marriage & (per 1,000 girls) & (per 1,000 girls) & (per 1,000 girls) \\
\hline
  & 0.070 & -0.020 & 2.774 & -8.536 \\
  & (0.353) & (0.139) & (6.411) & (6.737) \\
\hline
Mean dep. var. & 29.166 & 0.161 & 65.545 & 108.955 \\
Effective N. Obs. &   398 &   441 &   443 &   561 \\
Robust p-value &   0.683 &   0.746 &   0.756 &   0.179 \\
Optimal Bandwidth & 0.151 & 0.157 & 0.158 & 0.239 \\
\hline\hline
\end{tabular}

    \label{tab:marriages}
                \begin{minipage}{\textwidth}
{\footnotesize \textbf{Notes:} RDD estimates of Pentecostal electoral victory on civil marriage outcomes during the mayoral term (years 1--4 after election). Column (1): average age at marriage, computed as a weighted mean across all female age brackets using bracket midpoints. Columns (2)--(4): marriage rates per 1,000 girls in the corresponding age group (under 15, 15--19, and 20--24). Population denominators are interpolated from the 2000, 2010, and 2022 censuses. Source: IBGE \textit{Estat\'{i}sticas do Registro Civil}, 2004--2020. The running variable is the margin of victory in municipal elections, with positive values indicating Pentecostal victories. Standard errors in parentheses, clustered at the municipality level. All estimates use degree 1 polynomial fit, triangular kernel and include state and electoral cycle fixed effects and lagged dependent variable. Optimal bandwidth selected following \citet{calonico2014robust}. Robust p-values use bias-corrected inference.  $^{*}$ $p<0.10$, $^{**}$ $p<0.05$, $^{***}$ $p<0.01$.}
\end{minipage}
\end{table}

\begin{table}[ht!]
    \centering
    \small
        \caption{Effect of Pentecostal Mayors on Municipal Budget Allocation}
\begin{tabular}{lcccc}
\hline\hline
 & (1) & (2) & (3) & (4) \\
\hline
 & Health & Education & Social & Culture \\
 &  &  & Assistance &  \\
\hline
  & 0.193 & 0.057 & -0.106* & 0.037 \\
  & (0.235) & (0.300) & (0.076) & (0.048) \\
\hline
Mean dep. var. & 7.462 & 10.601 & 1.286 & 0.355 \\
Effective N. Obs. &   499 &   387 &   419 &   513 \\
Robust p-value &   0.422 &   0.830 &   0.088 &   0.528 \\
Optimal Bandwidth & 0.185 & 0.134 & 0.148 & 0.204 \\
\hline\hline
\end{tabular}

    \label{tab:expenditures}
                \begin{minipage}{\textwidth}
{\footnotesize \textbf{Notes:} RDD estimates of Pentecostal electoral victory on the share of total municipal expenditure allocated to each functional category during the mayoral term (years 1--4 after election). Expenditure shares are computed from SICONFI fiscal reports, averaged across years within each mayoral term. The running variable is the margin of victory in municipal elections, with positive values indicating Pentecostal victories. Standard errors in parentheses, clustered at the municipality level. All estimates use degree 1 polynomial fit, triangular kernel and include state and electoral cycle fixed effects and lagged dependent variable. Optimal bandwidth selected following \citet{calonico2014robust}. Robust p-values use bias-corrected inference.  $^{*}$ $p<0.10$, $^{**}$ $p<0.05$, $^{***}$ $p<0.01$.}
\end{minipage}
\end{table}

\begin{table}[ht!]
    \centering
    \small
        \caption{Effect of Pentecostal Mayors on CRAS Social Assistance Services}
\begin{tabular}{lcccc}
\hline\hline
 & (1) & (2) & (3) & (4) \\
\hline
 & SCFV & Families & Articulation & Articulation \\
 & Adolescents & in PAIF & with Health & with Education \\
 & 15--17 & (per 1,000 pop.) & Services & Services \\
\hline
  & 0.002 & -3.434 & -0.001 & 0.014** \\
  & (0.062) & (5.536) & (0.008) & (0.007) \\
\hline
Mean dep. var. & 0.855 & 22.318 & 0.994 & 0.991 \\
Effective N. Obs. &   308 &   489 &   347 &   277 \\
Robust p-value &   0.989 &   0.609 &   0.991 &   0.043 \\
Optimal Bandwidth & 0.157 & 0.155 & 0.176 & 0.129 \\
\hline\hline
\end{tabular}

    \label{tab:cras}
                \begin{minipage}{\textwidth}
{\footnotesize \textbf{Notes:} RDD estimates of Pentecostal electoral victory on municipal social assistance services provided through \textit{Centros de Refer\^{e}ncia de Assist\^{e}ncia Social} (CRAS). Column (1): share of CRAS directly offering \textit{Servi\c{c}o de Conviv\^{e}ncia e Fortalecimento de V\'{i}nculos} (SCFV) for adolescents aged 15--17. Column (2): number of families in PAIF follow-up (\textit{Prote\c{c}\~{a}o e Atendimento Integral \`{a} Fam\'{i}lia}) per 1,000 population. Columns (3)--(4): share of CRAS reporting any articulation with health or education services, respectively. The running variable is the margin of victory in municipal elections, with positive values indicating Pentecostal victories. Standard errors in parentheses, clustered at the municipality level. All estimates use degree 1 polynomial fit, triangular kernel and include state and electoral cycle fixed effects, lagged dependent variable, and log population. Optimal bandwidth selected following \citet{calonico2014robust}. Robust p-values use bias-corrected inference.  $^{*}$ $p<0.10$, $^{**}$ $p<0.05$, $^{***}$ $p<0.01$.}
\end{minipage}
\end{table}

\begin{table}[ht!]
    \centering
    \small
        \caption{Effect of Pentecostal Mayors on HPV Vaccination Coverage}
\begin{tabular}{lcccc}
\toprule
 & RD Estimate & SE & p-value & N \\
\midrule
\multicolumn{5}{l}{\textit{Panel A: Coverage by age at end of term}} \\
[4pt]
Age 9 & -18.30*** &  7.24 & 0.006 & 311 \\
Age 10 & -19.58* & 11.31 & 0.063 & 335 \\
Age 11 & -18.29** &  9.00 & 0.027 & 295 \\
Age 12 & -5.47 &  7.72 & 0.430 & 338 \\
Age 13 & -1.92 & 11.17 & 0.710 & 334 \\
Age 14 &  0.51 & 13.79 & 0.863 & 332 \\
[6pt]
\multicolumn{5}{l}{\textit{Panel B: Robustness of ages 9--11 average}} \\
[4pt]
Baseline & -18.98*** &  7.13 & 0.004 & 317 \\
Bandwidth $0.5h$ & -27.48*** &  8.51 & 0.001 & 172 \\
Bandwidth $1.5h$ & -19.53*** &  5.48 & 0.000 & 398 \\
Quadratic & -22.41*** &  8.57 & 0.009 & 406 \\
Uniform kernel & -18.32** &  7.88 & 0.020 & 241 \\
No covariates & -22.88*** &  7.81 & 0.001 & 345 \\
\bottomrule
\end{tabular}

    \label{tab:hpv}
                \begin{minipage}{\textwidth}
{\footnotesize \textbf{Notes:} Panel A reports RDD estimates for cumulative HPV vaccination coverage (percentage points) by single year of age, measured at the last year of each mayoral term (2016 for the 2012 cycle, 2020 for the 2016 cycle). Coverage for younger ages reflects mostly vaccination under the current mayor, while coverage for older ages includes doses administered under the previous administration. Panel B reports robustness checks for the average coverage of girls aged 9--11. All specifications use degree 1 polynomial fit and triangular kernel unless otherwise noted, and include state and electoral cycle fixed effects. Optimal bandwidth selected following \citet{calonico2014robust}. Robust p-values use bias-corrected inference. $^{*}$ $p<0.10$, $^{**}$ $p<0.05$, $^{***}$ $p<0.01$.}
\end{minipage}
\end{table}

\begin{figure}[ht!]
    \centering
    \caption{Placebo Cutoff Test: RD Estimates at Artificial Cutoffs}
    \label{fig:placebo_cutoff}
    \includegraphics[width=0.85\textwidth]{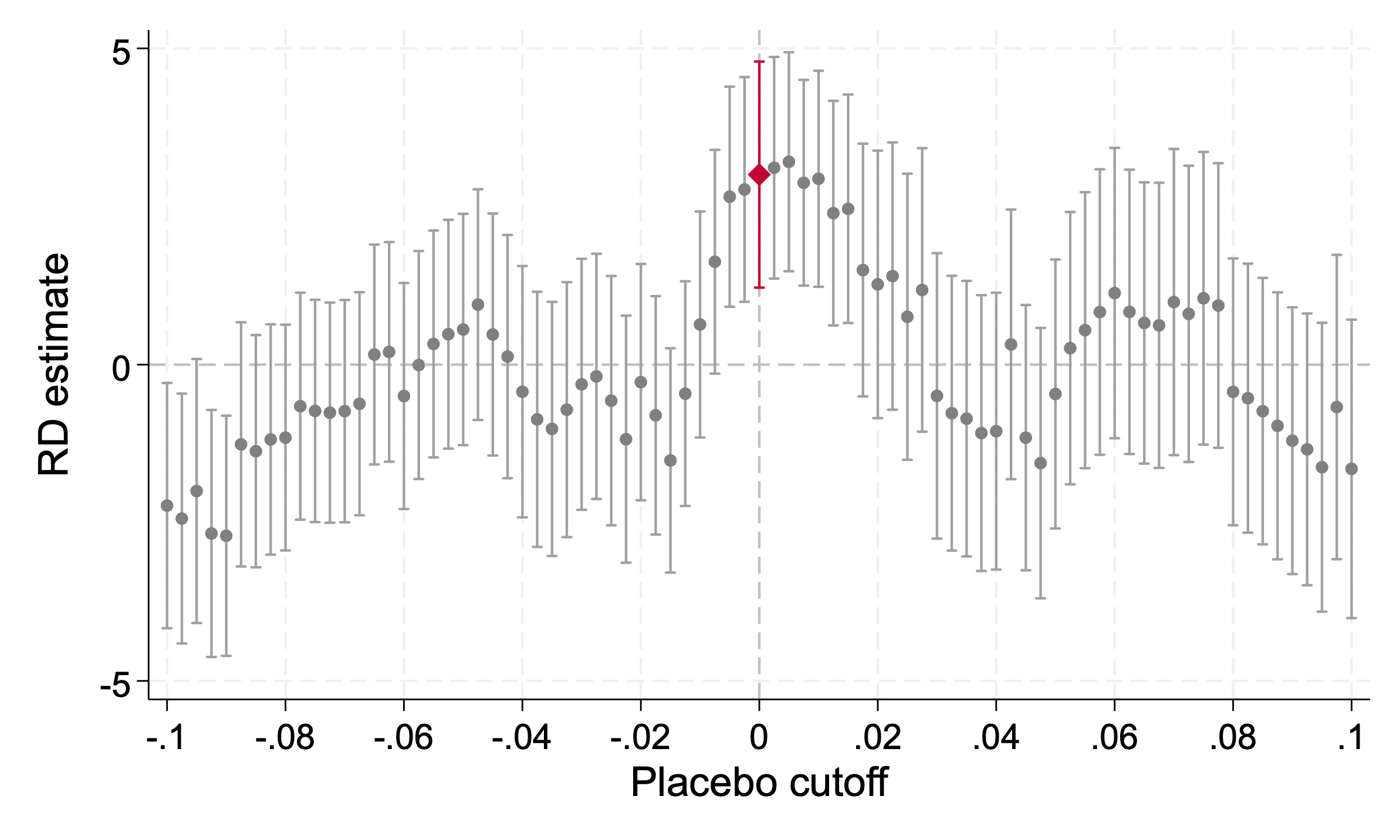}
    \begin{minipage}{0.95\textwidth}
    \vspace{0.2cm}
    \small
    \textbf{Notes:} RD point estimates and 95\% robust confidence intervals for the effect on teenage birth rates, estimated at artificial cutoffs of the running variable ranging from $-0.10$ to $+0.10$ in increments of 0.01. The full sample is used at each cutoff. The true cutoff at zero is highlighted in red. All specifications use the baseline controls (state and cycle fixed effects, lagged dependent variable) with automatic bandwidth selection and bias-corrected robust inference.
    \end{minipage}
\end{figure}

\begin{figure}[ht!]
    \centering
    \caption{Leave-One-Out Sensitivity: Syphilis}
    \label{fig:leave_one_out}
    \includegraphics[width=0.85\textwidth]{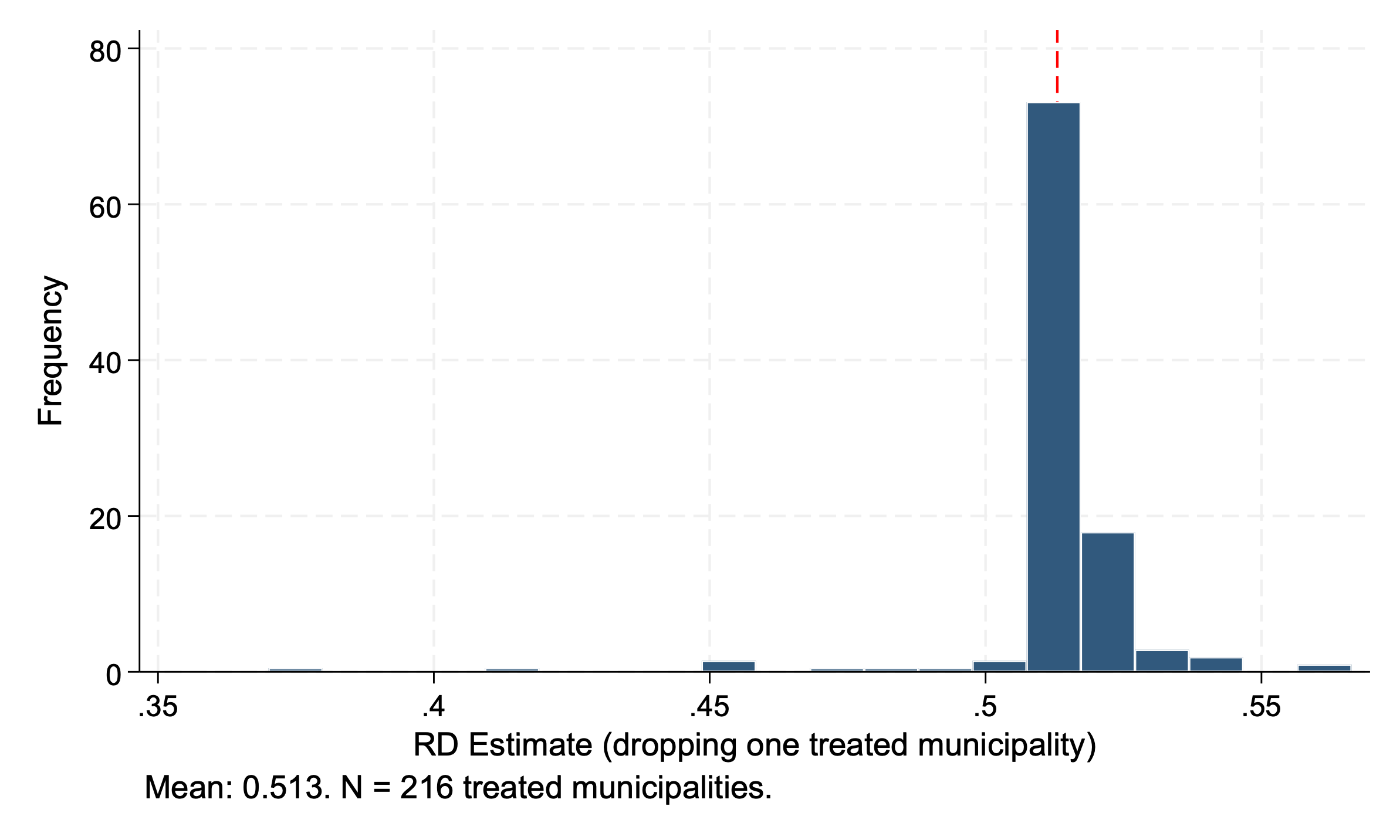}
    \begin{minipage}{0.95\textwidth}
    \vspace{0.2cm}
    \small
    \textbf{Notes:} Distribution of RDD estimates for the effect of Pentecostal mayors on the syphilis rate (per 1,000 girls), obtained by re-estimating the baseline specification while dropping each treated municipality one at a time. The dashed red line indicates the mean estimate across all leave-one-out iterations. The exercise uses the baseline bandwidth and preferred specification (degree 1 polynomial, triangular kernel, state and cycle fixed effects, lagged dependent variable).
    \end{minipage}
\end{figure}

\begin{figure}[ht!]
    \centering
    \caption{Heterogeneity by Baseline Sex Education Provision}
    \label{fig:het_sexed}
    \begin{subfigure}{0.48\textwidth}
        \centering
        \includegraphics[width=\textwidth]{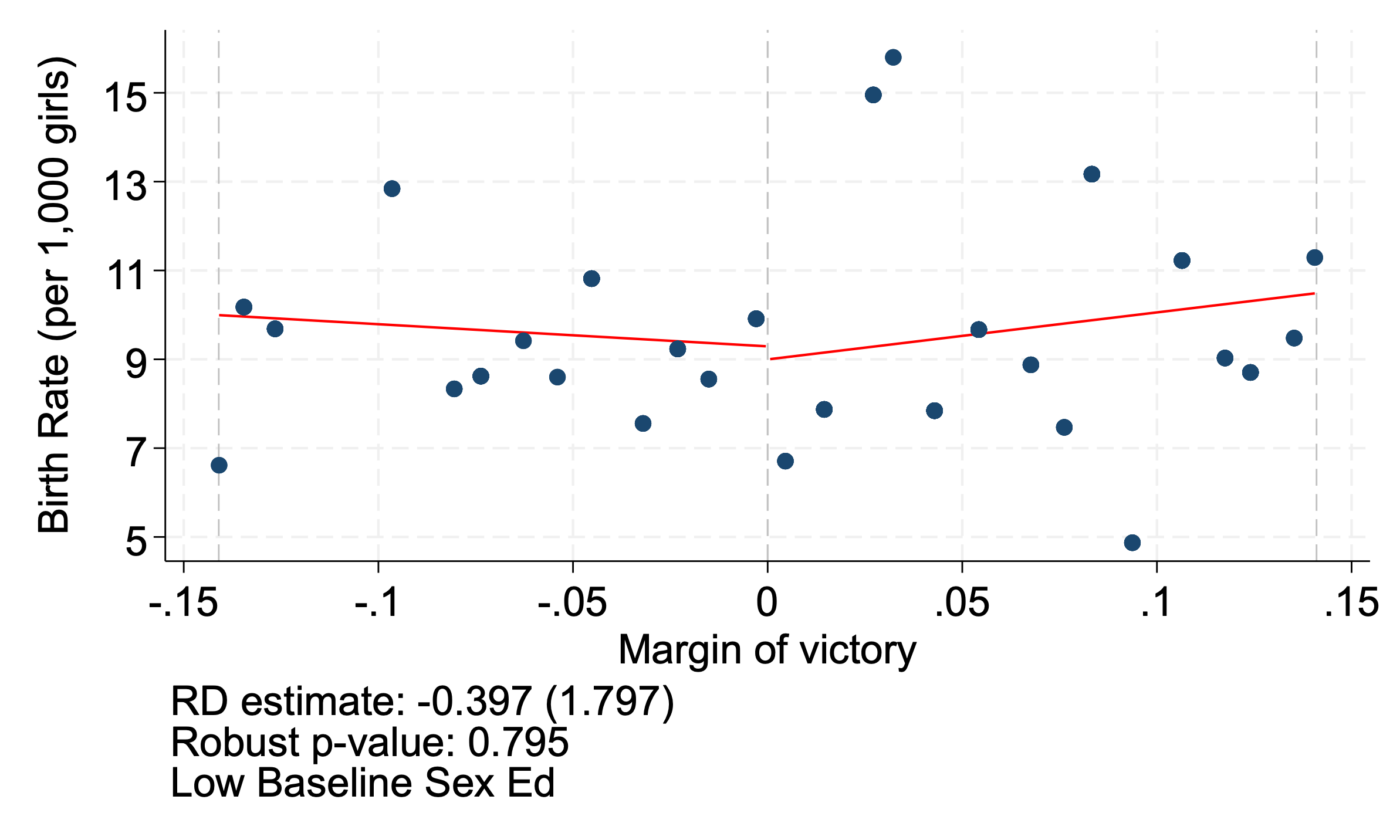}
    \end{subfigure}
    \hfill
    \begin{subfigure}{0.48\textwidth}
        \centering
        \includegraphics[width=\textwidth]{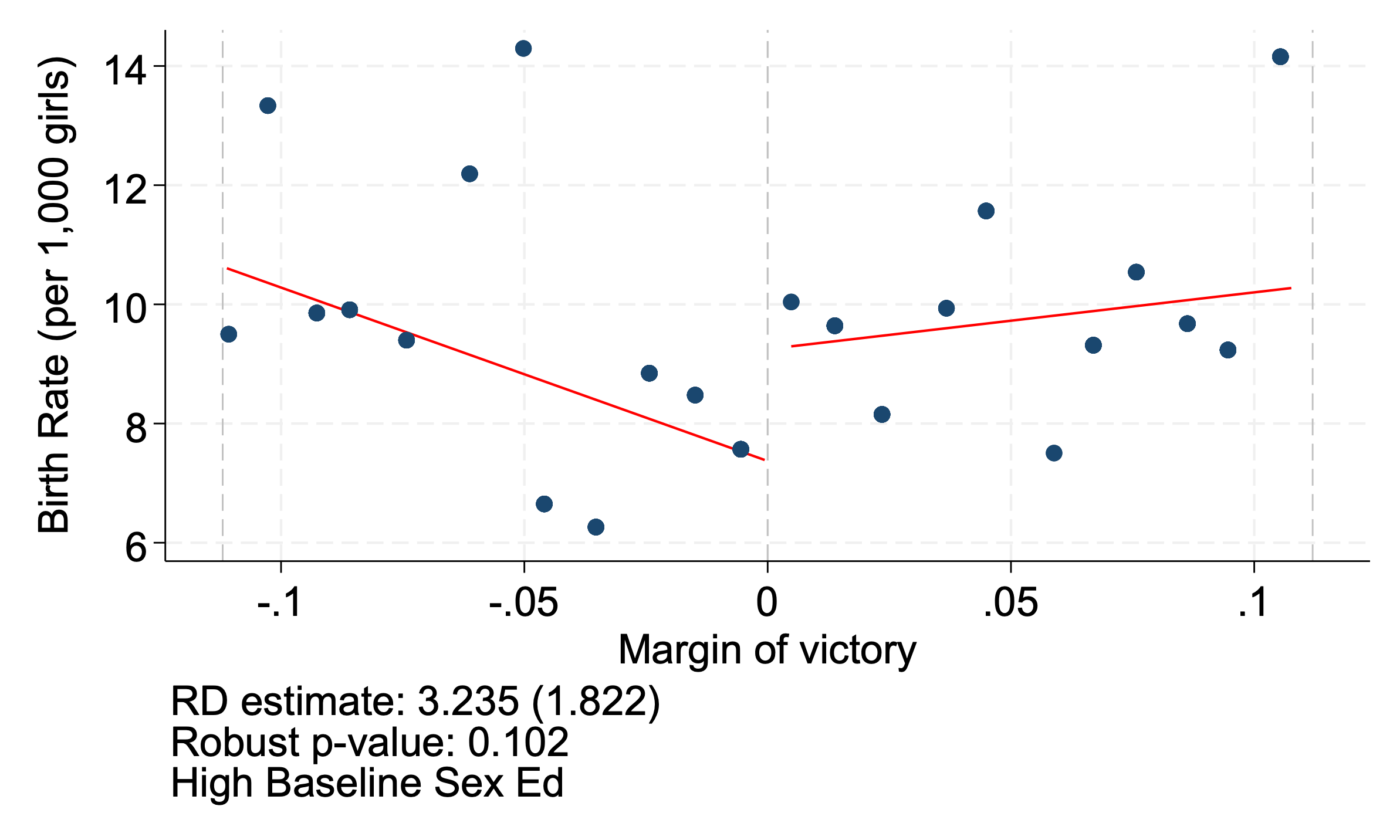}
    \end{subfigure}
    \begin{minipage}{0.95\textwidth}
    \vspace{0.2cm}
    \small
    \textbf{Notes:} RDD estimates of the effect of Pentecostal mayors on teen birth rates, split at the median of the municipality-level share of municipal schools offering sexual education in the previous electoral cycle (\textit{Prova Brasil}). The left panel shows municipalities with below-median baseline sex education provision; the right panel shows municipalities with above-median provision. Sample restricted to the 2012 and 2016 cycles, as no pre-treatment \textit{Prova Brasil} data are available for the 2008 cycle. All specifications use degree 1 polynomial fit, triangular kernel, and include state and cycle fixed effects, lagged dependent variable, and municipality controls.
    \end{minipage}
\end{figure}

\begin{figure}[ht!]
    \centering
    \caption{Heterogeneity by Local Evangelical Population Share}
    \label{fig:het_evangelical}
    \begin{subfigure}{0.48\textwidth}
        \centering
        \includegraphics[width=\textwidth]{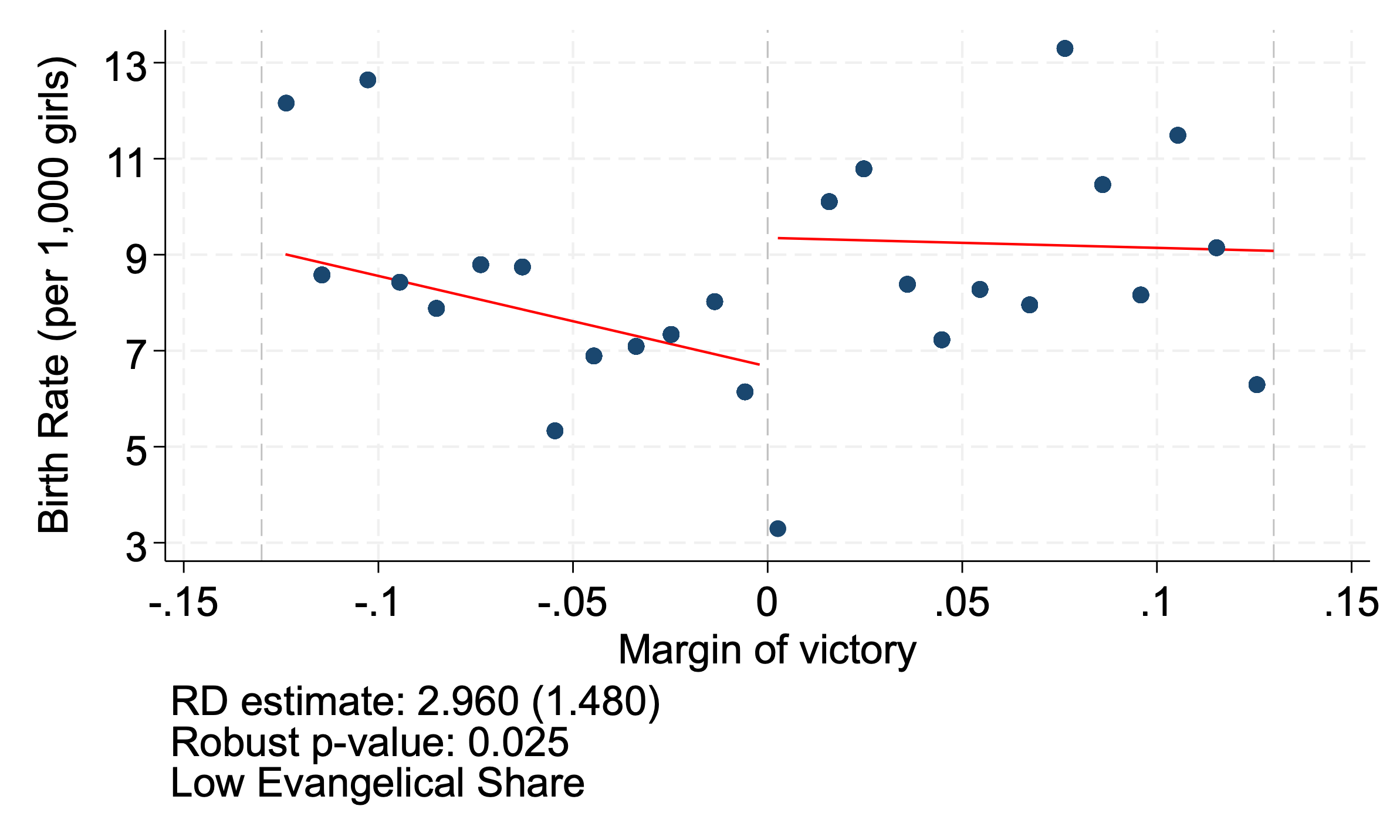}
    \end{subfigure}
    \hfill
    \begin{subfigure}{0.48\textwidth}
        \centering
        \includegraphics[width=\textwidth]{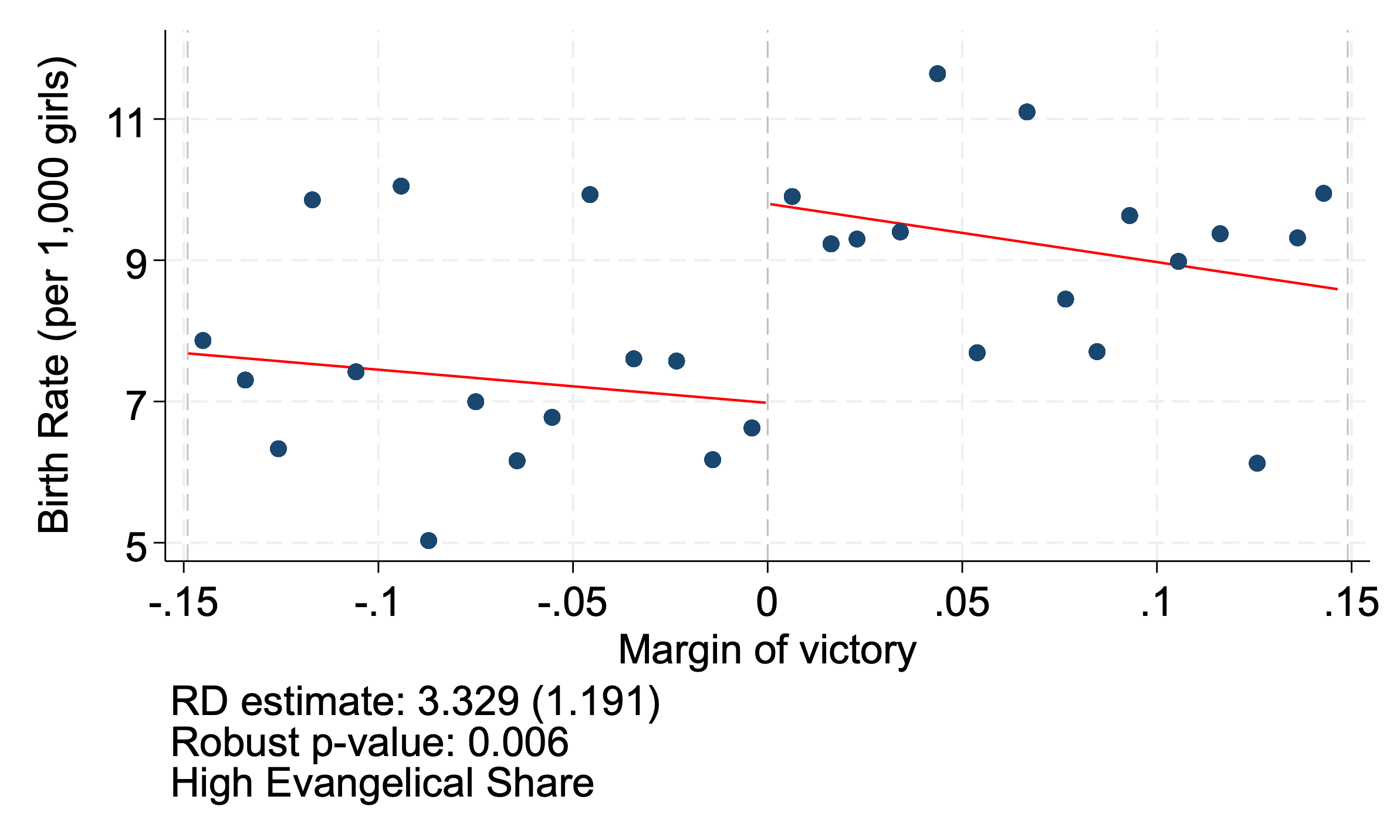}
    \end{subfigure}
    \begin{minipage}{0.95\textwidth}
    \vspace{0.2cm}
    \small
    \textbf{Notes:} RDD estimates of the effect of Pentecostal mayors on teen birth rates, split at the median of the municipal evangelical population share (2010 Census). The left panel shows municipalities with below-median evangelical share; the right panel shows municipalities with above-median share. All specifications use degree 1 polynomial fit, triangular kernel, and include state and cycle fixed effects, lagged dependent variable, and municipality controls.
    \end{minipage}
\end{figure}

\end{document}